\documentclass[]{aa}       
\usepackage{graphicx}
\usepackage{amssymb}
\usepackage{longtable}
\usepackage{supertabular}
\usepackage{natbib} 

\begin{document}

\title{Clouds and red giants interacting with the base of AGN jets}

\author{V. Bosch-Ramon \inst{1} \and
        M. Perucho\inst{2} \and
        M. V. Barkov \inst{3}
    }

\authorrunning{Bosch-Ramon et al.}

\titlerunning{Clouds and red giants interacting with AGN jets}

\institute{Dublin Institute for Advanced Studies, 31 Fitzwilliam Place, Dublin 2, Ireland; valenti@cp.dias.ie
\and
Dept. d'Astronomia i Astrof\'{\i}sica, Universitat de Val\`encia, C/ Dr. Moliner 50, 46100, Burjassot (Val\`encia), Spain; 
Manel.Perucho@uv.es 
\and
Max-Planck-Institut f\"ur Kernphysik, Saupfercheckweg 1, 69117 Heidelberg, Germany; bmv@mpi-hd.mpg.de
}

\offprints{V. Bosch-Ramon, \email{valenti@cp.dias.es}}

\date{Received <date> / Accepted <date>}

\abstract
{Extragalactic jets are formed close to supermassive black-holes in the center of galaxies. Large amounts of gas, dust, and stars cluster in the galaxy nucleus, and interactions between this ambient material and the jet base should be frequent, having dynamical as well as radiative consequences.}  
{This work studies the dynamical interaction of an obstacle, a clump of matter or the atmosphere of an evolved star,
with the innermost region of an extragalactic jet. Jet mass-loading and the high-energy outcome of this interaction are briefly discussed.} 
{Relativistic hydrodynamical simulations with axial symmetry have been carried out for homogeneous and inhomogeneous obstacles inside a relativistic jet. These obstacles may represent a medium inhomogeneity or the disrupted atmosphere of a red giant star.}
{Once inside the jet, an homogeneous obstacle expands and gets disrupted after few dynamical timescales, whereas in the inhomogeneous case, a solid core can smoothen the process, with the obstacle
mass-loss dominated by a dense and narrow tail pointing in the direction of the jet. In either case, matter is expected to accelerate and eventually get incorporated to
the jet. Particles can be accelerated in the interaction region, and produce variable gamma-rays in the ambient matter, magnetic and photon fields.}
{The presence of matter clumps or red giants into the base of an extragalactic jet likely implies significant jet mass-loading and slowing down. Fast flare-like gamma-ray events, and some level of persistent emission, are expected due to these interactions.} 

\maketitle

\section{Introduction} \label{intro}

Jets of active galactic nuclei (AGN) are collimated outflows originated at the core of galaxies, very likely in the vicinity of a  supermassive black hole (SMBH) that accretes matter from
its environment \citep[e.g.][]{beg84}. The inner regions of galaxies contain  large amounts of gas, dust, and stars \citep[e.g.][]{bur70}, so there is plenty of material that could interact
with the AGN jet. About a 1\% of the present stars are evolved objects \citep[e.g.][]{you77} whose external layers, close to the  SMBH, can be tidally distorted \citep[see,
e.g.,][and references therein]{bar10}. Stellar collisions could also release matter to the  environment \citep[e.g.][]{you77a}. In AGN with significant star formation in their cores,
massive stars with strong winds will be also present. In bright AGN, hot gas embedding denser and cooler clouds is expected to surround the  SMBH \citep[the Broad Line Region -BLR-;
e.g.][]{kro81,ree87}. In these bright AGN, X-ray heating can also perturb the external layers  of stars even if tidal forces are negligible \citep[e.g.][]{shu83,pen88}. 
Wandering dark clouds could also be present in the region close to the SMBH, as it has been recently discovered in our own galaxy \citep{gil12}.
It seems therefore
unavoidable that stellar matter 
and medium  inhomogeneities will frequently interact with, and in some cases penetrate into, the AGN innermost jet regions. Entrainement of medium matter leads to jet
mass-loading, invoked by several authors \citep[e.g.][]{blk96,lb02} to explain the deceleration of FRI jets  between pc-scales, where they show relativistic speeds, and kpc-scales, where
their velocities are sub-relativistic as shown by the jet and counter-jet symmetric emission. 

Studies of the dynamics  of the interaction between clouds or red giants (RG) after entering into an AGN jet have been carried out  \citep[e.g.][]{ara10,bar10,bar11}. However, previous
works were of analytical nature, mainly focused on the subsequent high-energy  emission. Other works have studied the global impact of stars or clouds on the jet propagation and content
\citep[e.g.][]{kom94,ste97,cho05,hub06,sut07,jey09}, but numerical studies of the consequences of obstacles being inside the innermost regions of AGN jets are rare.

In this work, we aim at studying numerically the evolution of a dense cloud and the disrupted layers of an RG inside an AGN jet. Homogeneous and power-law (plus a solid core) density profiles
have been adopted, the former to simulate the simplest cloud or RG scenario, and the latter what might be a more realistic situation  for a disrupted RG atmosphere. This can allow a 
characterization of the dynamical evolution of obstacles inside jets more accurate than in previous works, as well as a better understanding of the impact on the jet itself, and the
resulting high-energy radiation. At this stage, we neglect the role of the magnetic field in the jet, accounting only for its ram pressure, and assume an axisymmetric (2D) interaction, with
the star/obstacle at rest in the laboratory frame. Magnetic fields and 3-dimensional (3D) effects will be included in future work.

\section{Physical scenario} \label{phys}

Unlike smooth media, which interacts with the jet through a shear layer preventing direct mixing, medium inhomogeneities like
matter clumps, stars, etc., could penetrate effectively into the jet. The requirement for the obstacle to enter into the jet
is to have a jet perpendicular speed $v_{\rm o}\ga v_{\rm sc}$, where $v_{\rm sc}$ is roughly the shocked obstacle sound
speed, i.e. $v_{\rm sc}\sim c\,\sqrt{\rho_{\rm j}\Gamma_{\rm j}/\rho_{\rm o}}$, $\Gamma_{\rm j}=1/(1-(v_{\rm j}/c)^2)^{1/2}$
the jet Lorentz factor, and $\rho_{\rm j}$ and $\rho_{\rm o}$ the jet and obstacle densities in the reference frame of the
latter. This assumes that the relevant jet pressure comes from bulk motion, which implies that the jet power $L_{\rm j}=\pi
R_{\rm j}^2 (\Gamma_{\rm j}-1)\rho_{\rm j}v_{\rm j}c^2$, where $R_{\rm j}$ is the jet radius. Once the obstacle is inside the
jet, neglecting magnetic field effects, a bow-shaped shock forms in the jet. Another shock with speed $v_{\rm sc}$ crosses
the obstacle, or in the case of an RG, its external layers \citep[e.g.][and references therein]{ara10,bar10}. Being
much faster, the jet is to be much lighter than the obstacle to entrain the latter. Therefore, the obstacle shock crossing
(or dynamical) time $t_{\rm d}\sim R_{\rm o}/v_{\rm sc}$, with $R_{\rm o}$ as the obstacle radius, will be much longer than
the bow-shock formation time, $\sim R_{\rm o}/c$. After few $t_{\rm d}$, the obstacle has been accelerated by the jet up to a
velocity $\sim v_{\rm j}$ \citep[e.g.][]{bla79,bar10,bar11}. It follows that if the time of jet crossing, $t_{\rm j}\sim R_{\rm j}/v_{\rm o}$, is smaller than $t_{\rm d}$, the obstacle will be able to leave the jet; otherwise, it will be
dragged by the jet and presumably disrupted and mixed with its matter.  

For obstacles remaining long enough in the jet, and to have a dynamically strong impact on the jet, the
obstacles have to affect significantly the total jet mass, momentum and energy fluxes. In the present  context, it can mean also that
enough mass will be entrained to enforce significant jet deceleration and energy dissipation, i.e. the obstacle
mass-injection flux $\dot{M}\sim \dot{M}_{\rm cr}=L_{\rm j}/(\Gamma_{\rm j}-1)\,c^2$ \citep{hub06}. In addition, $t_{\rm j}$
should be significantly longer than $t_{\rm d}$, to allow the jet to drag the whole obstacle matter. The value of $\dot{M}$
can be estimated as $\sim N_{\rm o}\,\chi^2\,M_{\rm o}/4t_{\rm j}(z_{\rm o})$, where $N_{\rm o}$ is the number of obstacles
(clouds, stars, etc.) within a sphere of radius $z_{\rm o}$, the jet characteristic height, $\chi=R_{\rm j}(z_{\rm o})/z_{\rm
o}\sim 0.1$, and $M_{\rm o}=4\pi\,R_{\rm o}^3\,\rho_{\rm o}/3$ the detachable obstacle mass. The value of $z_{\rm o}$ depends
strongly on the scenario under study. It might be the size of the BLR \citep{ara10}, a region in which SMBH tidal forces
affect the RG atmosphere \citep{bar10}, or the distance at which accretion disk X-rays can affect the star external layers.
Note that for $t_{\rm d}\ll t_{\rm j}$, the obstacles will initially affect just the outermost jet shell, but eventually a
strong shear layer will develop reaching the jet core. For $t_{\rm d}\la t_{\rm j}$, the obstacle mass will be more
distributed over the jet cross section, likely speeding up the process of mass-loading and energy dissipation (e.g.
emission). Note that here only sudden mass-loss by shocked obstacles is treated; continuous mass-loss via, e.g., stellar
winds is not considered (see \citealt{ara12}; Perucho et al., in preparation). A significant dynamical impact will likely have its observable counterpart in the form of non-thermal radiation. 
However, we remark also that, even when providing little mass-loading, the occasional penetration of dense
and large obstacles within the jet could have observable consequences \citep{ara10,bar10,bar11}.

\begin{figure}[!h]
   \centering
\includegraphics[clip,angle=0,width=\columnwidth]{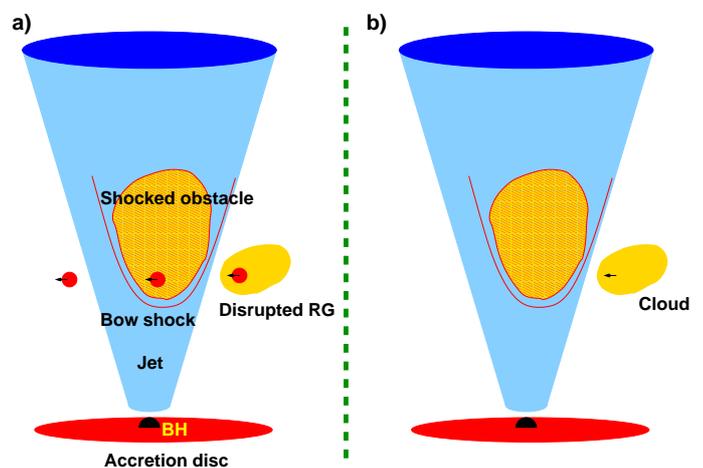}
\caption{Sketch of the two scenarios contemplated here. 
{\bf a)} Penetration of an RG, with the external layers detached (to different degrees)
due to gravitational disruption, into the jet. 
{\bf b)} Penetration of a massive clump of matter into the jet.}
\label{fig1}
\end{figure}

Figure~\ref{fig1} shows a sketch of the scenarios simulated here. The simulations account for the phase when the obstacle is
well inside the innermost region of an AGN jet. Two types of obstacles have been considered. The first type consists on an
homogeneous gas sphere, which may represent a dense cloud in the central AGN regions (e.g. from the BLR), or a homogeneous
bulk of material from the external RG layers detached by tidal SMBH forces. The second case considers a power-law profile
with a radial dependence $\propto R^{-2}$ for the density (plus a solid core), which may represent the RG external layers
more realistically, or just slightly affected by tidal forces\footnote{It is worth noting that the power-law case may also
correspond to an RG with its wind confined by external pressure outside the jet, forming a bubble that would be brought by the
RG into the jet.}.

\section{Simulations}\label{sim}

We have performed three numerical simulations. In simulation~1 (S1), we have calculated the interaction between a homogeneous spherical obstacle and the jet; in simulation~2 (S2) we have
introduced a constant density core plus a power-law profile for the radial variation of the density out of this core ($R^{-2}$, see Section~\ref{phys}); in simulation~3 (S3) we have
reproduced S2 with double resolution. The three simulations have 2D axial symmetry and have been carried out in cylindrical coordinates. This is justified by the symmetric nature of the
problem solved (initial interaction and obstacle ablation), which may only be broken by inhomogeneities in the jet that lead, at larger scales than studied here, to instabilities in the
cometary tail formed by the obstacle material and subsequent mixing with jet matter. It should be noted that these simulations represent a simplification, and a necessary initial step, of
the more realistic situation, which would require the obstacle to penetrate into the jet with transversal motion. This can only be solved via 3D-simulations and is beyond the
scope of this initial work.  

We have used the finite-volume code {\it Ratpenat}, which solves the equations of relativistic hydrodynamics in conservation form using high-resolution-shock-capturing methods. {\it
Ratpenat} is a hybrid parallel code  -- MPI + OpenMP -- extensively and intensively tested \citep[e.g.,][]{pe10}. The code includes the Synge equation of state \citep{sy56} with two
populations of particles, namely, leptons (electrons and positrons) and baryons (protons). The simulations were performed in Tirant, at the \emph{Servei d'Inform\`atica} of the
\emph{Universitat de Val\`encia}, using 16 processors in S1 and S2, and 32 processors in S3. 

In S1, the numerical grid box expands to $10^{15}\,{\rm cm}$ in the radial direction, with a homogeneous grid up to
$2\times10^{14}\,{\rm cm}$ including 400 cells, plus an extended grid with 200 cells covering the remaining $8\times
10^{14}\,{\rm cm}$.  In the direction of the jet flow (axial direction), the grid covers $6\times 10^{14}\,{\rm cm}$. The
(homogeneous) obstacle is located at  $z=10^{14}\,{\rm cm}$, with radius $R_{\rm o}=10^{13}\,{\rm cm}$, and the rest of the
grid is filled by the jet flow. The resolution used in S1 is 20 cells/$R_{\rm o}$, with a total of $600\times 1200$ cells. In
S2, the grid expands to $3\times 10^{14}\,{\rm cm}$, with a homogeneous region up to $10^{14}\,{\rm cm}$ using 320 cells, and
200 cells extending the grid $2 \times 10^{14}\,{\rm cm}$ further. Along the symmetry axis, the grid is $3\times
10^{14}\,{\rm cm}$ long. The obstacle is located at $z=2.5\times10^{13}\,{\rm cm}$, it has a radius $R_{\rm o}=10^{13}\,{\rm
cm}$, with an inner core of constant density with radius of $2.5 \times 10^{12}\,{\rm cm}$, and a surrounding
region with decreasing density (see Sect.~\ref{phys}) up to $R_{\rm o}$. As in S1, the rest of the grid is filled by jet
flow. The resolution used in S2 is 32 cells/$R_{\rm o}$, with a total of $520\times960$ cells. In the case of S3, the
simulation is focused on a smaller region of S2, using a resolution 64 cells/$R_{\rm o}$ with a total of $320\times960$ cells
that cover a grid size of  $5\times10^{13}\,{\rm cm}$ in the radial direction and $1.5\times10^{14}\,{\rm cm}$ in the axial
direction. 

The jet is modelled in all three simulations as a matter dominated flow with proton-electron composition. The  properties of the jet in S1 have been chosen to result in a total jet kinetic
power $L_{\rm j,s1}=2\times10^{44}\,{\rm erg/s}$ for a jet radius $R_{\rm j}=10^{15}\,{\rm cm}$, with jet velocity $v_{\rm j}=0.866\,c=2.6\times10^{10}\,{\rm cm/s}$, density $\rho_{\rm
j,s1}=1.35\times 10^{-18}\,{\rm g/cm^3}$, temperature $T_{\rm j}=10^{10}\,{\rm K}$, adiabatic exponent $\gamma_{\rm j}=1.45$, and Mach number $M_{\rm j}=16.8$. With these parameters, the
kinetic power of the jet within the homogeneous simulated grid is $L_{\rm j,g1}=8\times10^{42}\,{\rm erg/s}$ and the mass flux is $\dot{M}_{\rm j,g1}=2.8\times10^{21}\,{\rm g/s}$. In S2 and
S3, the jet density used is slightly larger $\rho_{\rm j,s23}=2.4\times 10^{-18}\,{\rm g/cm^3}$, giving a total jet power within a radius $R_{\rm j}$ of $L_{\rm
j,s23}=3.5\times10^{44}\,{\rm erg/s}$. In the case of S2, $L_{\rm j,g2}=3.5\times10^{42}\,{\rm erg/s}$ and $\dot{M}_{\rm j,g2}=1.25\times10^{21}\,{\rm g/s}$, and, for S3, $L_{\rm
j,g3}=8.75\times10^{41}\,{\rm erg/s}$ and $\dot{M}_{\rm j,g3}=3.12\times10^{20}\,{\rm g/s}$.

The obstacle in S1 has density $\rho_{\rm o}=1.35\times10^{-12}\,{\rm g/cm^{3}}$ and temperature $T_{\rm o}=10^4\,{\rm K}$, with a total mass $M_{\rm o}\simeq 6\times10^{27}\,{\rm g}$. In
the case of S2, the obstacle has a core density $\rho_{\rm o}=1.25\times10^{-10}\,{\rm g/cm^{3}}$ within the inner $2.5\times10^{12}\,{\rm cm}$, and a decreasing density $\propto R^{-2}$ ($R$ is
the spherical radius here), starting with $\rho=1.7\times10^{-11}\,{\rm g/cm^{3}}$ at $2.5\times10^{12}\,{\rm cm}$.

The material of the shocked obstacle can have rather complex properties. Once in pressure equilibrium with the jet ram pressure, the obstacle is radiation dominated, and radiative cooling through
thermal Bremsstrahlung may be important. For simplicity, we have modelled the obstacle as a non-relativistic proton-electron plasma. This has some implications for the compression and
stability of the obstacle, since the adiabatic coefficient may not be accurate with respect to realistic cases, and shocked matter densities (and instabilities) could be enhanced by
radiative cooling. However, at this stage of the study the calculations can capture the relevant aspects of the simulated process. More accurate simulations will be performed in the
future.

\section{Results}\label{res}

\subsection{Homogeneous case}\label{res1}

In S1, the interaction region rapidly develops the expected structure, with a bow shock in the jet, and a contact discontinuity separating both the shocked media and the obstacle shocked by a slow shock. The eroded material from the obstacle is  dragged downstream, forming a cylindrical tail surrounded by rarefied jet flow that has crossed the bow shock.

\begin{figure*}[!t]
  \includegraphics[clip,angle=0,width=0.48\textwidth]{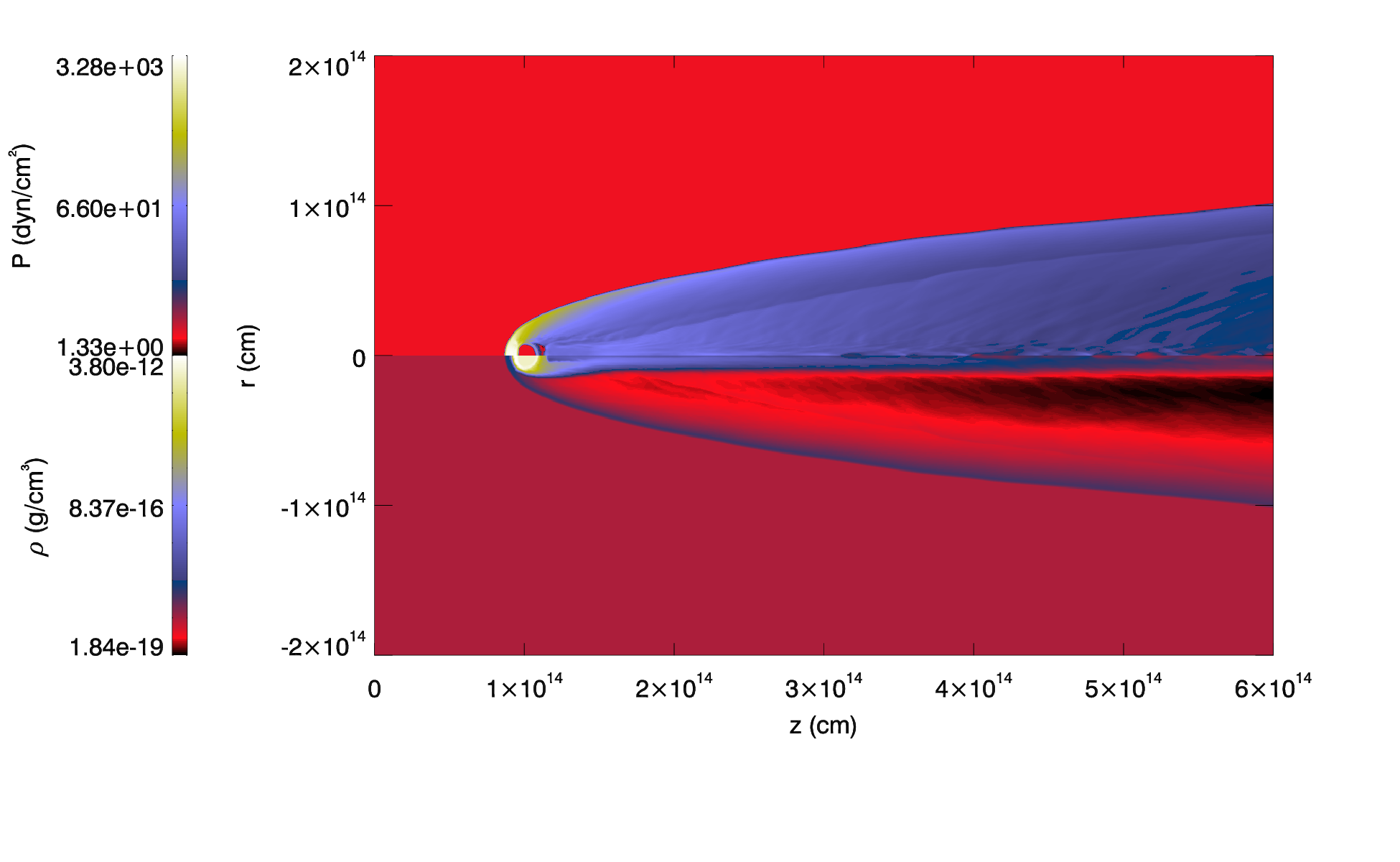}
  \includegraphics[clip,angle=0,width=0.48\textwidth]{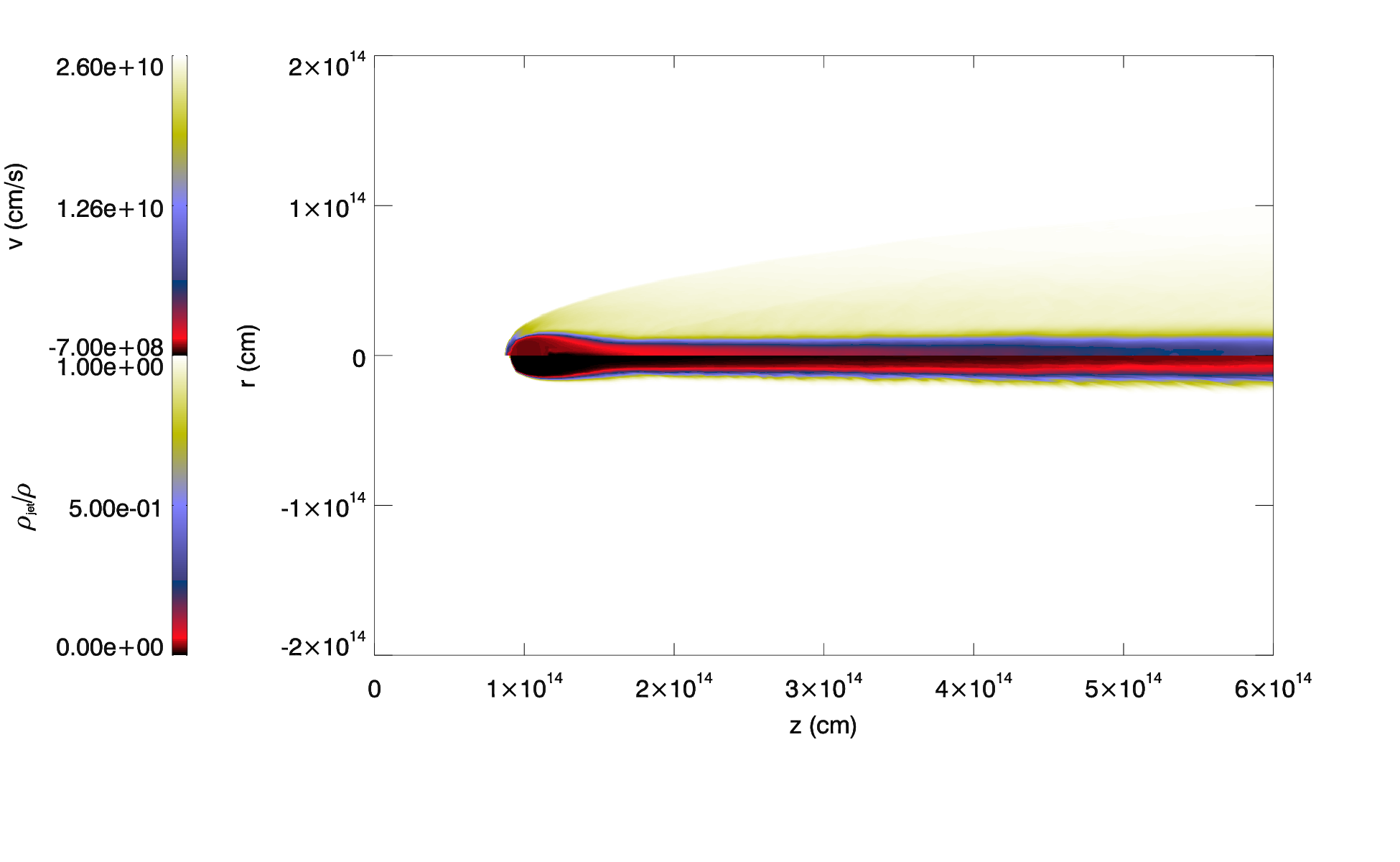}
  \caption{Combined maps of logarithm of pressure (left panel, upper half), density (left panel, lower half), 
  axial velocity (right panel, upper half), and jet mass fraction (right panel, lower half) at $t\simeq9\times10^4\,$s for S1. 
The plots make use of the axisymmetric 
nature of the simulation. Jet mass fraction equals 1 for pure jet material, 0 for pure obstacle material, and values in between
indicate mixing. Units are cgs.}
  \label{fig:maps1.1}
  \end{figure*}

Figure~\ref{fig:maps1.1} shows a snapshot of  the simulation at $t\simeq 9\times10^4\,$s, which corresponds to the initial phase. 
After $t=2.3\times10^5\,$s, the shock has completely crossed the obstacle. The pressure at the apex of the bow shock oscillates between $10^3$ and
$1.5\times10^3\,{\rm dyn/cm^2}$, and the temperature of the post-shock gas is at this point $\simeq10^{12}\,{\rm K}$\footnote{The $e^\pm$ temperature could be significantly lower due to pair production \citep[e.g.][]{{bkzs71}}.}. After the obstacle matter is heated by the shock, it expands, increasing the
cross-section of the interaction with the jet flow.  Up to this point, the mass-load by obstacle material is modest (see Sect.~\ref{ml}) and mainly due to continuous ablation 
from its outer layers. This gas accumulates in a cylindrical region with radius of the order of $R_{\rm o}$, forming a cometary-like tail of obstacle gas separated from the rarefied,
shocked jet flow by a smooth shear layer in which little mixing occurs, as no instabilities are triggered in the layer during this phase. 

\begin{figure*}[!t]
  \includegraphics[clip,angle=0,width=0.48\textwidth]{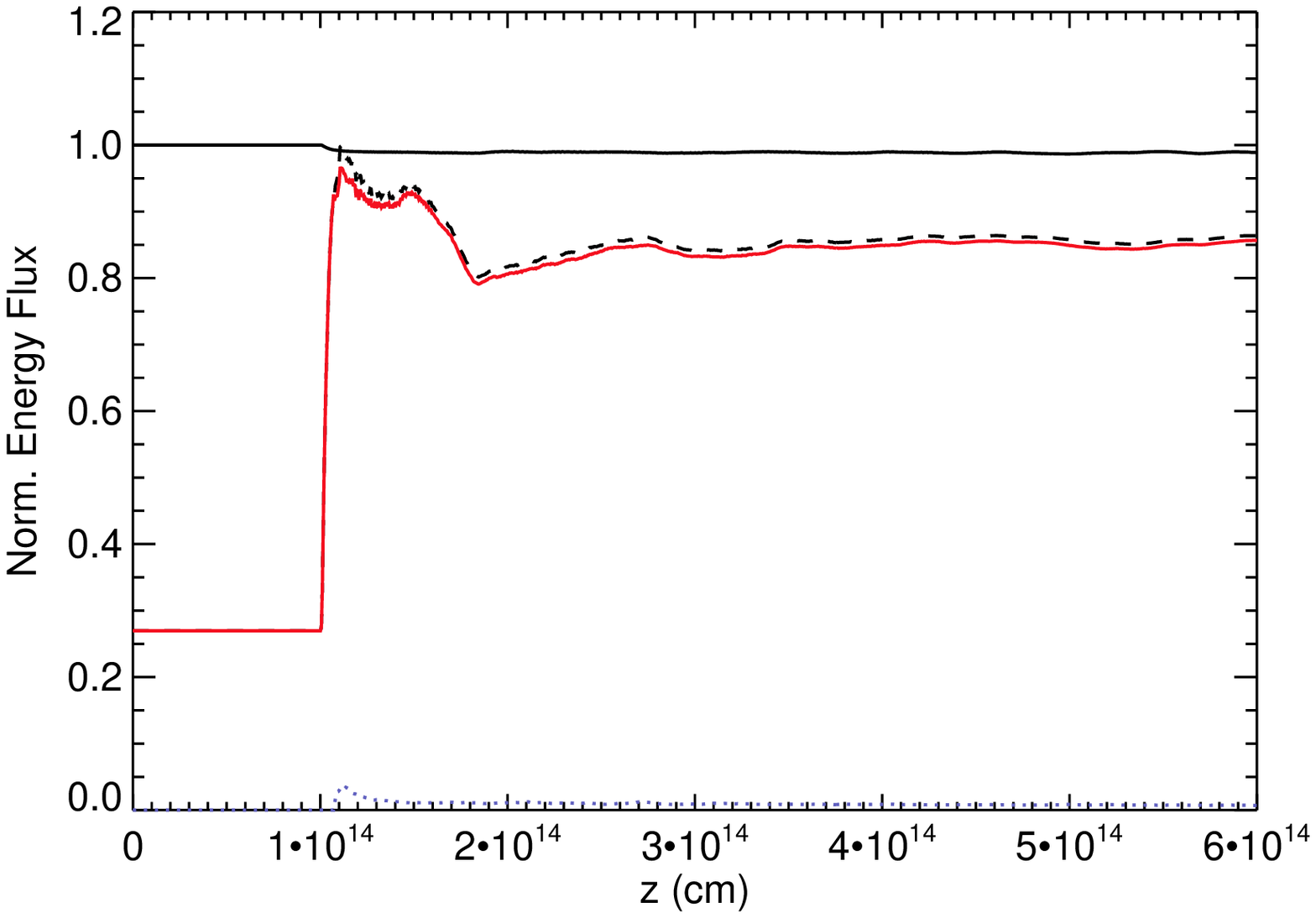}
  \includegraphics[clip,angle=0,width=0.48\textwidth]{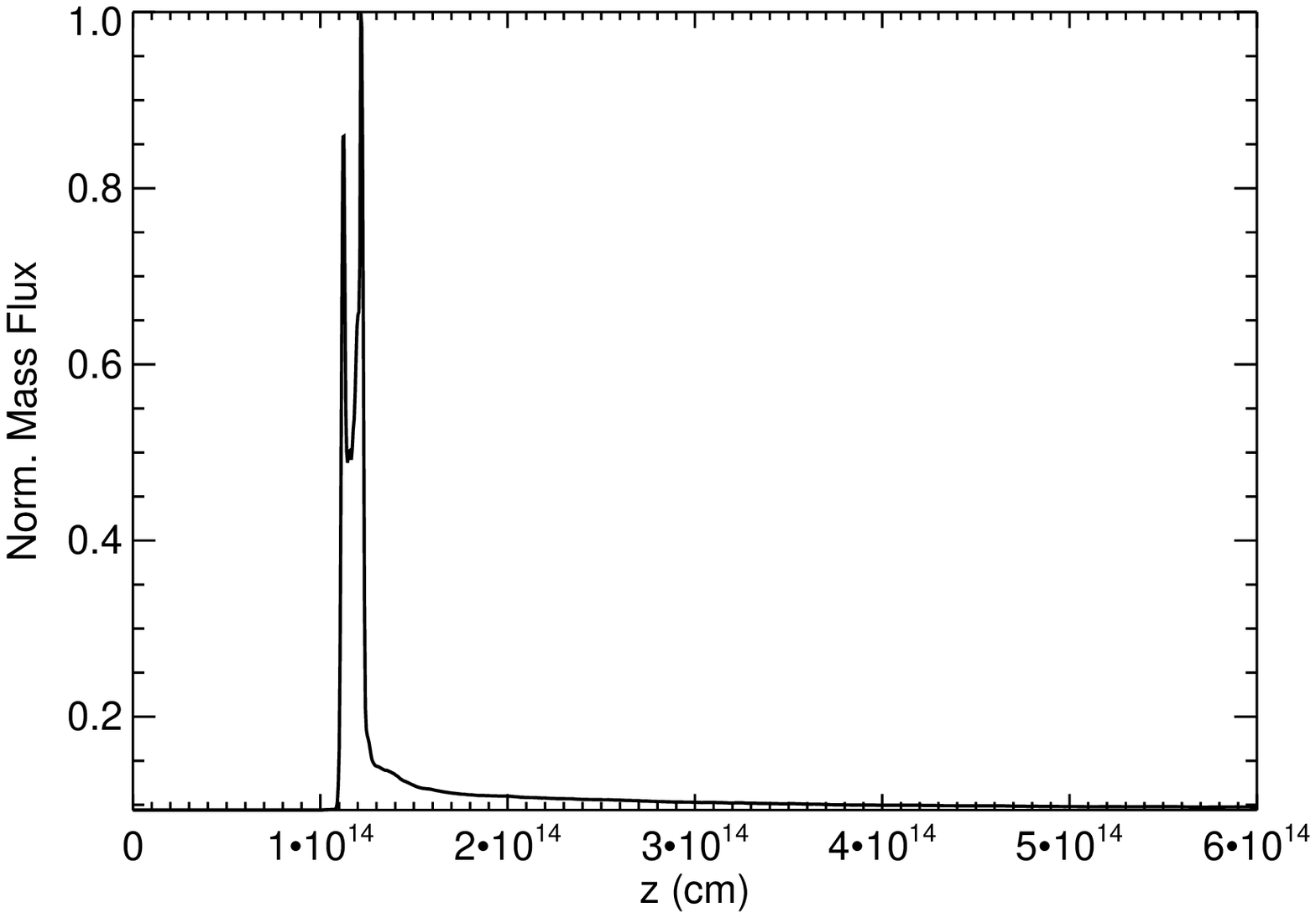}\\
  \includegraphics[clip,angle=0,width=0.48\textwidth]{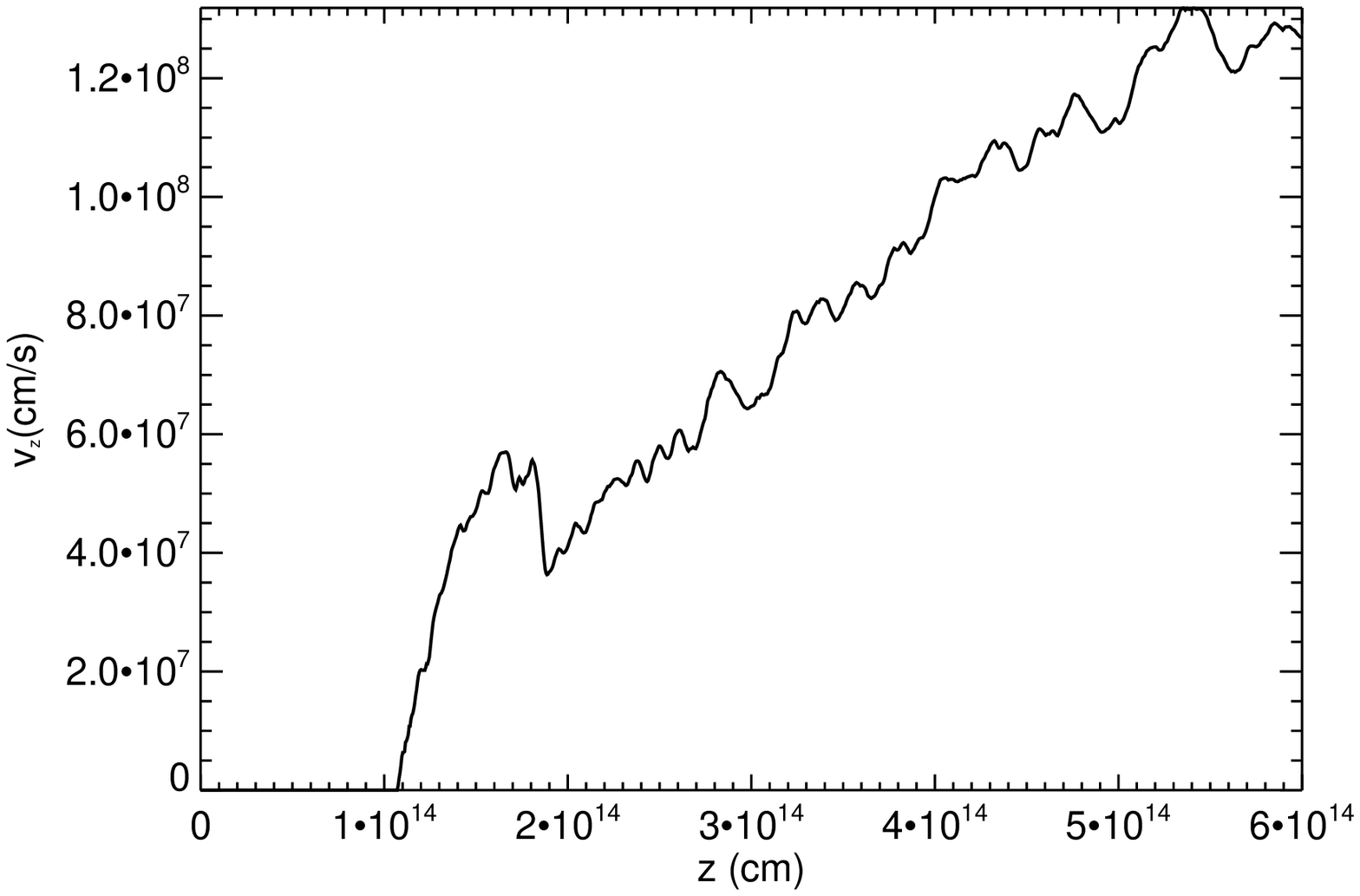}
  \includegraphics[clip,angle=0,width=0.48\textwidth]{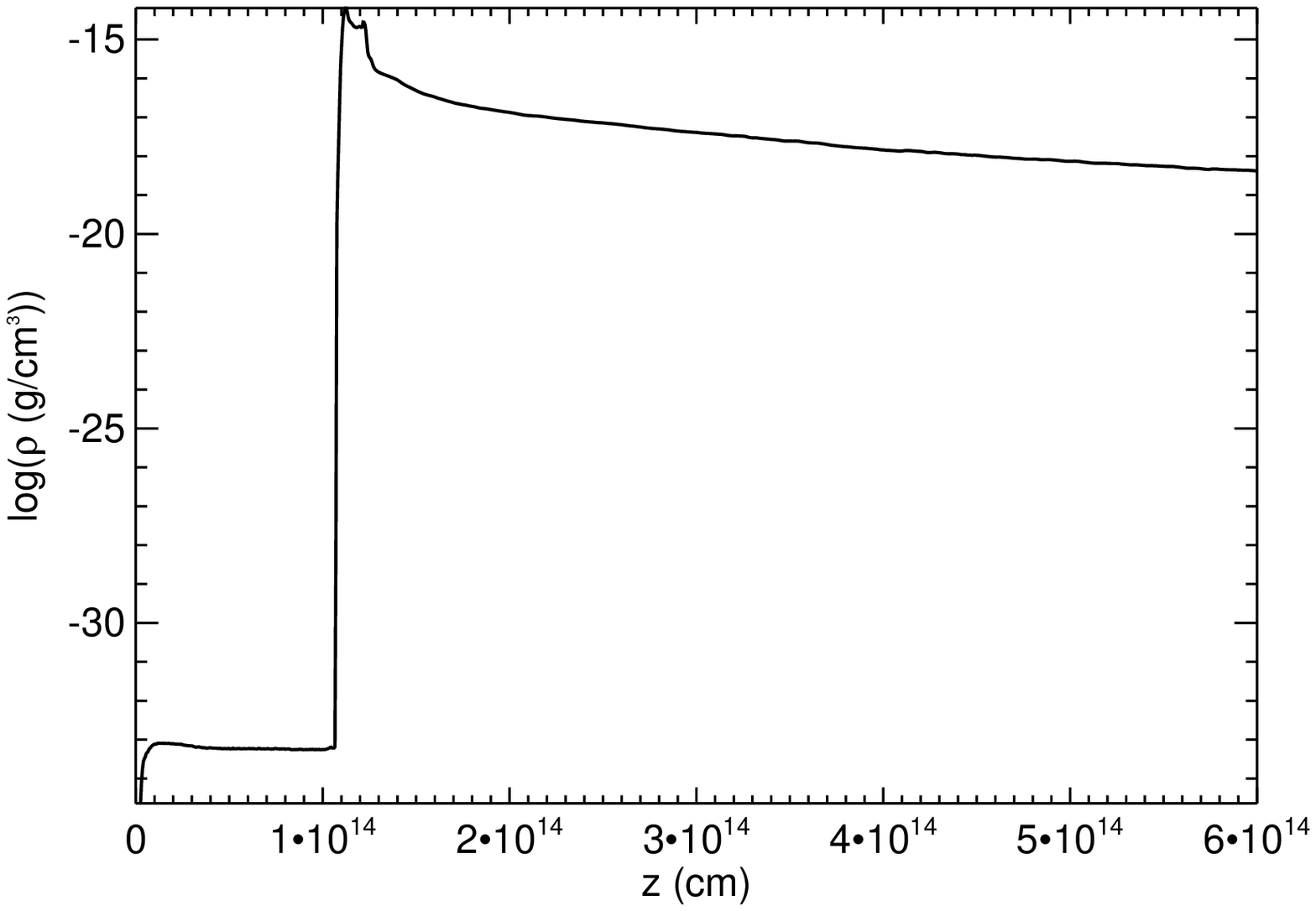}
  \caption{Axial cuts of different quantities for S1 at $t\simeq 6\times 10^5\,$s, before the instabilities completely destroy the obstacle. 
The top left panel shows $L_{\rm j,g1}-L_{\rm j,int,g1}$ for the jet material (solid line) 
normalized to its injection value $8\times10^{42}\,{\rm erg/s}$, and the internal 
energy luminosity $L_{\rm j,int,g1}$ normalized to its maximum value, $8\times10^{40}\,{\rm erg/s}$ (dashed line); 
the red solid line stands for the internal energy luminosity in the jet material, and the blue dotted line (very close to the bottom) for the internal 
energy luminosity in the obstacle material. The top right panel shows the mass flux, normalized to its maximum value 
($4.7\times10^{22}\,{\rm g/s}$). The bottom left and right panels show the mean velocity (averaged over the grid section) and density in the obstacle gas.}
  \label{fig:S1.1}
  \end{figure*}

Figure~\ref{fig:S1.1} shows the trends in energy
flux, mass flux, obstacle mean density (averaged over the grid area $A_{\rm j}$) and velocity along the axial direction for a snapshot at a relatively early stage. In this plot, the energy flux is separated into kinetic plus rest-mass energy flux 
($L_{\rm j,g1}-L_{\rm j,int,g1}=(\rho (h W -1) W - \rho W \varepsilon) v_{\rm j} A_{\rm j}$, with $h$ being the enthalpy, $W$ the Lorentz factor, and
$\varepsilon$ the specific internal energy), and internal energy flux  ($L_{\rm j,int,g1}=\rho W \varepsilon) v_{\rm j} A_{\rm j}$). These variables
give an idea of the amount of energy flux in the form of internal energy or kinetic and rest-mass energy.
A fraction of $L_{\rm j,g1}$ in this phase is mainly invested in heating the shocked jet material, with only a small amount transferred to the obstacle gas (upper panels in Fig.~\ref{fig:S1.1}). Some fraction of this energy may go to non-thermal particles. The gas in the tail expands and accelerates along the $z$ coordinate, reaching a mean axial speed $v_{\rm z}\simeq 1.2\times10^8\,{\rm cm/s}$ (bottom left panel in Fig.~\ref{fig:S1.1}), with $v_{\rm z}\simeq 10^9\,{\rm cm/s}$ right on the symmetry axis.

 \begin{figure*}[!t]
  \includegraphics[clip,angle=0,width=0.48\textwidth]{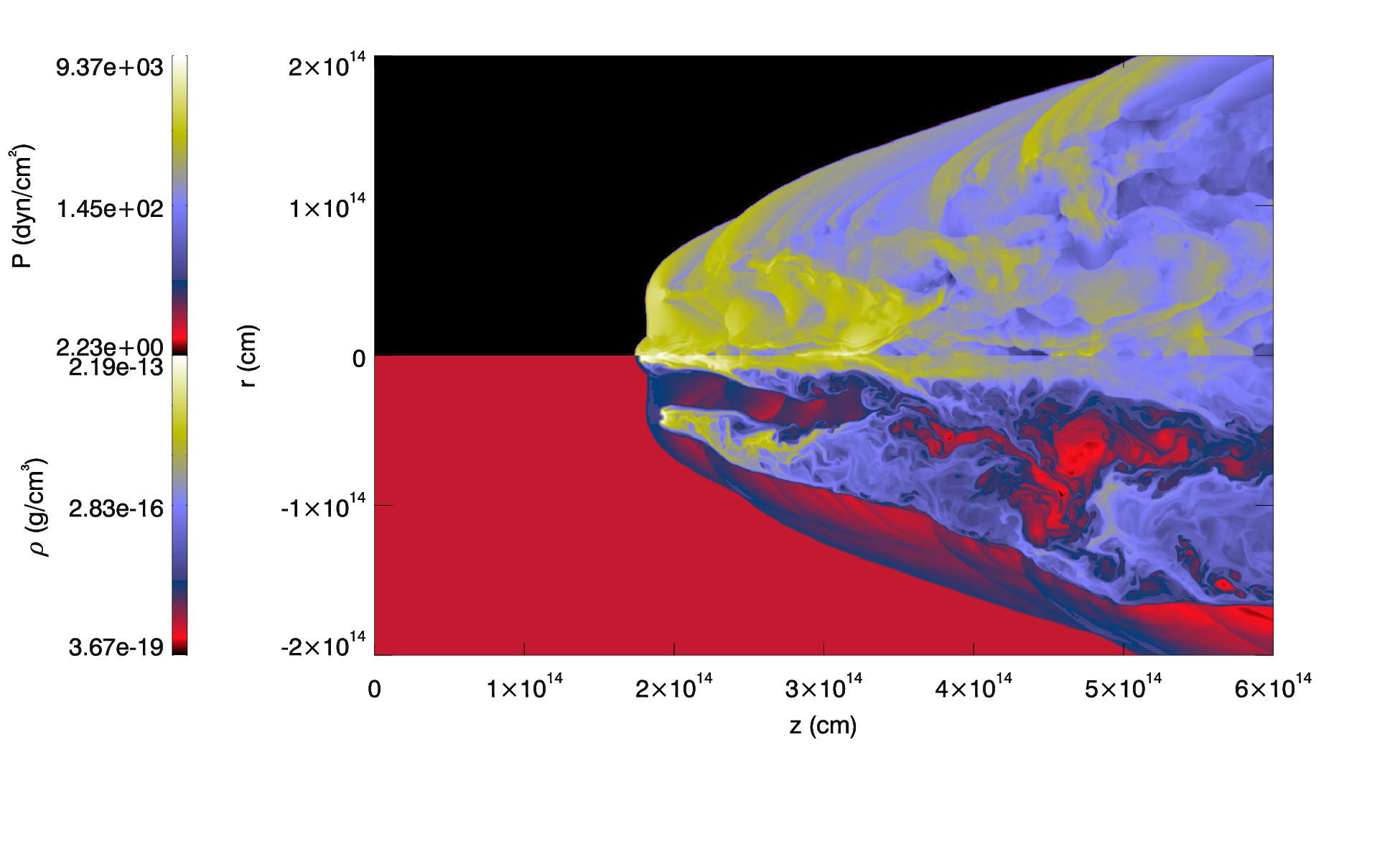}
  \includegraphics[clip,angle=0,width=0.48\textwidth]{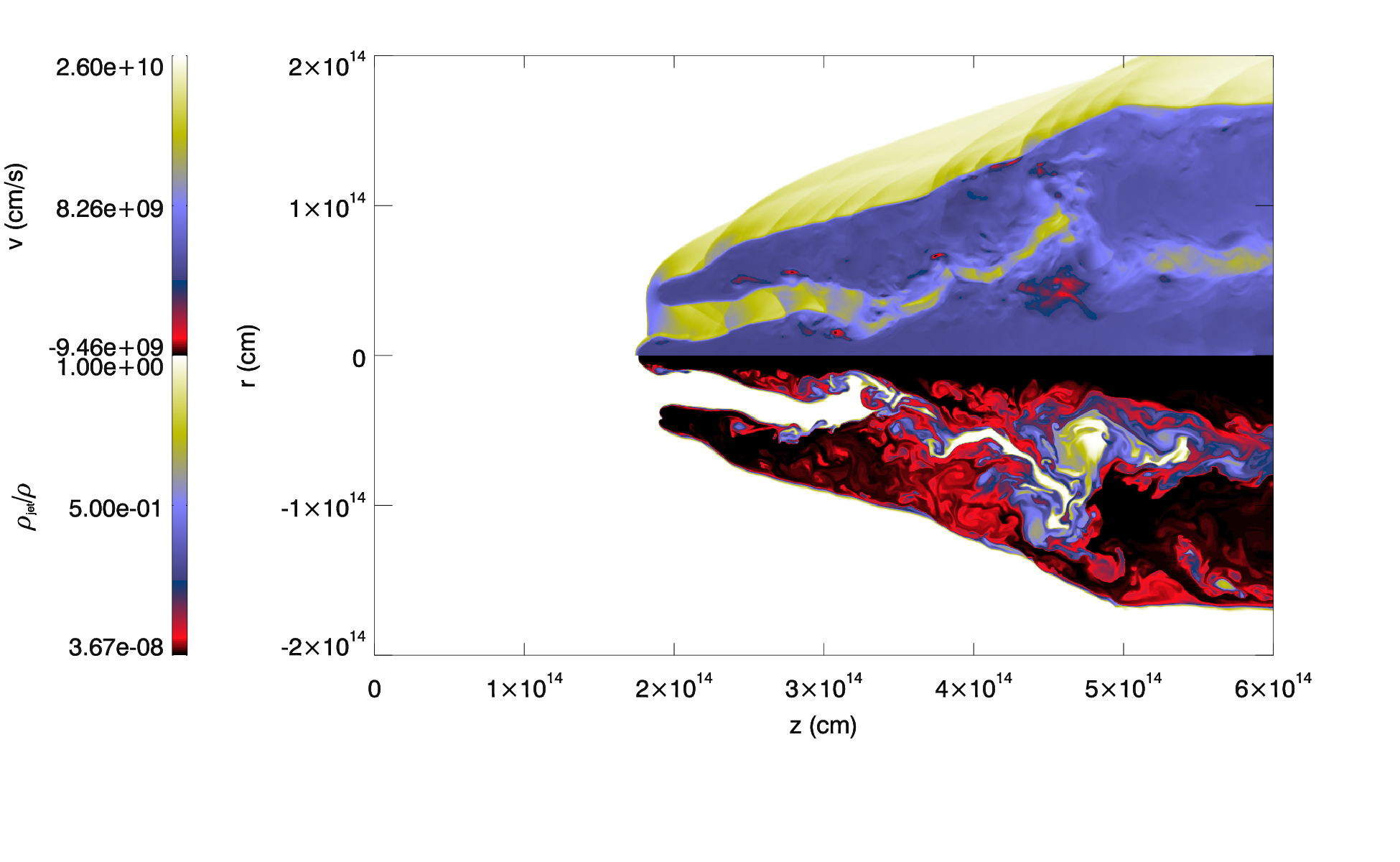}
  \caption{Same as Fig~\ref{fig:maps1.1} at $t\simeq1.1\times10^6\,$s}
  \label{fig:maps1.2}
  \end{figure*}  
 
\begin{figure}[!t]
  \includegraphics[clip,angle=0,width=0.48\textwidth]{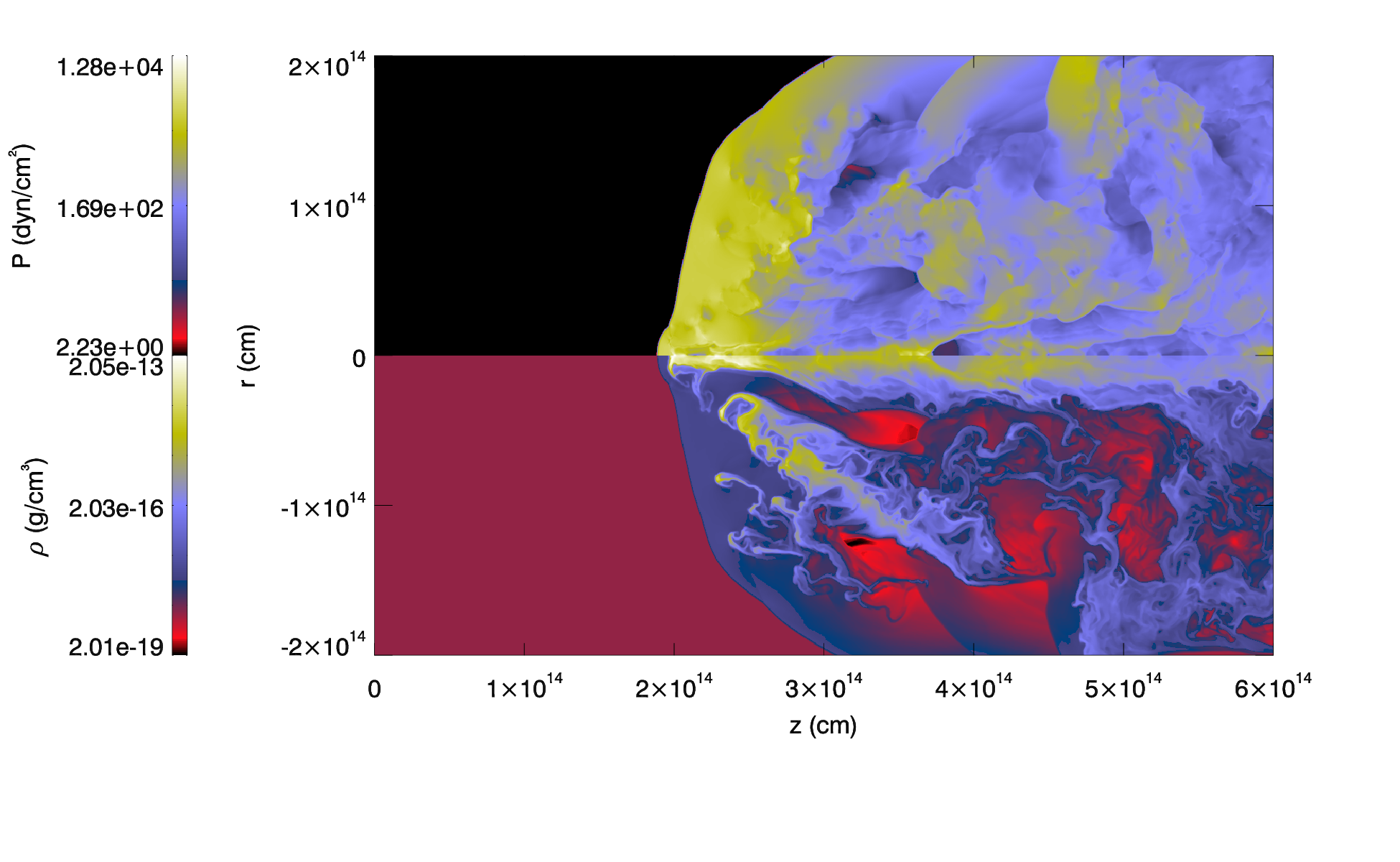}
  \includegraphics[clip,angle=0,width=0.48\textwidth]{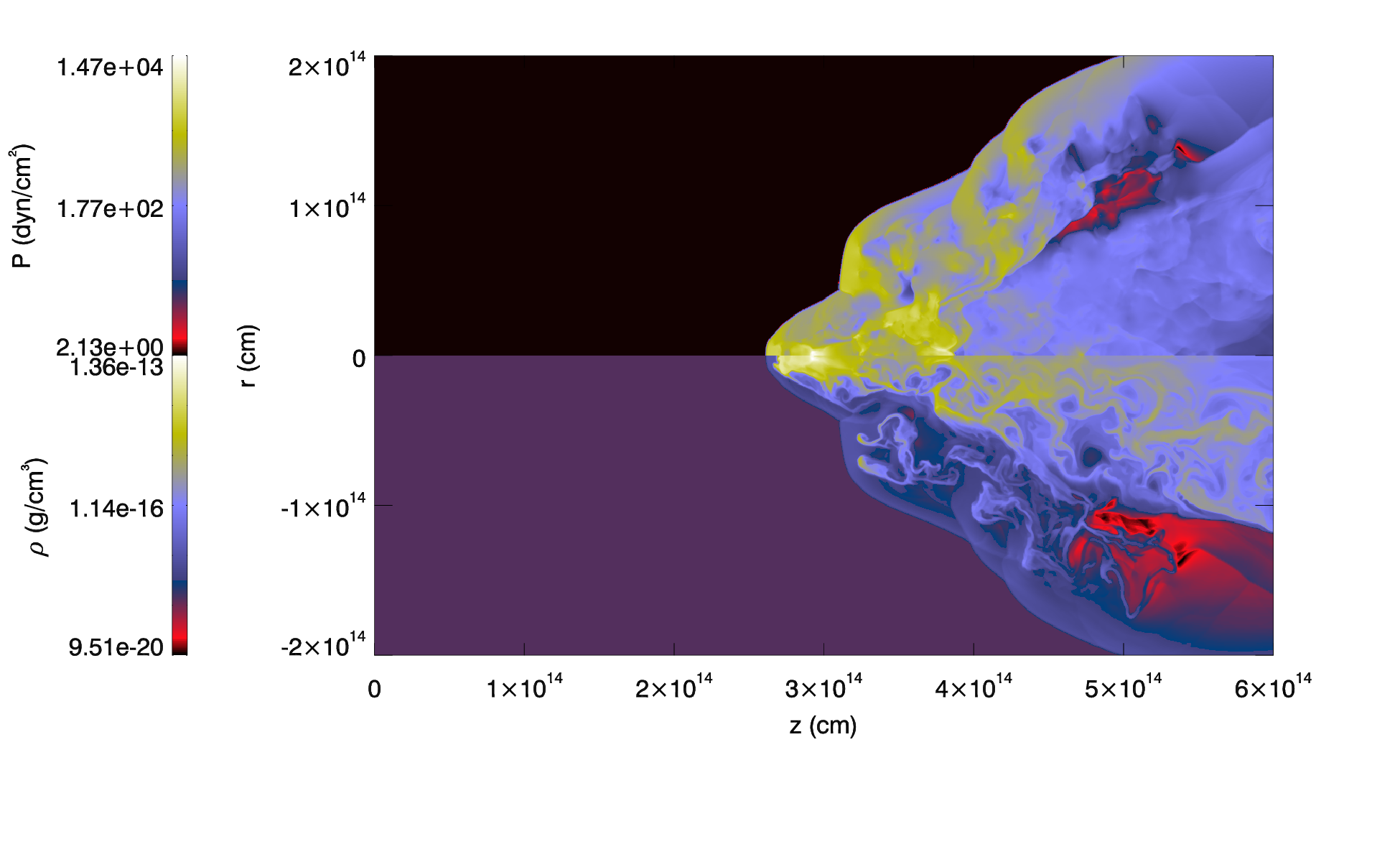}
  \includegraphics[clip,angle=0,width=0.48\textwidth]{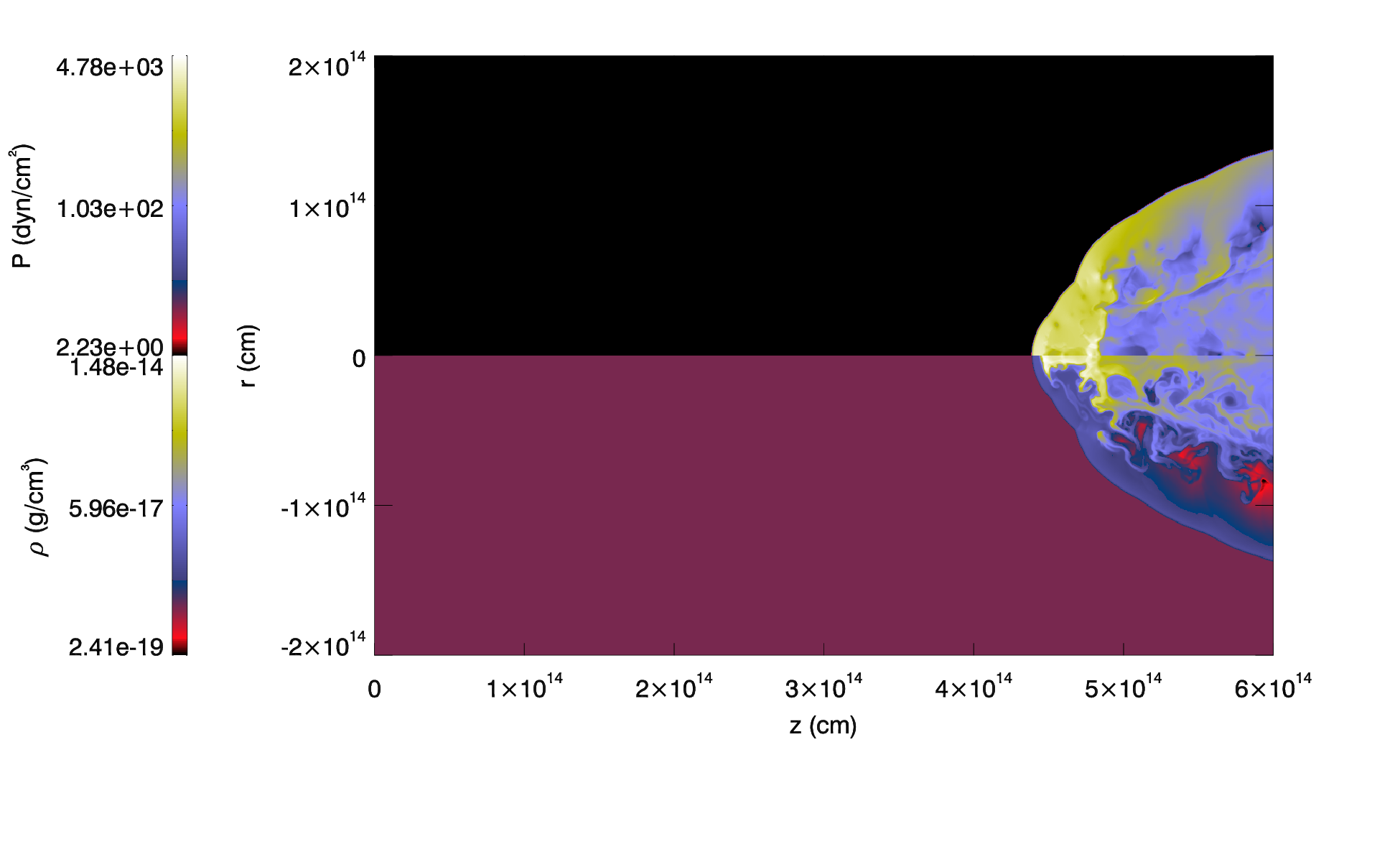}
  \caption{Same as left panel in Fig.~\ref{fig:maps1.1} at $t\simeq1.15\times10^6\,$s, $t\simeq1.3\times10^6\,$s and
  $t\simeq1.45\times10^6\,$s.}
  \label{fig:maps1.3}
  \end{figure}

\begin{figure*}[!t]
  \includegraphics[clip,angle=0,width=0.32\textwidth]{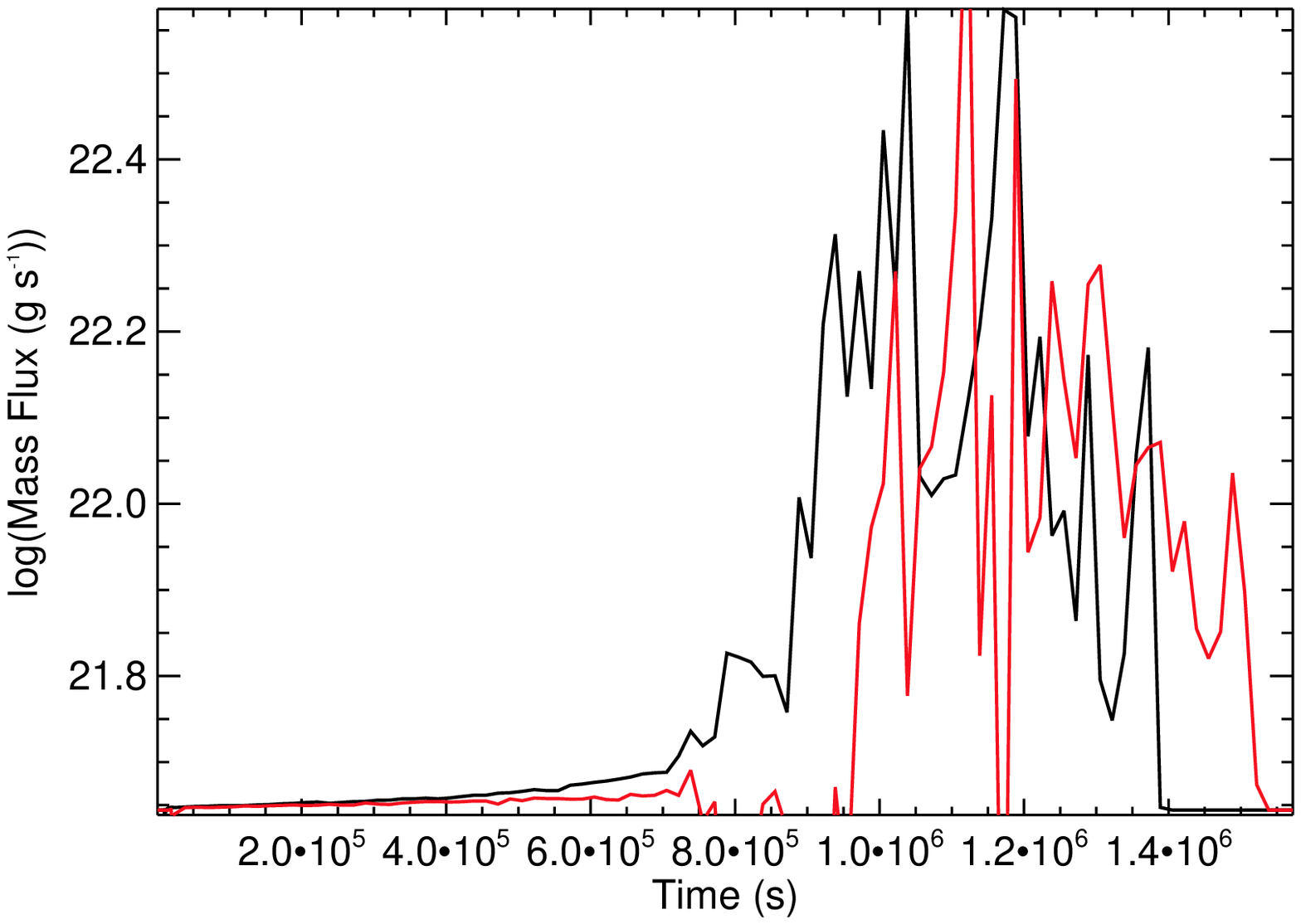}
  \includegraphics[clip,angle=0,width=0.32\textwidth]{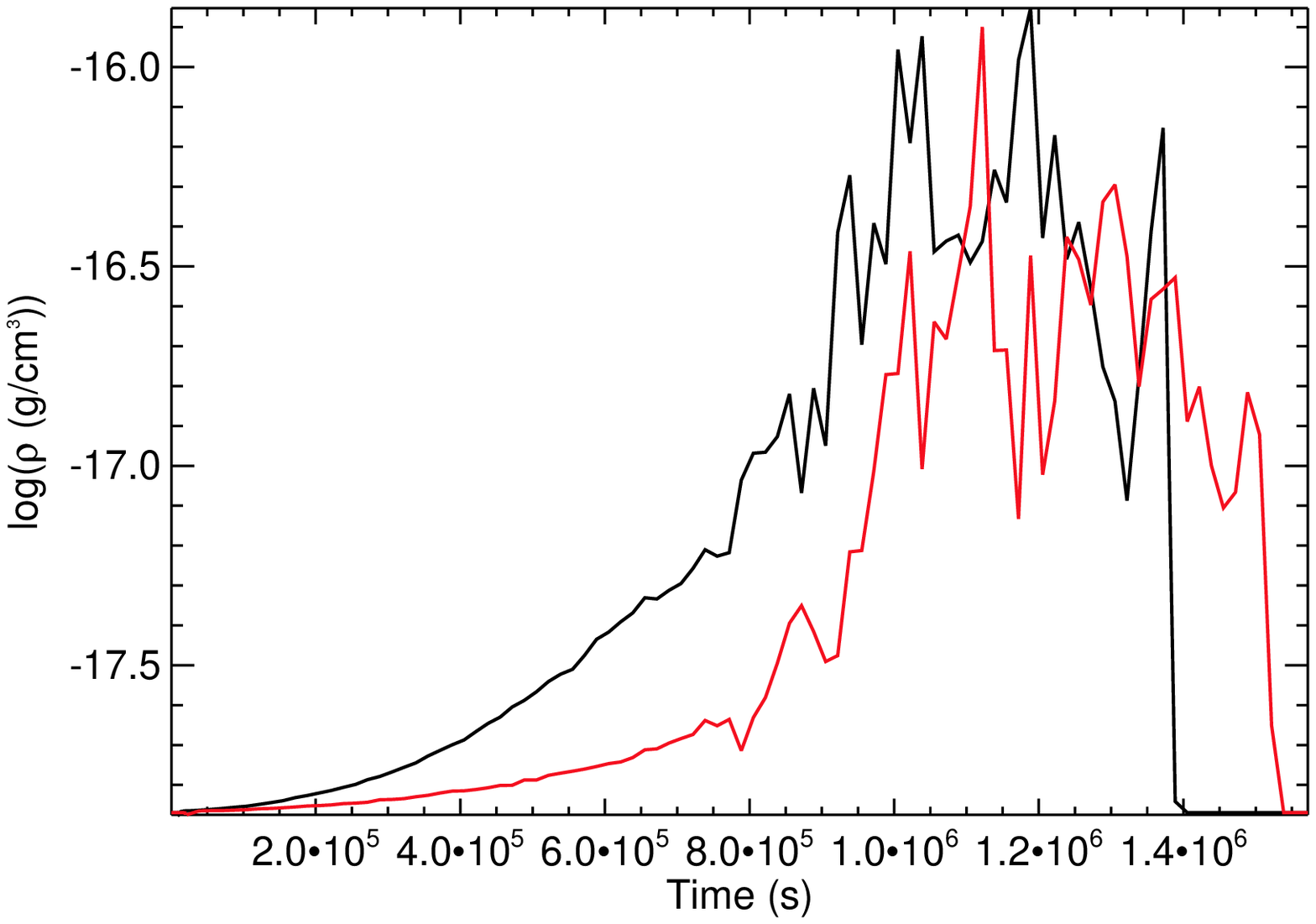}
  \includegraphics[clip,angle=0,width=0.32\textwidth]{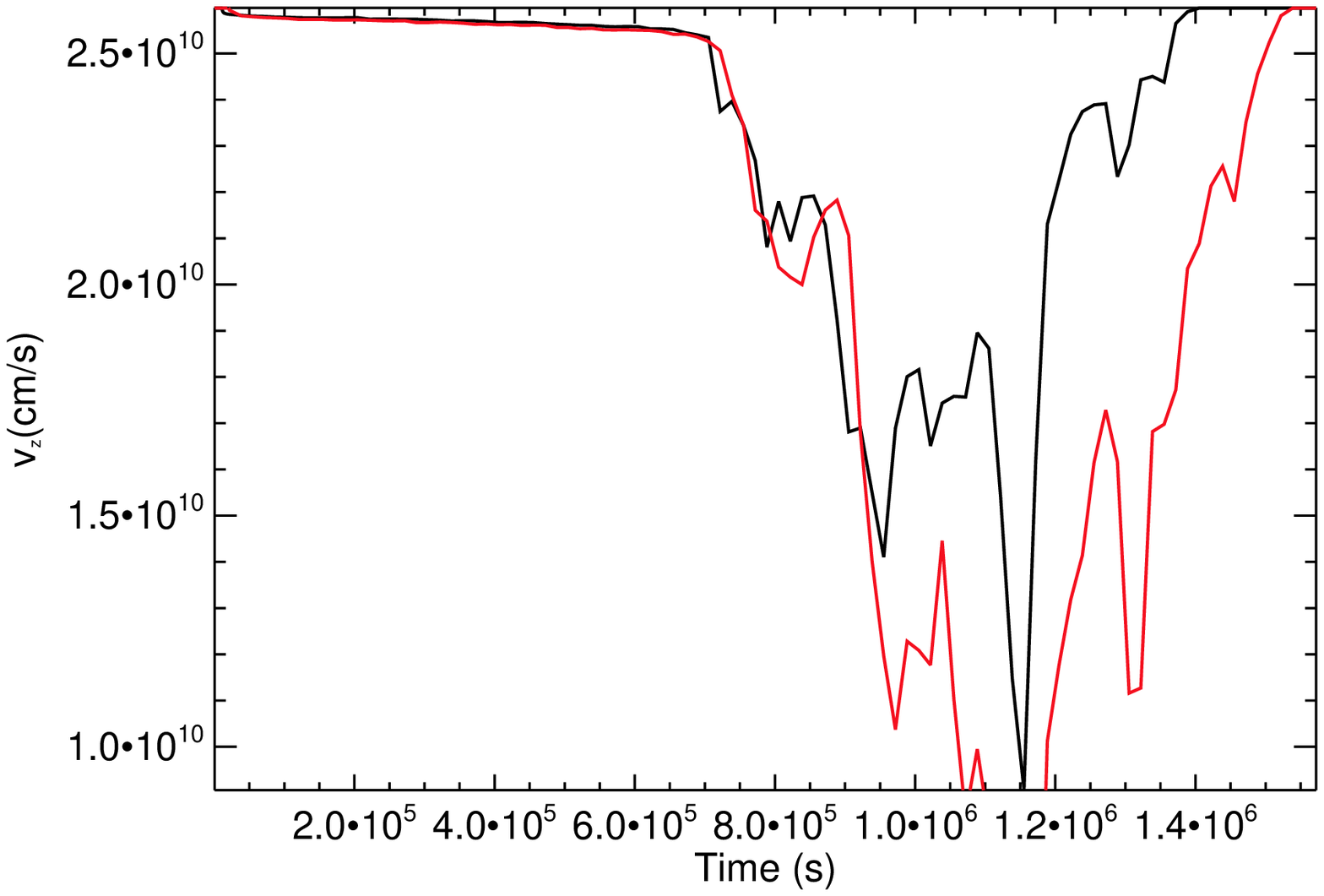}
  \caption{Mass flux (left panel), mean density (central panel), and mean velocity (right panel) versus time for S1 at $z=3\times10^{14}$ 
(half grid, black lines) and $6\times10^{14}\,{\rm cm}$  (end of the grid, red lines). 
}
  \label{fig:S1.2}
  \end{figure*}

After the obstacle is shocked, instabilities develop inside and disrupt it, detaching clumps of material and enhancing turbulent mixing with the jet flow in the cometary-like tail. The
instabilities manage to tear the external portions of obstacle gas at $t\simeq 7\times 10^5\,$s, as the jet flow penetrates through the obstacle. This process repeats itself in smaller
clumps that have been detached from the obstacle and in the material remaining at the symmetry axis, all of them also torn by the jet (see Figs.~\ref{fig:maps1.2} and \ref{fig:maps1.3}).
These clumps move radially and are accelerated when entering the region of rarefied (post-shock) jet material, reaching mean velocities $\simeq 3\times10^9\,{\rm cm/s}$. At the same time,
the jet ram-pressure starts to effectively push the obstacle remains from its original location along the $z$-direction, dragging them downstream to the end of the grid. The left and
central panels of Fig.~\ref{fig:S1.2} show the increase in mass flux and mean density when the rest of the obstacle crosses through the axial locations where these values are computed (half
and end of the grid, along the axis). At the same time, the right panel shows a  strong decrease in the mean velocity at the plotted locations, which is accompanied by a drop of the mean
Mach number from $M_{\rm j}=16$ to values $<5$. 

\begin{figure*}[!t]
  \includegraphics[clip,angle=0,width=0.48\textwidth]{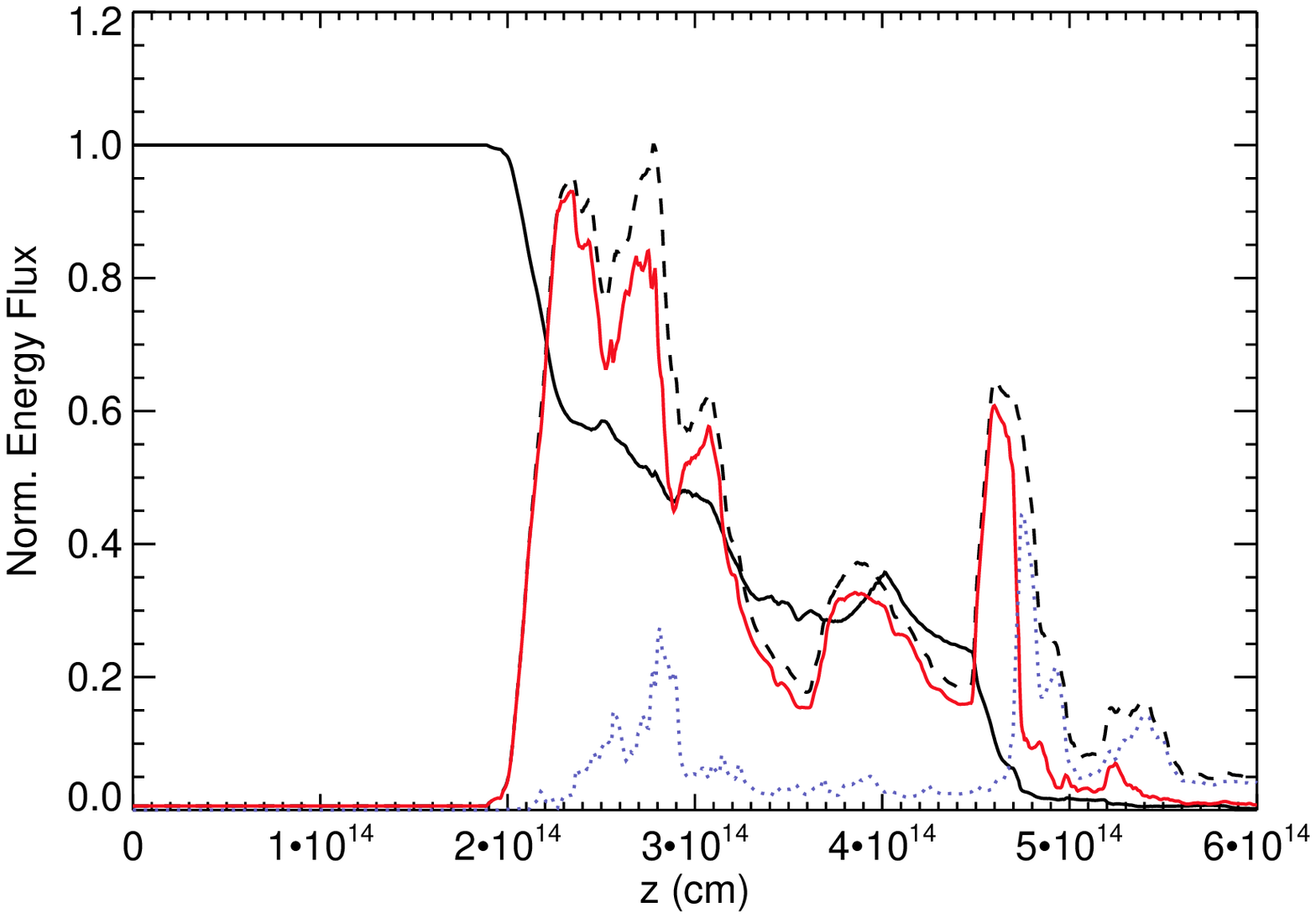}
  \includegraphics[clip,angle=0,width=0.48\textwidth]{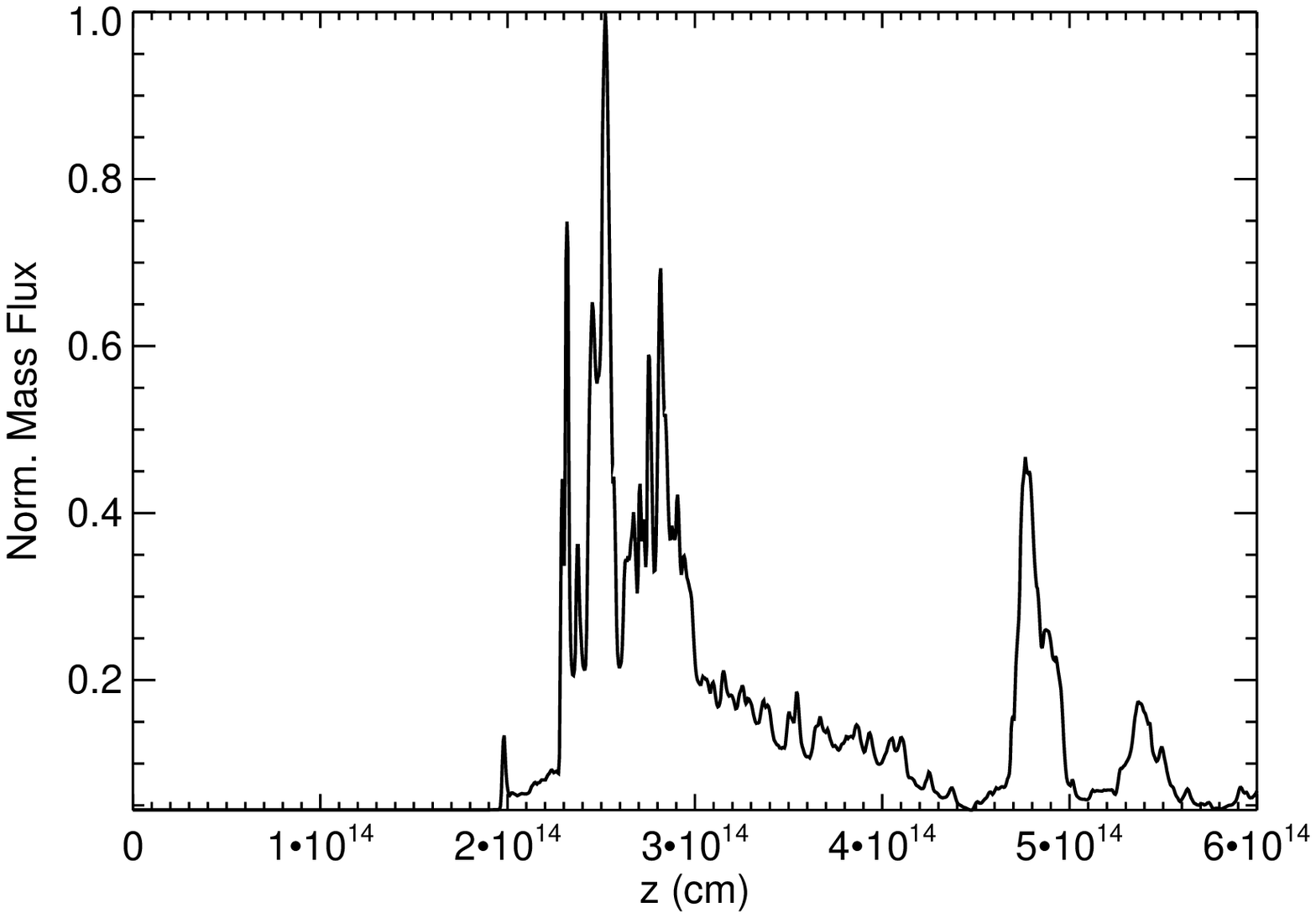}\\
  \includegraphics[clip,angle=0,width=0.48\textwidth]{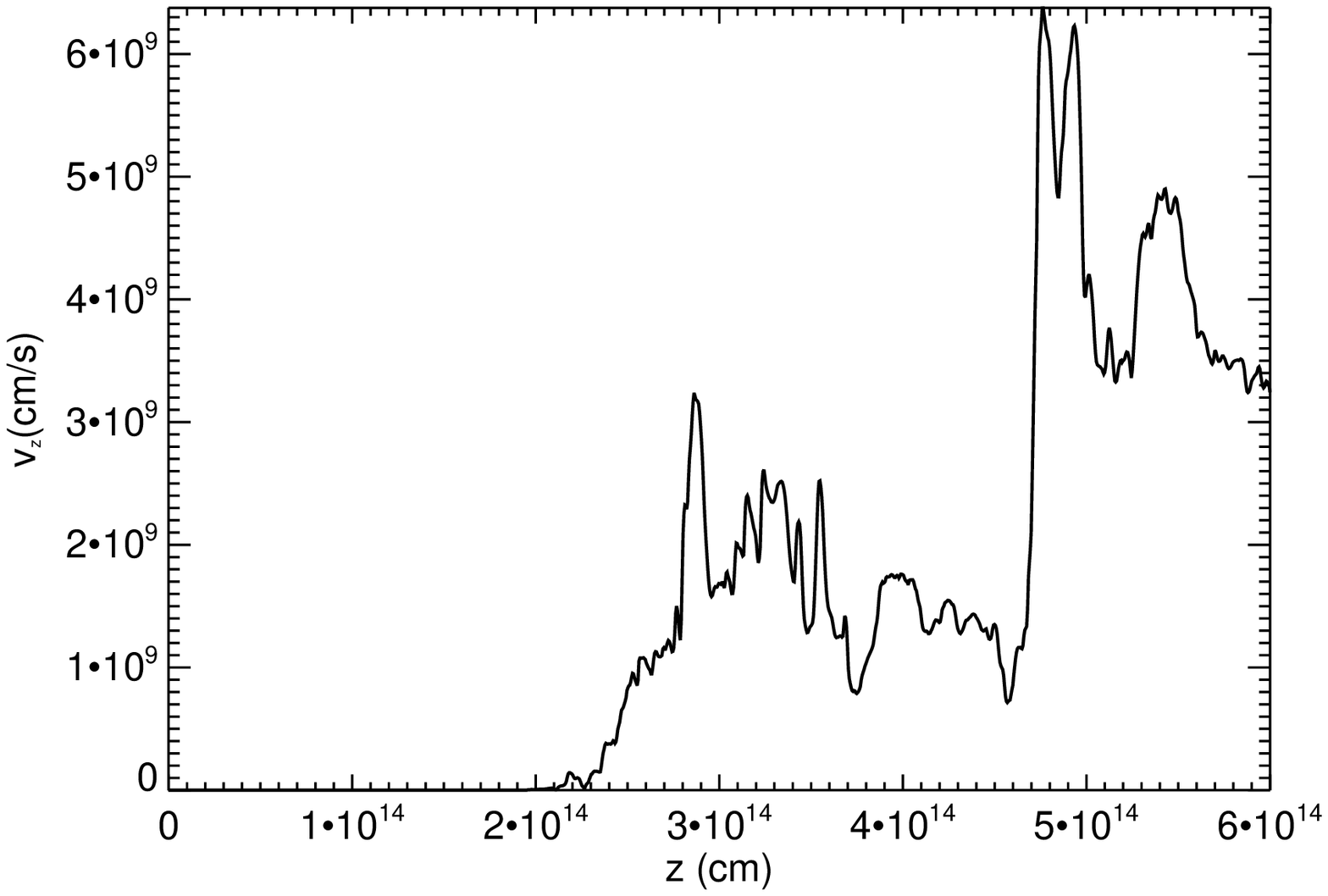}
  \includegraphics[clip,angle=0,width=0.48\textwidth]{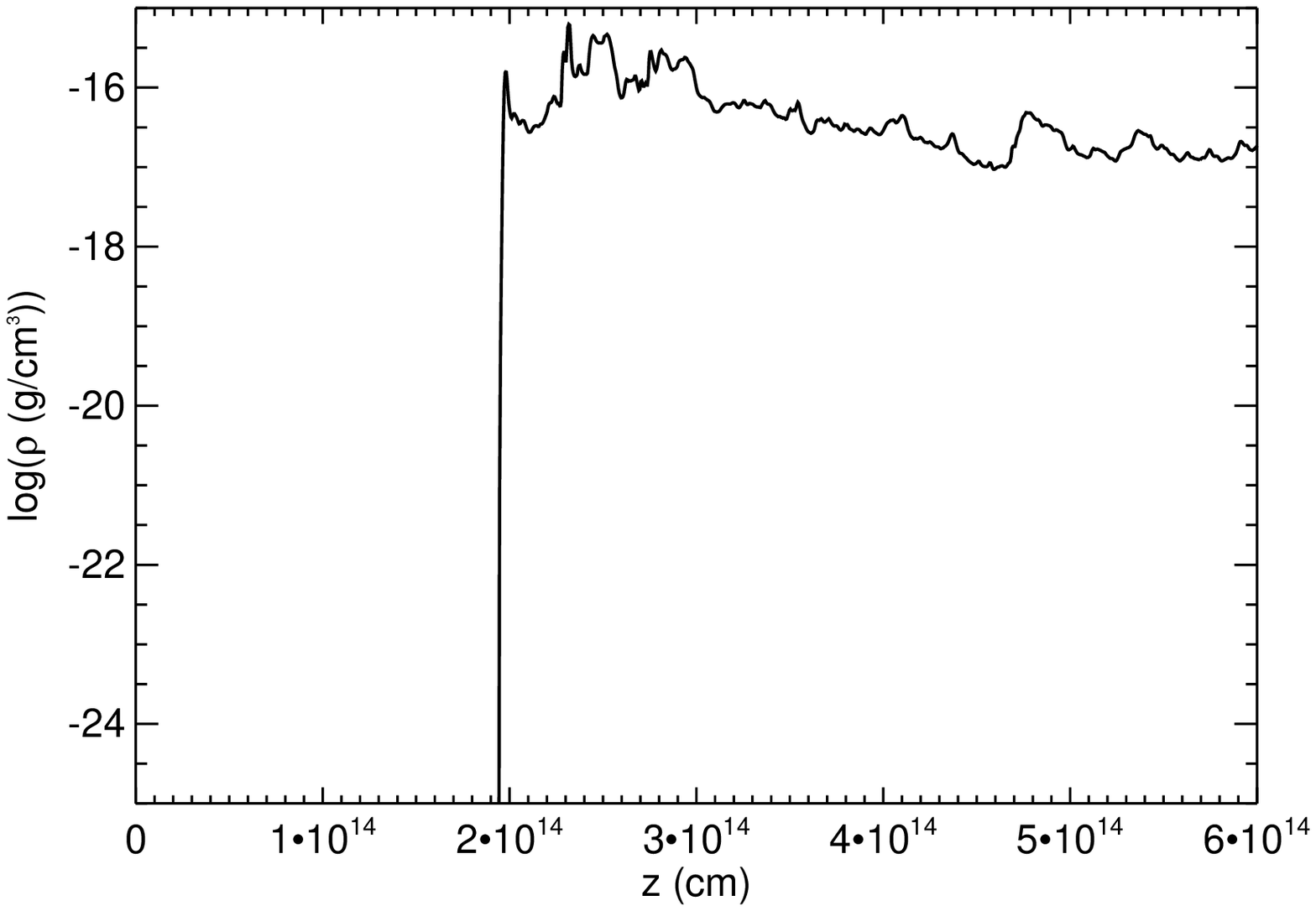}
  \caption{Axial cuts of different quantities for S1 at $t\simeq1.15\times10^6\,$s, 
 the time of maximum cross-section of the interaction between the obstacle and the jet (see Fig.~\ref{fig:maps1.3}).
 The top left panel shows $L_{\rm j,g1}-L_{\rm j,int,g1}$ for the jet material (solid line) 
normalized to its injection value $8\times10^{42}\,{\rm erg/s}$, and the internal 
energy luminosity $L_{\rm j,int,g1}$ normalized to its maximum value, $3.6\times10^{42}\,{\rm erg/s}$ (dashed line); 
the red solid line stands for the internal energy luminosity in the jet material, and the blue dotted line for the internal 
energy luminosity in the obstacle material. The top right panel shows the mass flux, normalized to its maximum value 
($10^{23}\,{\rm g/s}$). 
The bottom left and right panels show the mean velocity and density in the obstacle gas.}
  \label{fig:S1.3}
  \end{figure*}

Figure~\ref{fig:S1.3} shows the same plots as Figure~\ref{fig:S1.1} but at $t\simeq1.15\times10^6\,$. At this time the cross-section of the obstacle is maximum, as shown by the upper panel in
Figure~\ref{fig:maps1.3}. The plots show that the transfer of energy from the jet material to the obstacle is maximal in this situation, with the jet losing almost all of the initial
kinetic and mass energy flux, which is put into internal energy flux of the jet and the obstacle (and potentially into non-thermal particles), plus kinetic and mass energy flux of the
obstacle material (not shown). The latter is revealed by the large mean velocities reached by the obstacle material (bottom left panel). The mass flux shows an increase by nearly two orders
of magnitude with respect to the original jet flux in the simulated  region, and the obstacle mean density along the grid is also larger by several orders of magnitude, depending on the
position along the axis, than that shown in Figure~\ref{fig:S1.1}.

\subsection{Inhomogeneous case}\label{res2}

In this case, the interaction of the jet with the obstacle is steadier than in the homogeneous case. The pressure at the apex of the bow shock oscillates between $2\times10^3$ and
$2.5\times10^3\,{\rm dyn/cm^2}$, and the temperature of the post-shock gas there is again $\simeq10^{12}\,{\rm K}$. The ablation of mass from the outer, more dilute layers of the obstacle
occurs at a smoothly increasing rate,  until $t\simeq8\times10^5\,$s, when a significant portion of those layers is detached from the denser core. After that, almost the whole obstacle
beyond $2.5\times 10^{12}$~cm has been shocked, and the tail is loaded with this gas. At obstacle radii $<2.5\times 10^{12}$~cm the density is so high that the material is not affected by
the jet during the simulation (as expected for the stellar layers below the photosphere). 
Unlike in S1, shocking the obstacle external layers does not lead to an enlargement of the bow shock
cross-section, since the radial drop in density smoothens the process to some extent. As shown later, this effect is enhanced by a low resolution (see Sect.~\ref{res3}).

\begin{figure*}[!t]
  \includegraphics[clip,angle=0,width=0.48\textwidth]{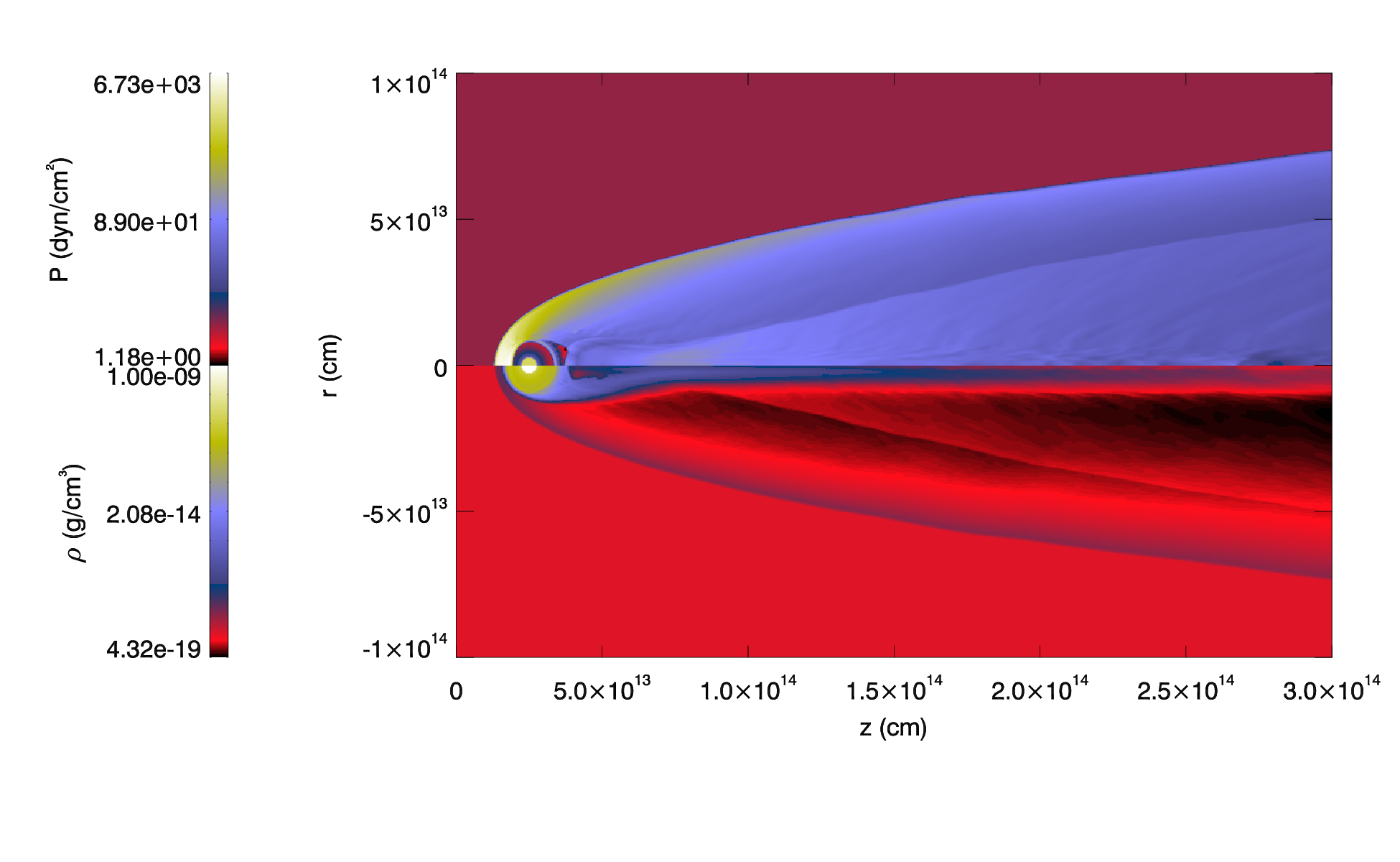}
  \includegraphics[clip,angle=0,width=0.48\textwidth]{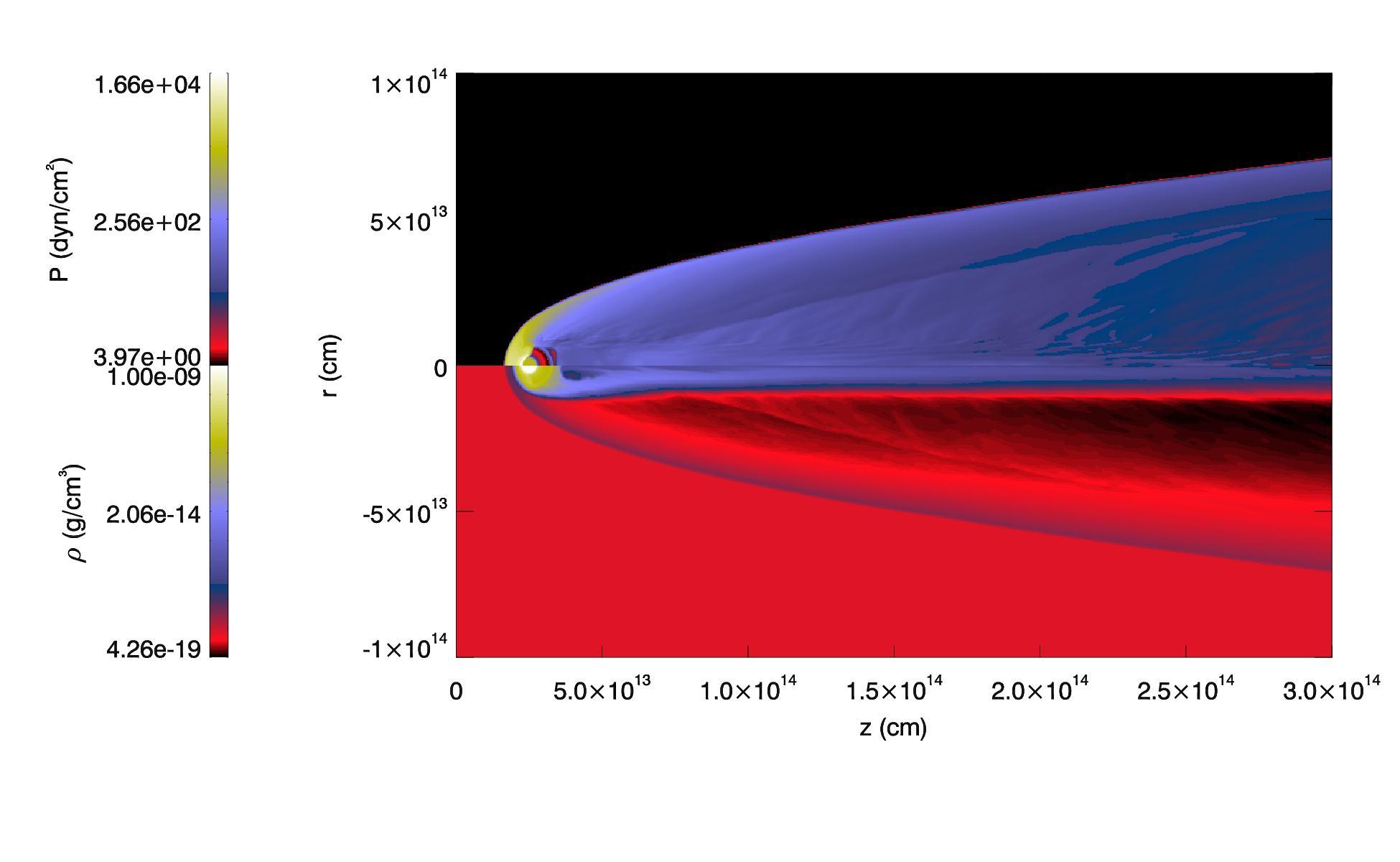}\\
  \includegraphics[clip,angle=0,width=0.48\textwidth]{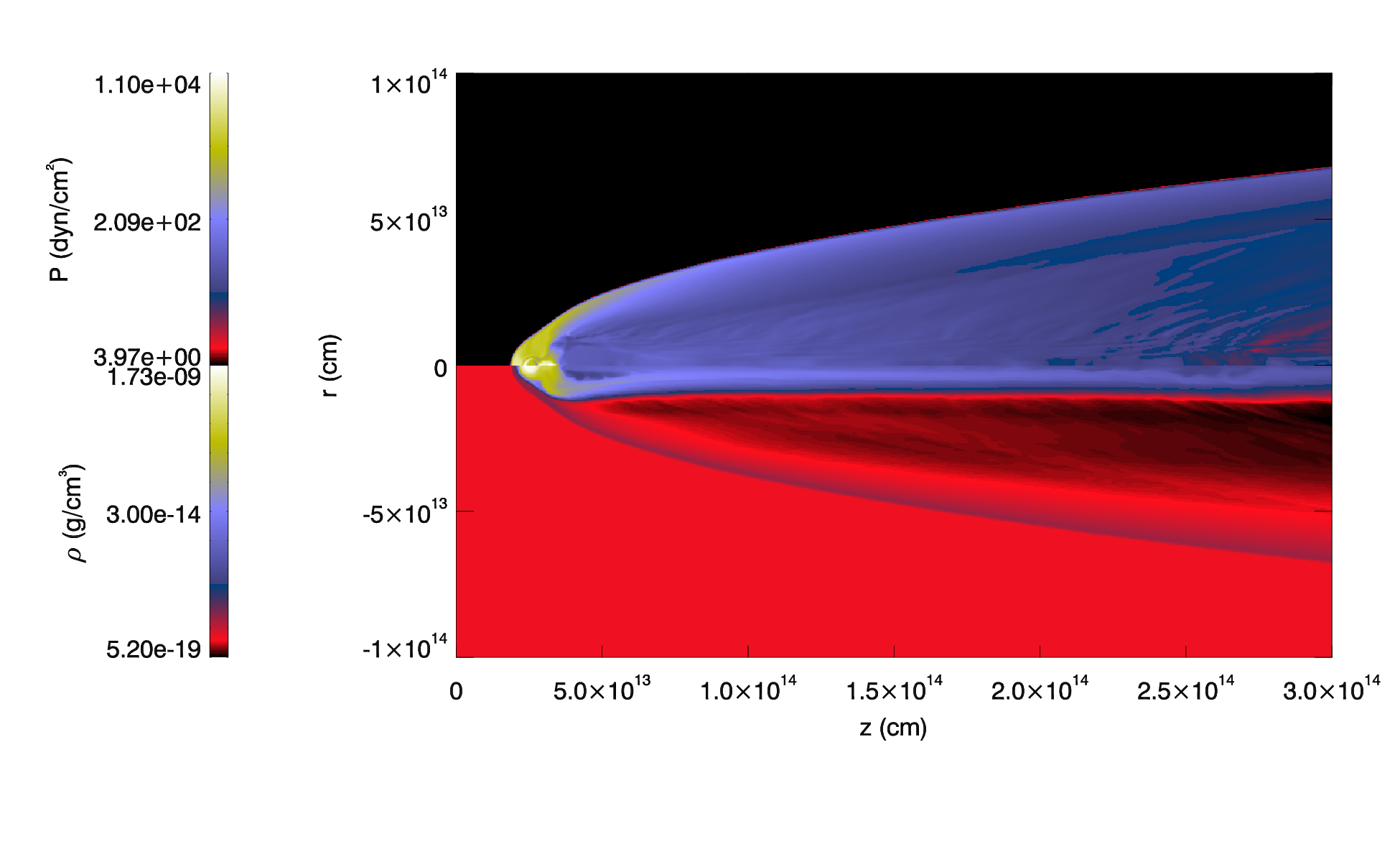}
 \includegraphics[clip,angle=0,width=0.48\textwidth]{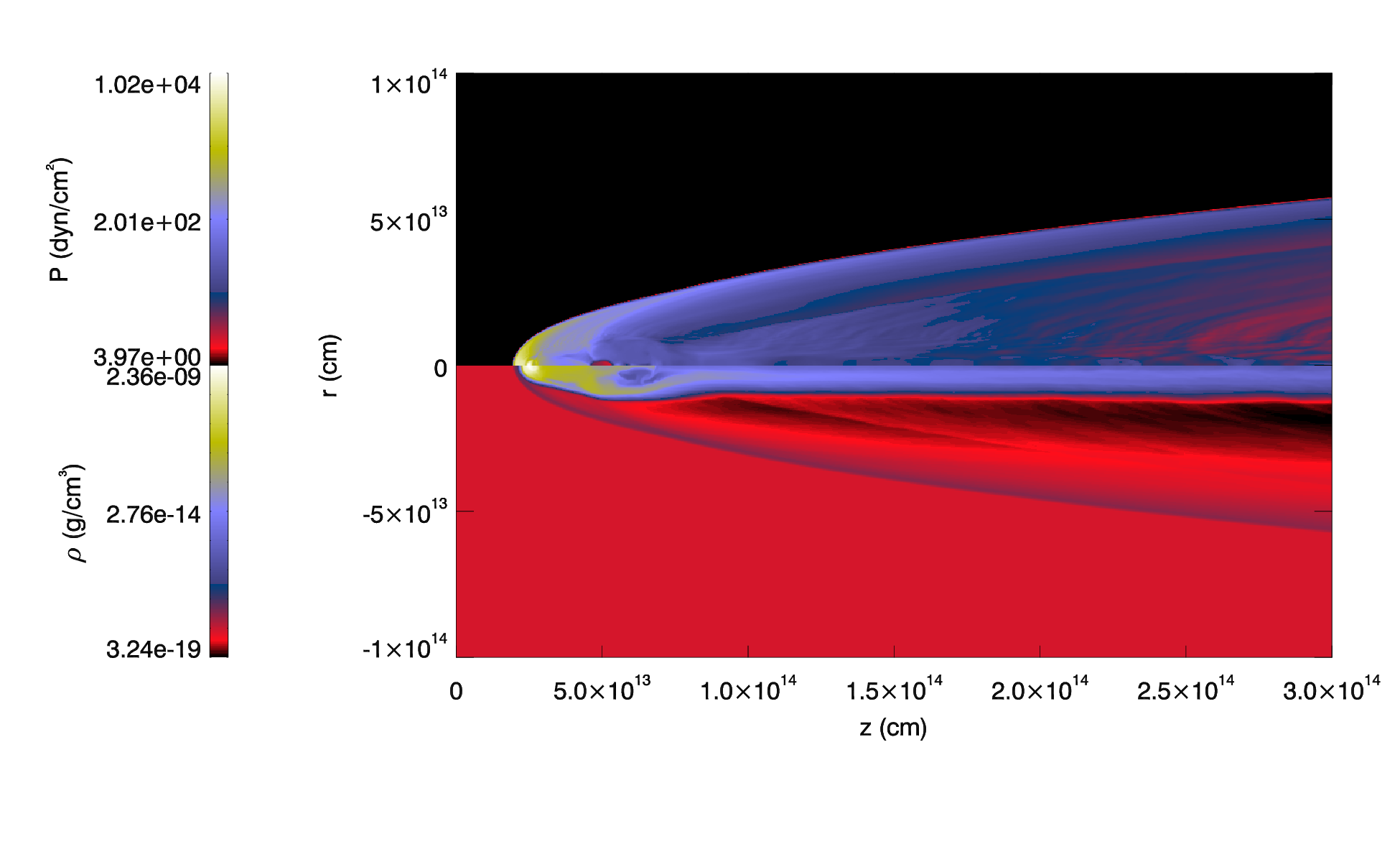}\\
  \includegraphics[clip,angle=0,width=0.48\textwidth]{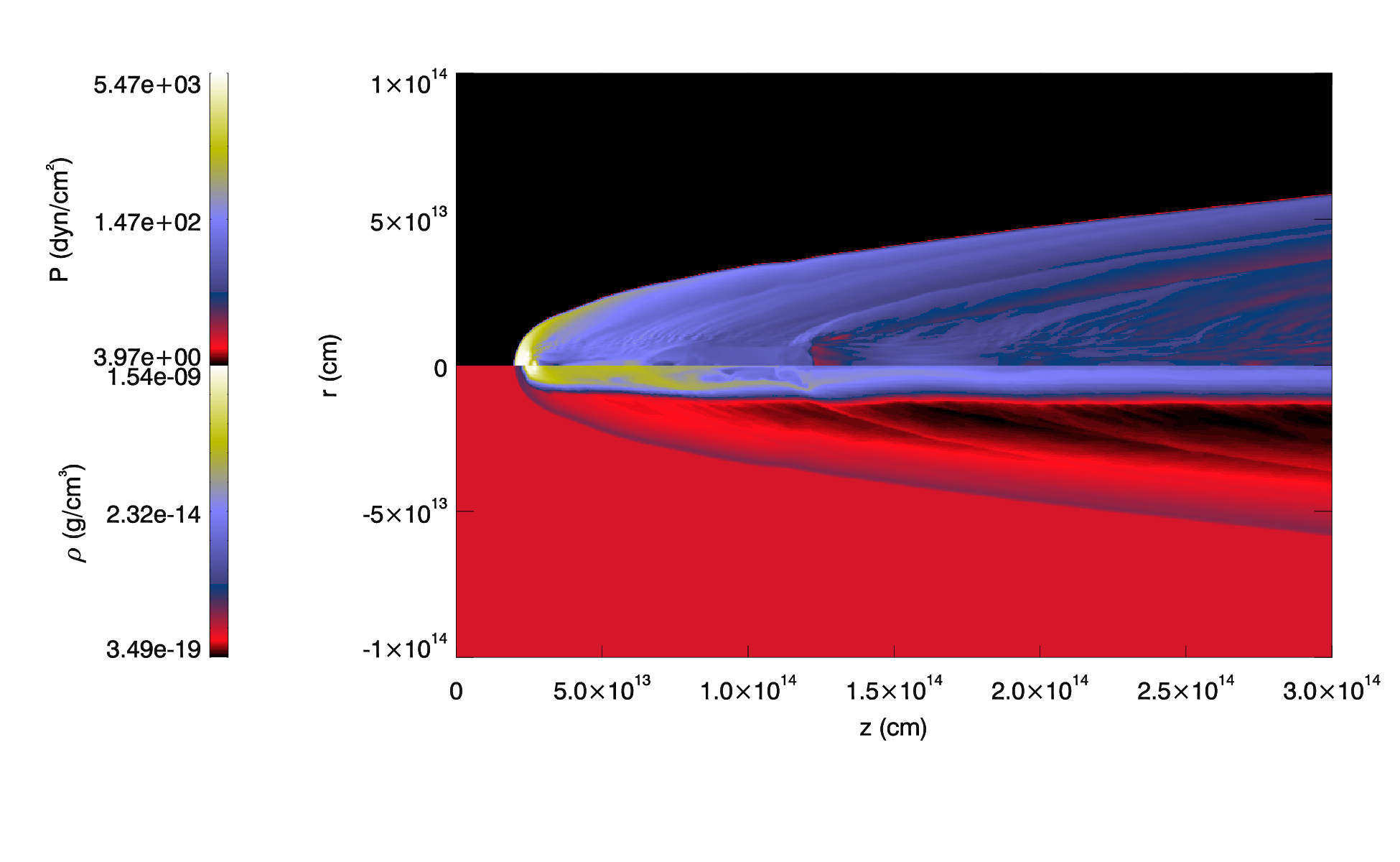}
 \includegraphics[clip,angle=0,width=0.48\textwidth]{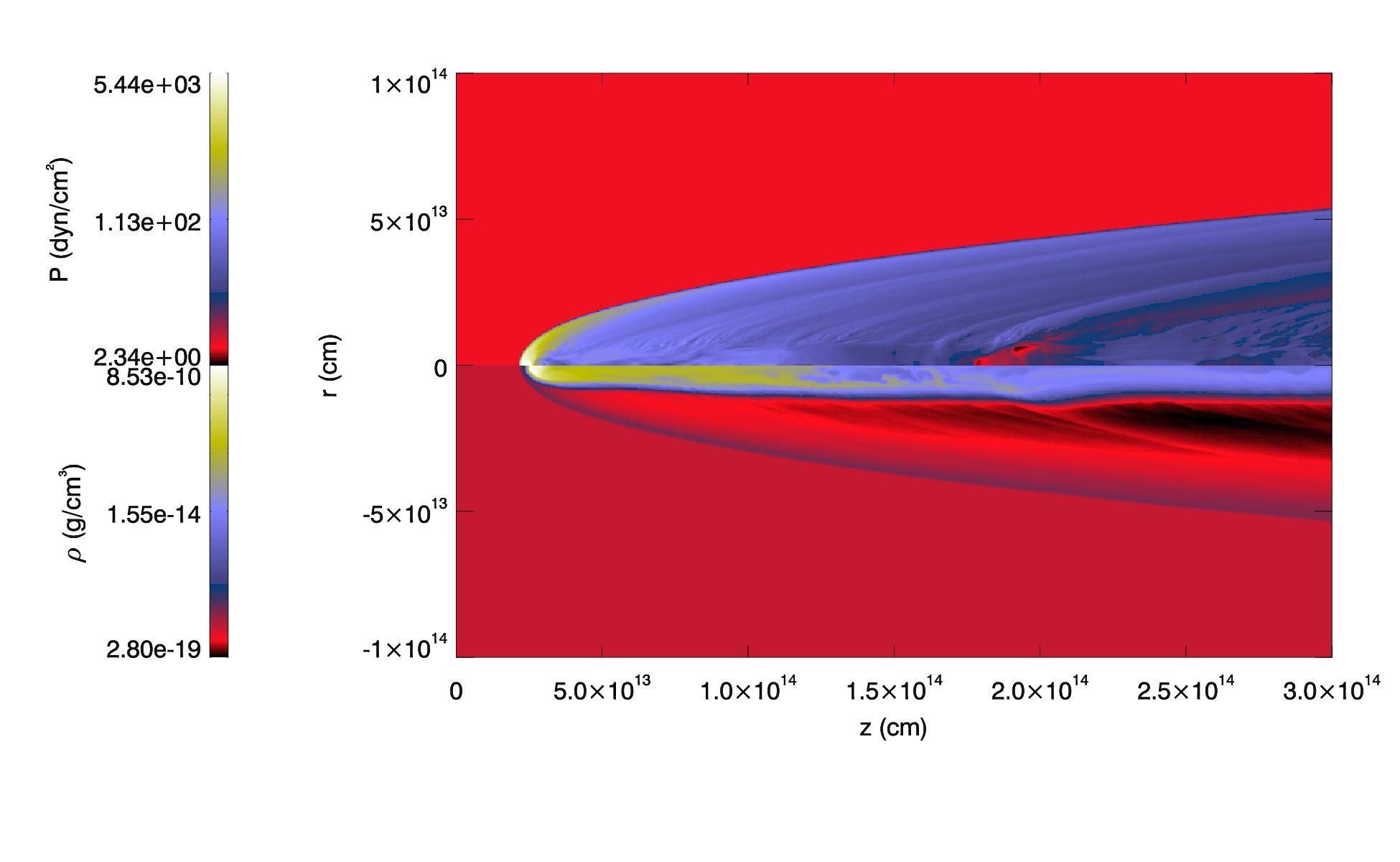}
  \caption{Same as left panel in Fig.~\ref{fig:maps1.1} for simulation S2 at times 
$t\simeq5\times10^4\,$s, $t\simeq1.6\times10^5\,$s, $t\simeq4.3\times10^5\,$s, $t\simeq7.9\times10^5\,$s,
 $t\simeq1.25\times10^6\,$s and $t\simeq1.8\times10^6\,$s from top to bottom and left to right.}
  \label{fig:maps2.1}
  \end{figure*}

Figure~\ref{fig:maps2.1} shows a series of maps of the whole process. The upper left panel corresponds to the initial phase. The next two panels (top right and central left), show the shock
travelling around the core and through the outer layers of the obstacle. In the central right panel, these layers have already been completely shocked and a large amount of gas that has
been detached from the core is visible as a bump in the tail between $z=5\times10^{13}\,{\rm cm}$ and $z=7\times10^{13}\,{\rm cm}$. The bottom panels show this bulk of mass propagating
downstream.     

\begin{figure*}[!t]
  \includegraphics[clip,angle=0,width=0.48\textwidth]{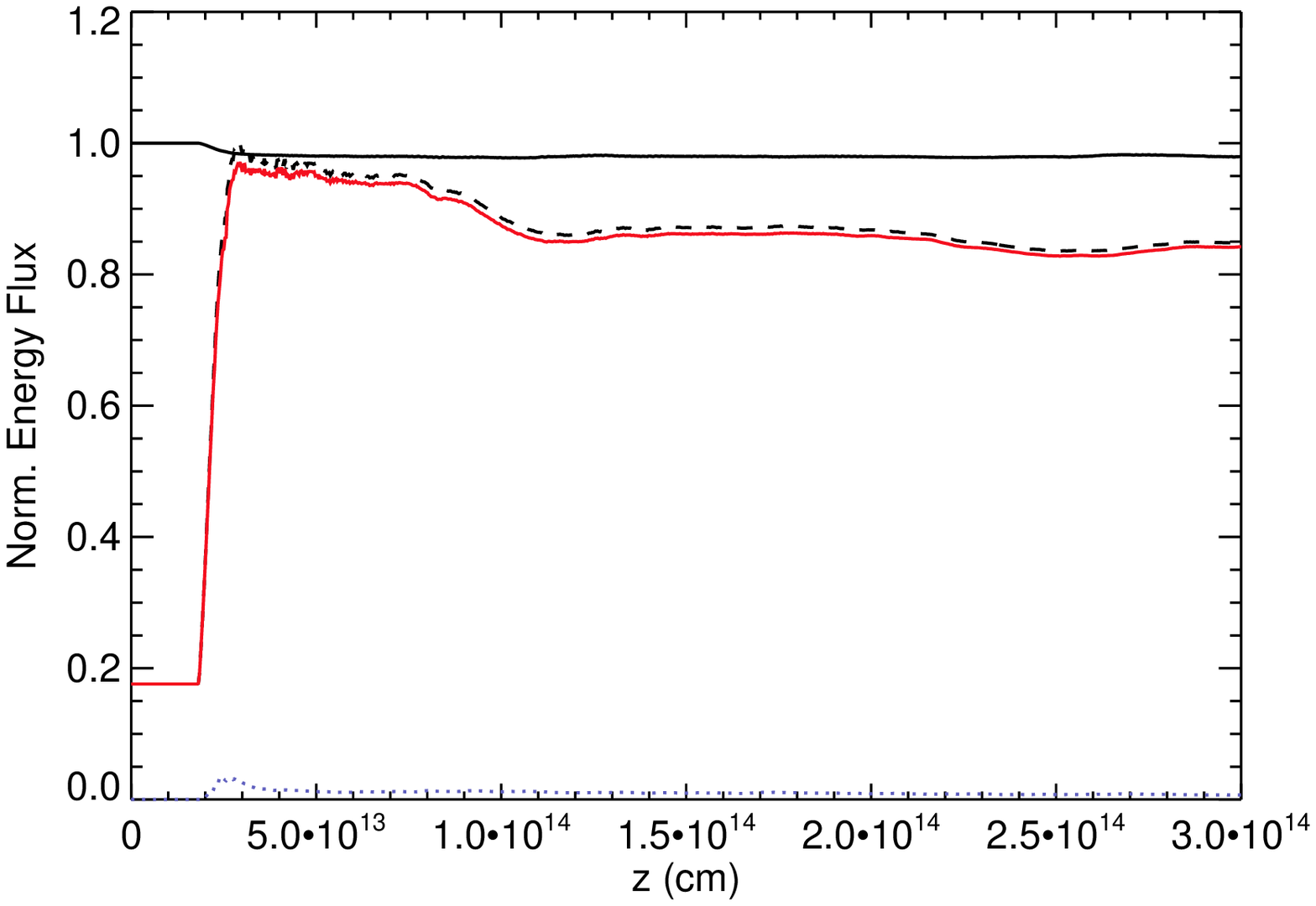}
  \includegraphics[clip,angle=0,width=0.48\textwidth]{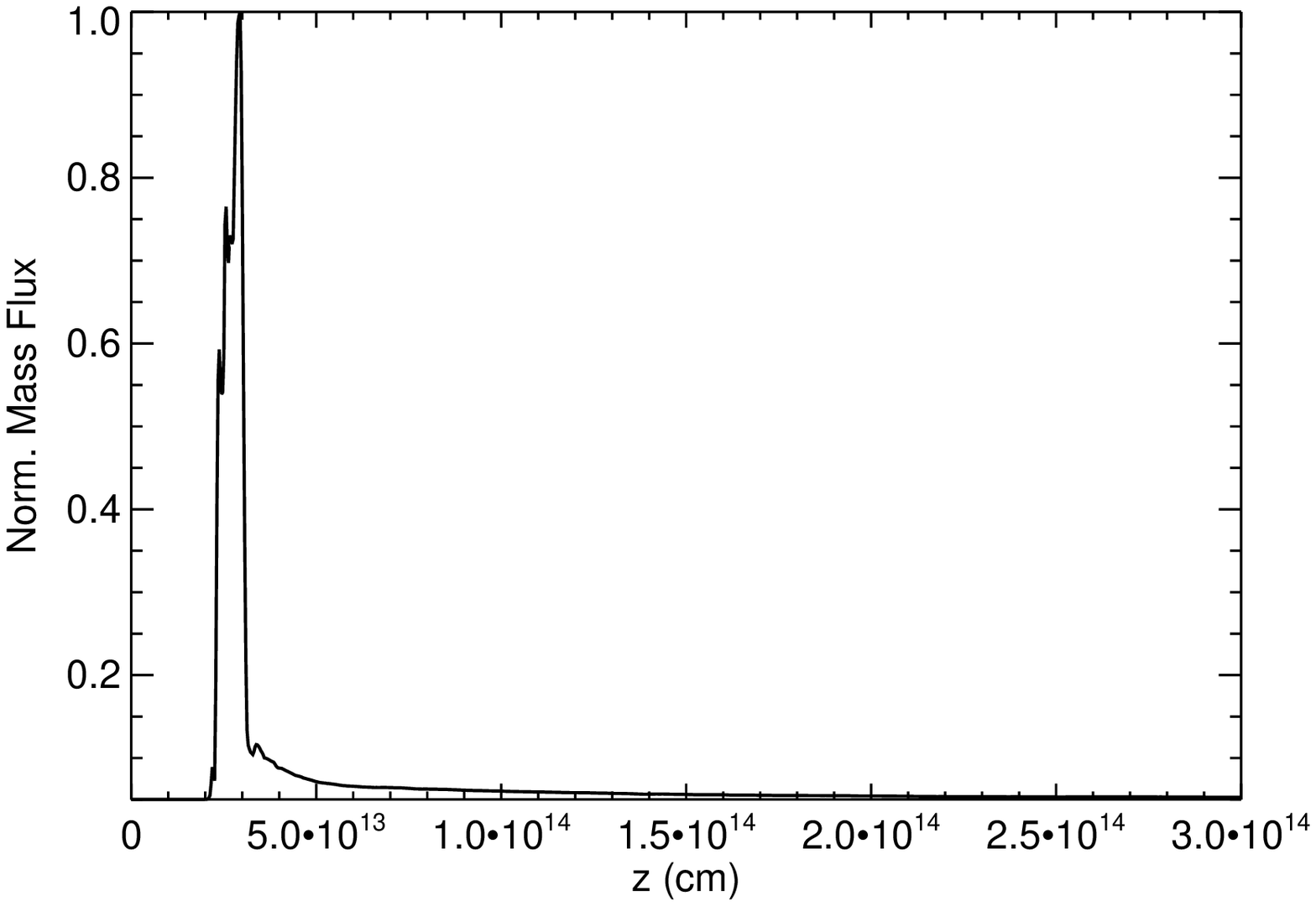}\\
  \includegraphics[clip,angle=0,width=0.48\textwidth]{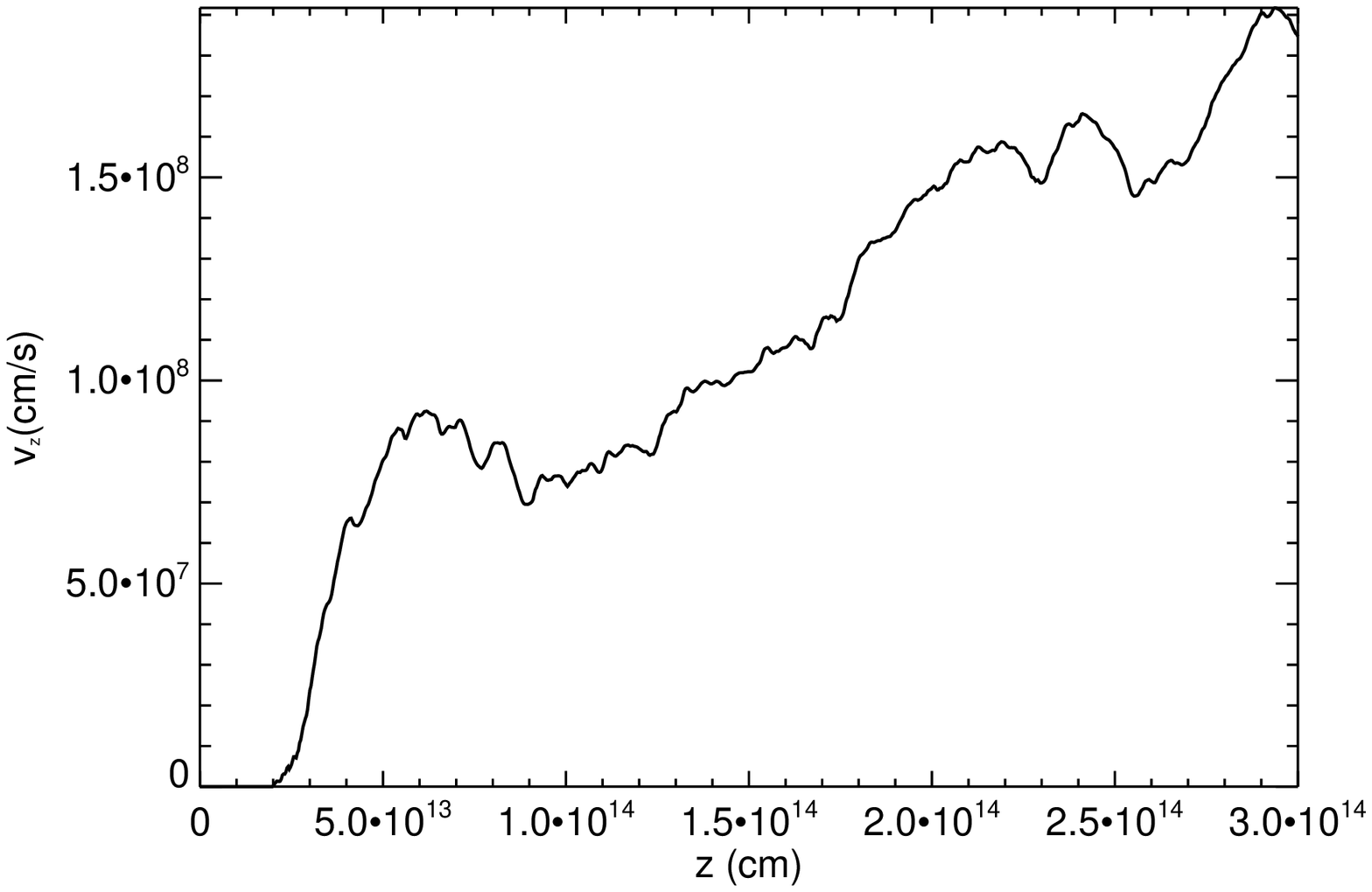}
  \includegraphics[clip,angle=0,width=0.48\textwidth]{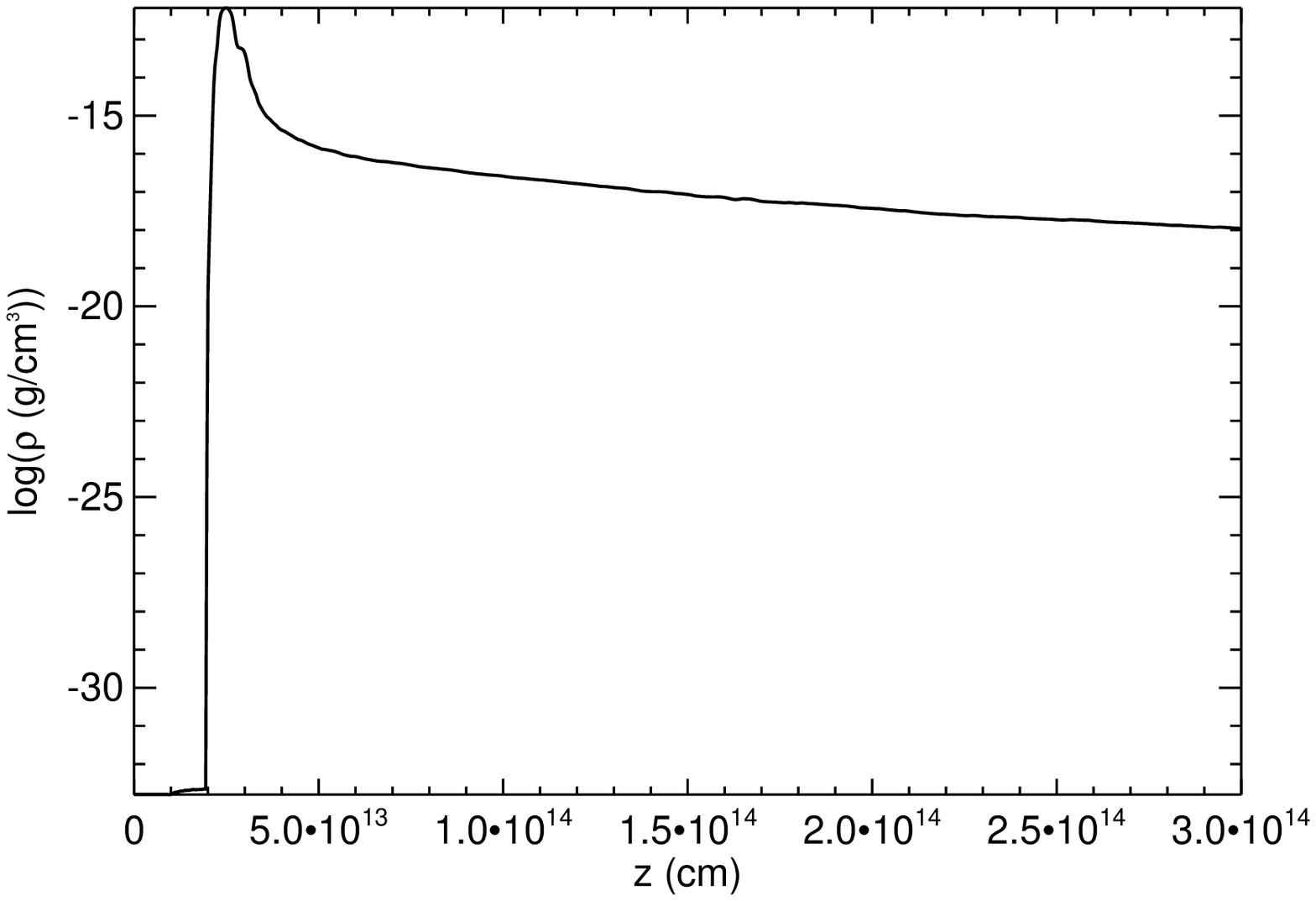}
  \caption{Axial cuts of different quantities for simulation S2 at $t\simeq 3.1\times 10^5\,$s, 
during the initial phase, before the shock completely crosses the obstacle. 
The top left panel shows $L_{\rm j,g1}-L_{\rm j,int,g1}$ for the jet material (solid line) 
normalized to its injection value $3.5\times10^{42}\,{\rm erg/s}$, and the internal 
energy luminosity $L_{\rm j,int,g1}$ normalized to its maximum value, $5.6\times10^{40}\,{\rm erg/s}$ (dashed line); 
the red solid line stands for the internal energy luminosity in the jet material, and the blue dotted line for the internal 
energy luminosity in the obstacle material. The top right panel shows the mass flux, normalized to its maximum value 
($4\times10^{22}\,{\rm g/s}$). The bottom left and right panels show 
the mean velocity and density in the obstacle gas.}
  \label{fig:S2.1}
  \end{figure*}

\begin{figure*}[!t]
  \includegraphics[clip,angle=0,width=0.48\textwidth]{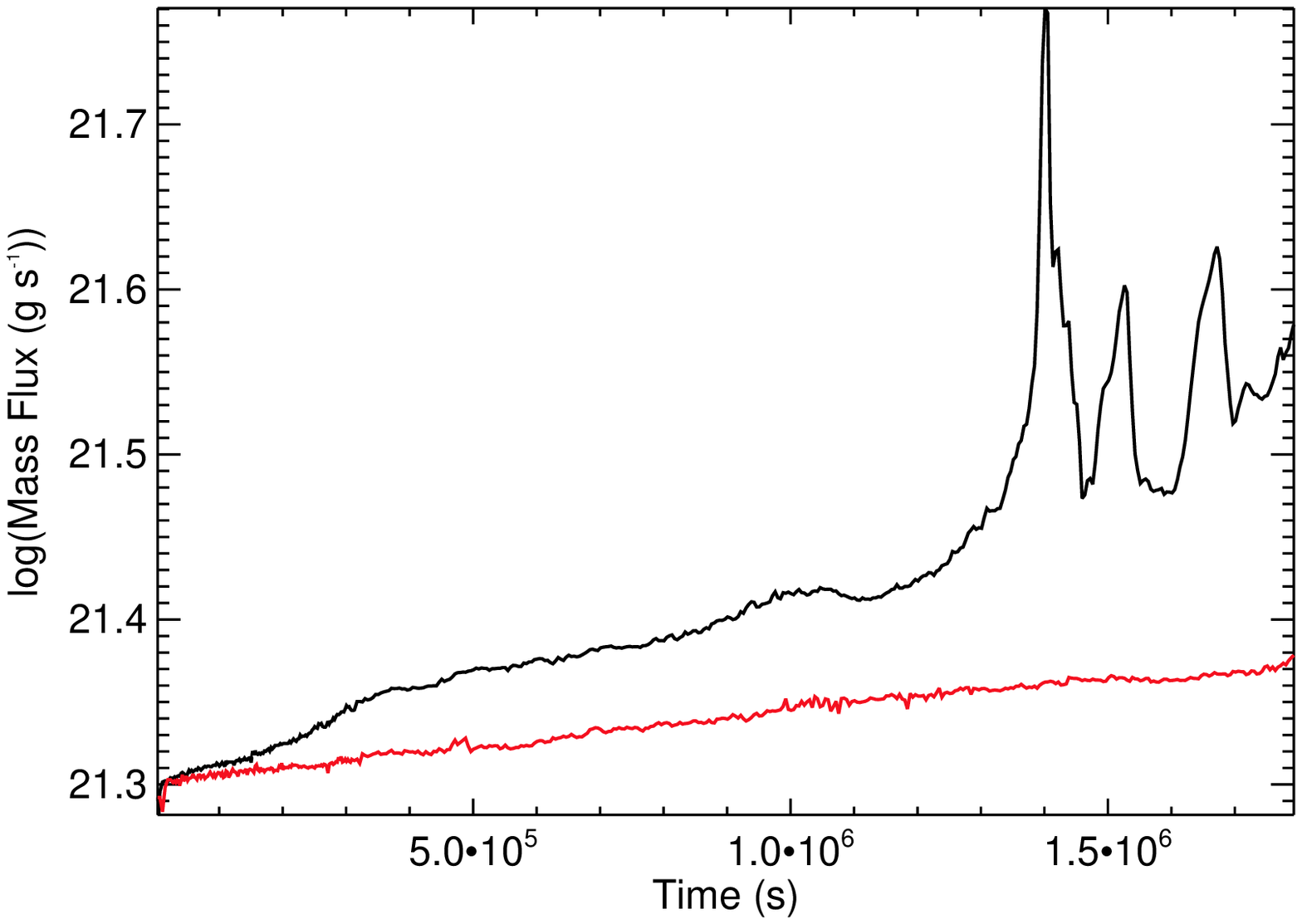}
  \includegraphics[clip,angle=0,width=0.48\textwidth]{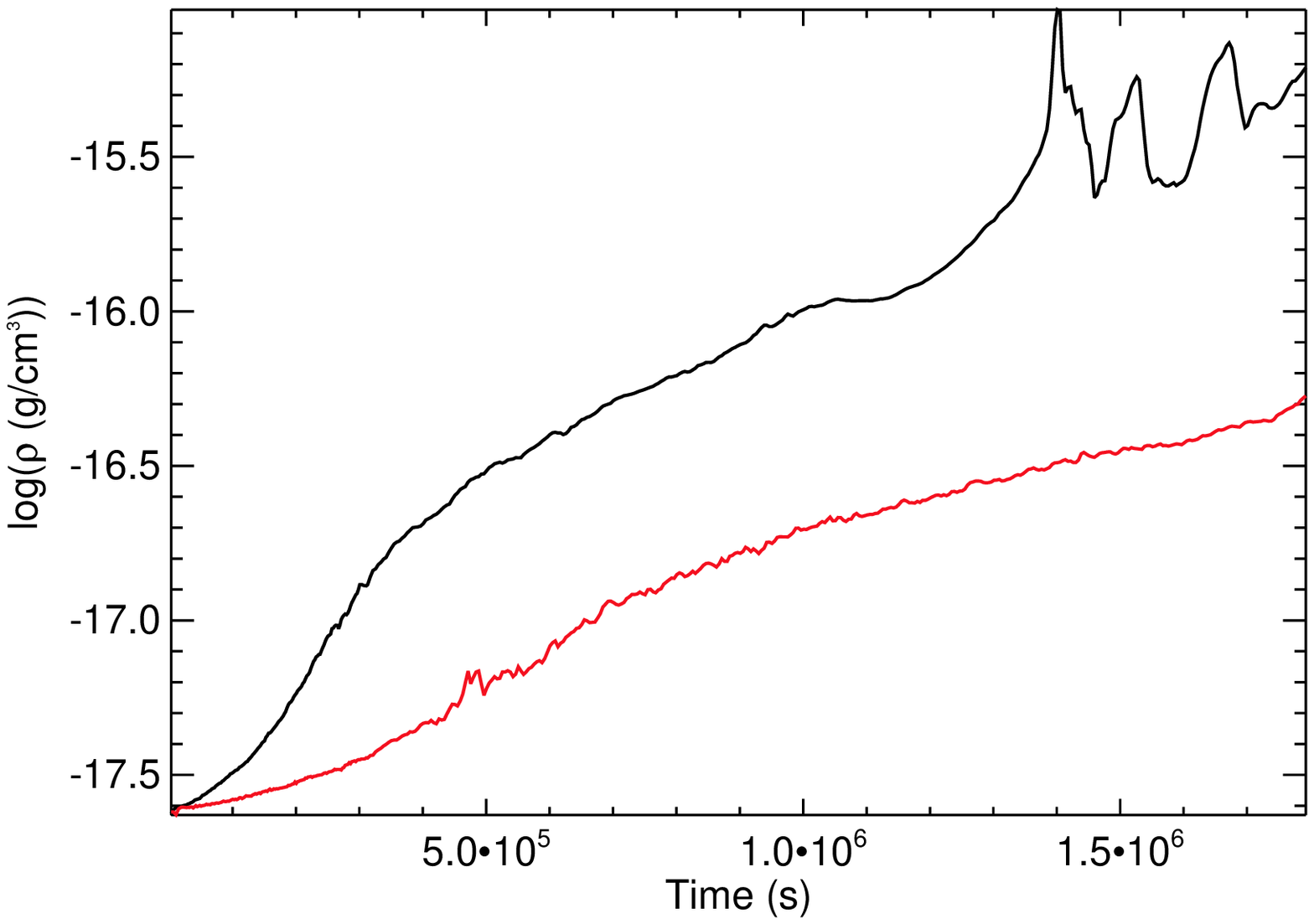}
  \caption{Mass flux (left) and mean density (right) versus time for simulation S2 at 
$z=1.5\times10^{14}$ 
(half grid, black lines) and $3\times10^{14}\,{\rm cm}$ (end of the grid, red lines).}
  \label{fig:S2.2}
  \end{figure*}

The upper panels in Fig.~\ref{fig:S2.1} show the normalized $L_{\rm j,g2}-L_{\rm j,int,g2}$ and mass flux versus distance along the axis at $t\simeq 3.1\times 10^5\,$s, i.e.
still during the initial phase. As seen in the figure, the jet loses a small amount of kinetic energy flux, which is used to heat the jet material and to heat and accelerate obstacle material.
The mass flux falls a factor of ten from the obstacle itself to its tail, where it has a constant value. The lower panels in Fig.~\ref{fig:S2.1} show the values of mean velocity and density
of the obstacle gas versus $z$ at the same time. The values obtained for the density in this initial quasi-steady regime of S2 are similar to those obtained
for S1. Regarding velocity, the obstacle material is accelerated to $v\simeq1.9\times10^8\,{\rm cm/s}$ at the end of the grid, which is a factor of 3 larger than the velocity reached by the
obstacle material in S1 at the same axial distance. This difference is in part due to the smaller total mass flux in S2 (see left panels in Figs.~\ref{fig:S1.2} and
\ref{fig:S2.2}), caused by the lower density of the external layers. On the symmetry axis, the maximum axial velocity is $v_{\rm z}\simeq 10^9\,{\rm cm/s}$ at the end of the grid, as in the case of S1, but at $z=3\times10^{14}\,{\rm cm}$,
i.e., half the distance needed in S1 to reach the same velocity. 

Figure~\ref{fig:S2.2} shows the mass flux within the grid and the mean density for two different $z$-values, namely at half ($z=1.5\times10^{14}\,{\rm cm}$) and at the end of the grid 
($z=3\times10^{14}\,{\rm cm}$), versus time. These plots show that the mass load grows smoothly with time at  $z=1.5\times10^{14}\,{\rm cm}$ until $t\simeq1.4\times10^6\,$s, which
corresponds to the time of passage of the clump of mass ablated around $t\simeq8\times10^5\,$s. By the end of the simulation, this clump has not still reached the end of the grid. The mean
velocity of the simulated portion of the jet is not significantly reduced during the whole simulation.     

\subsubsection{Inhomogeneous case: high resolution}\label{res3}

\begin{figure*}[!t]
  \includegraphics[clip,angle=0,width=0.48\textwidth]{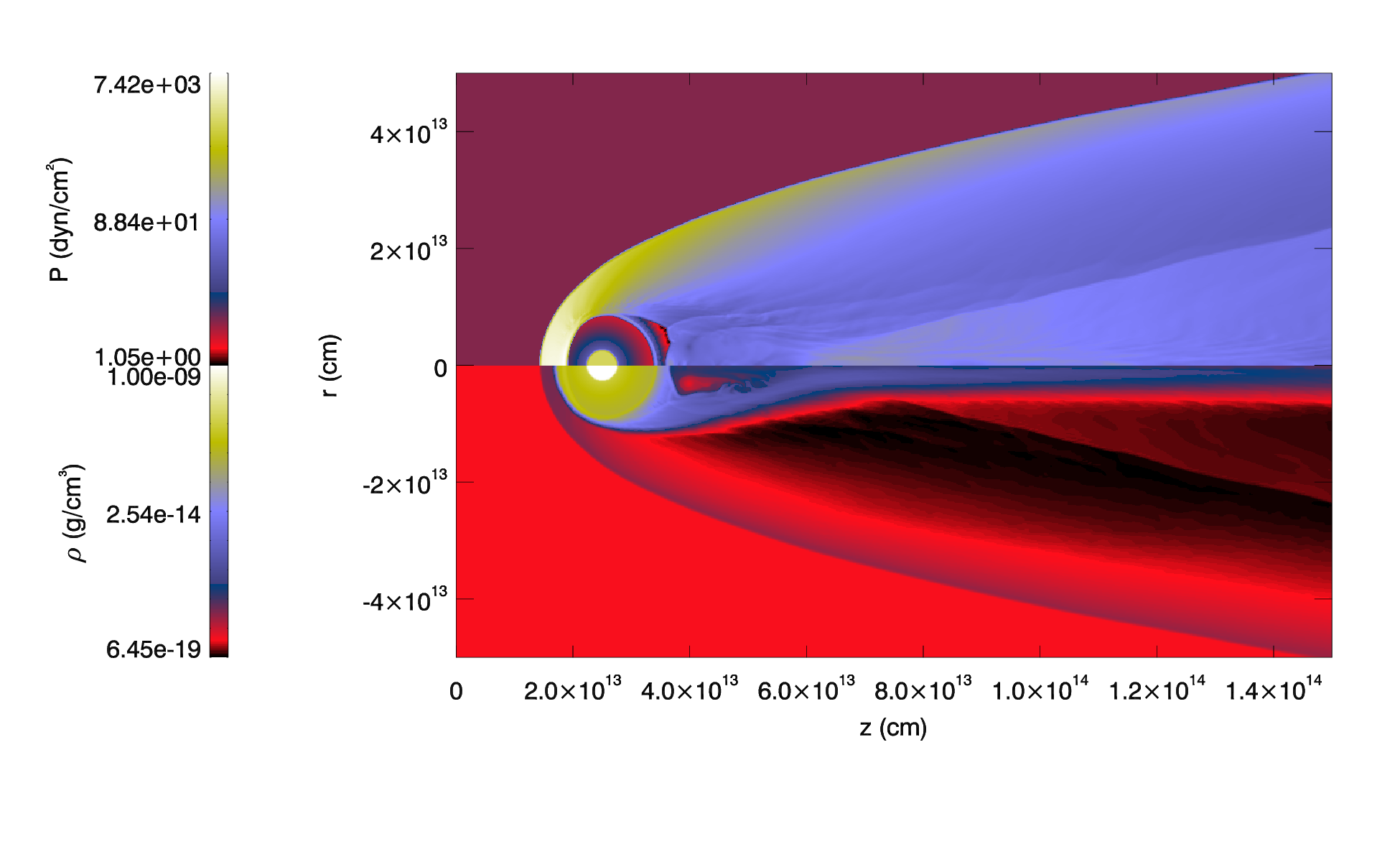}
  \includegraphics[clip,angle=0,width=0.48\textwidth]{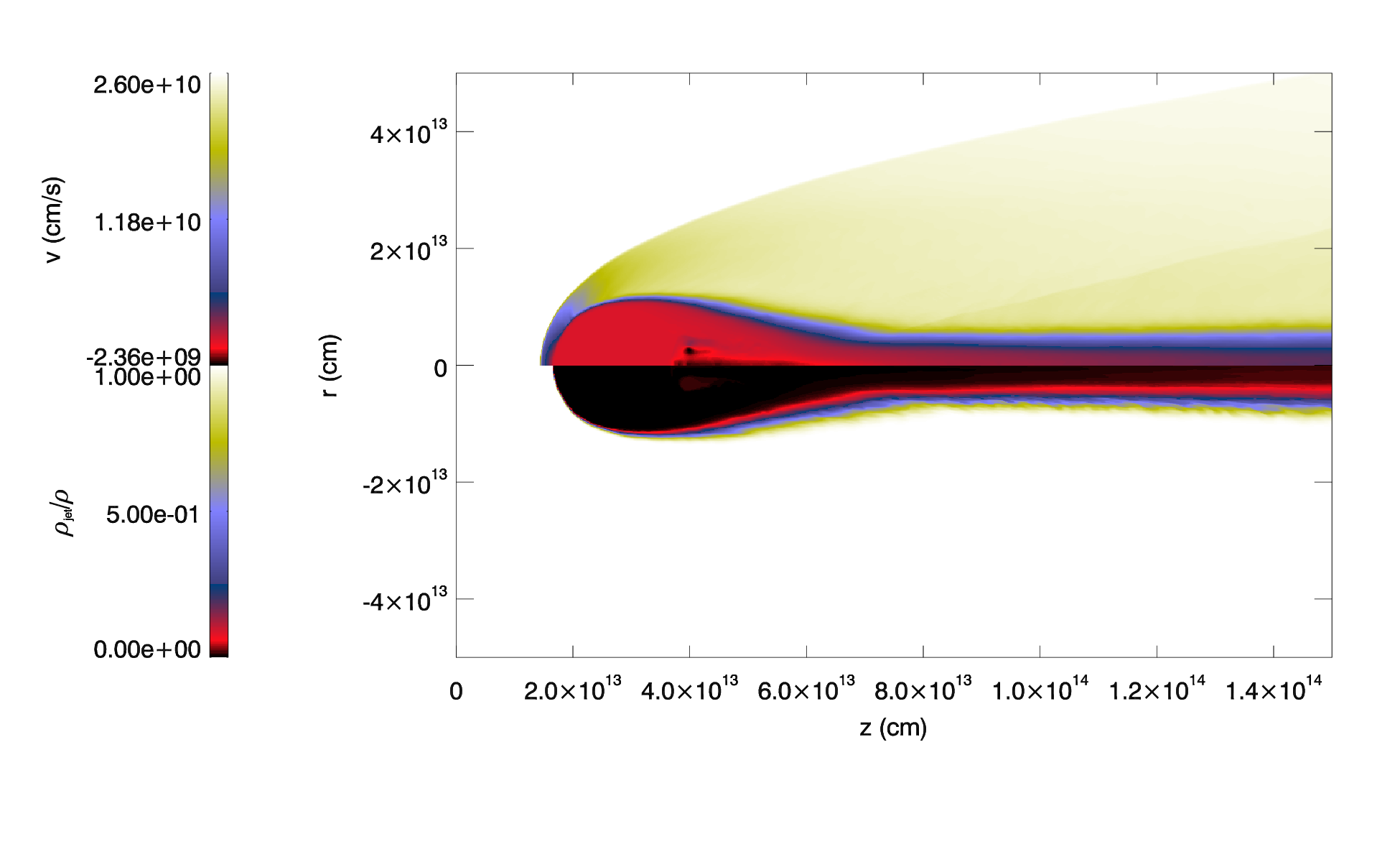}\\
  \includegraphics[clip,angle=0,width=0.48\textwidth]{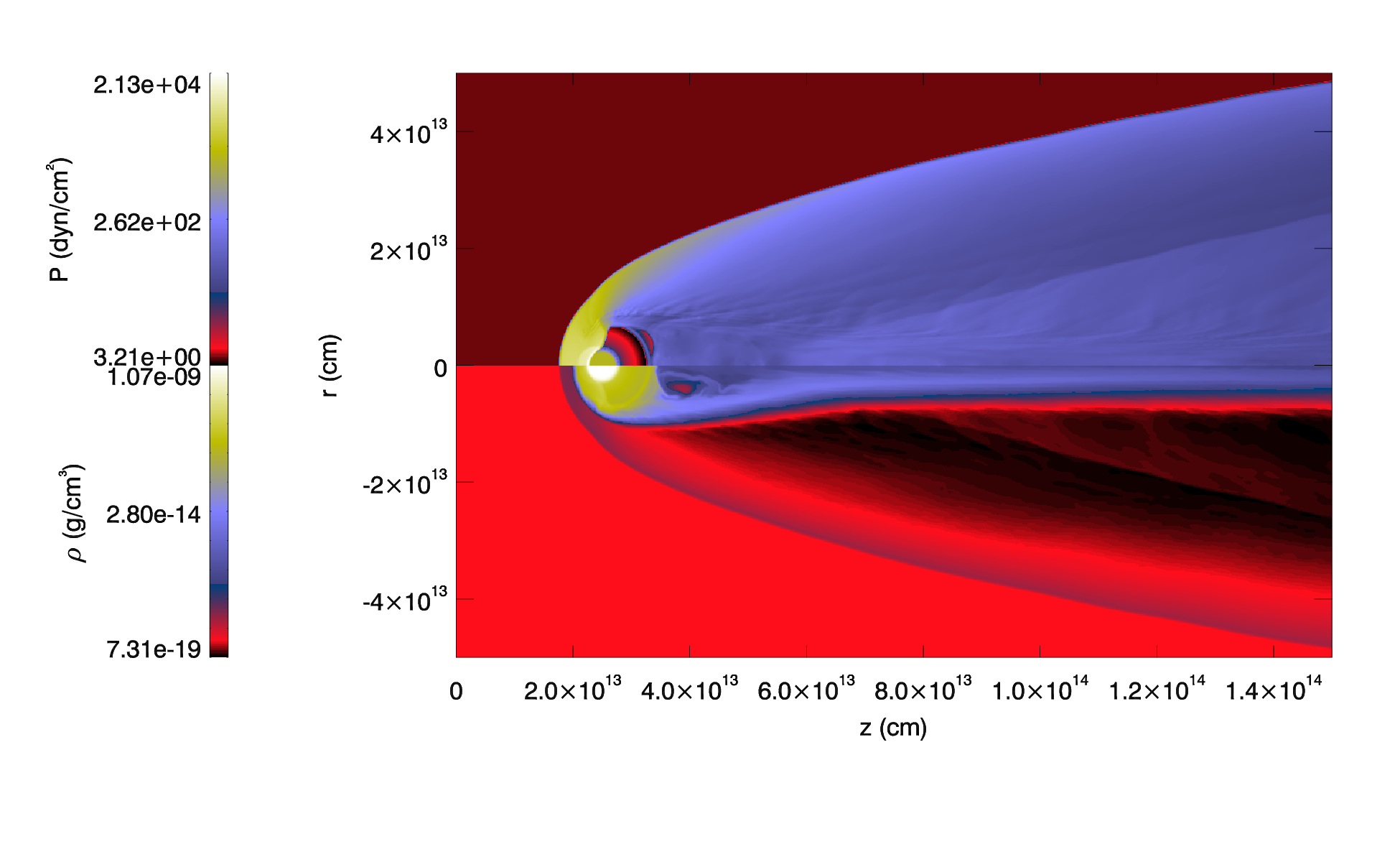}
 \includegraphics[clip,angle=0,width=0.48\textwidth]{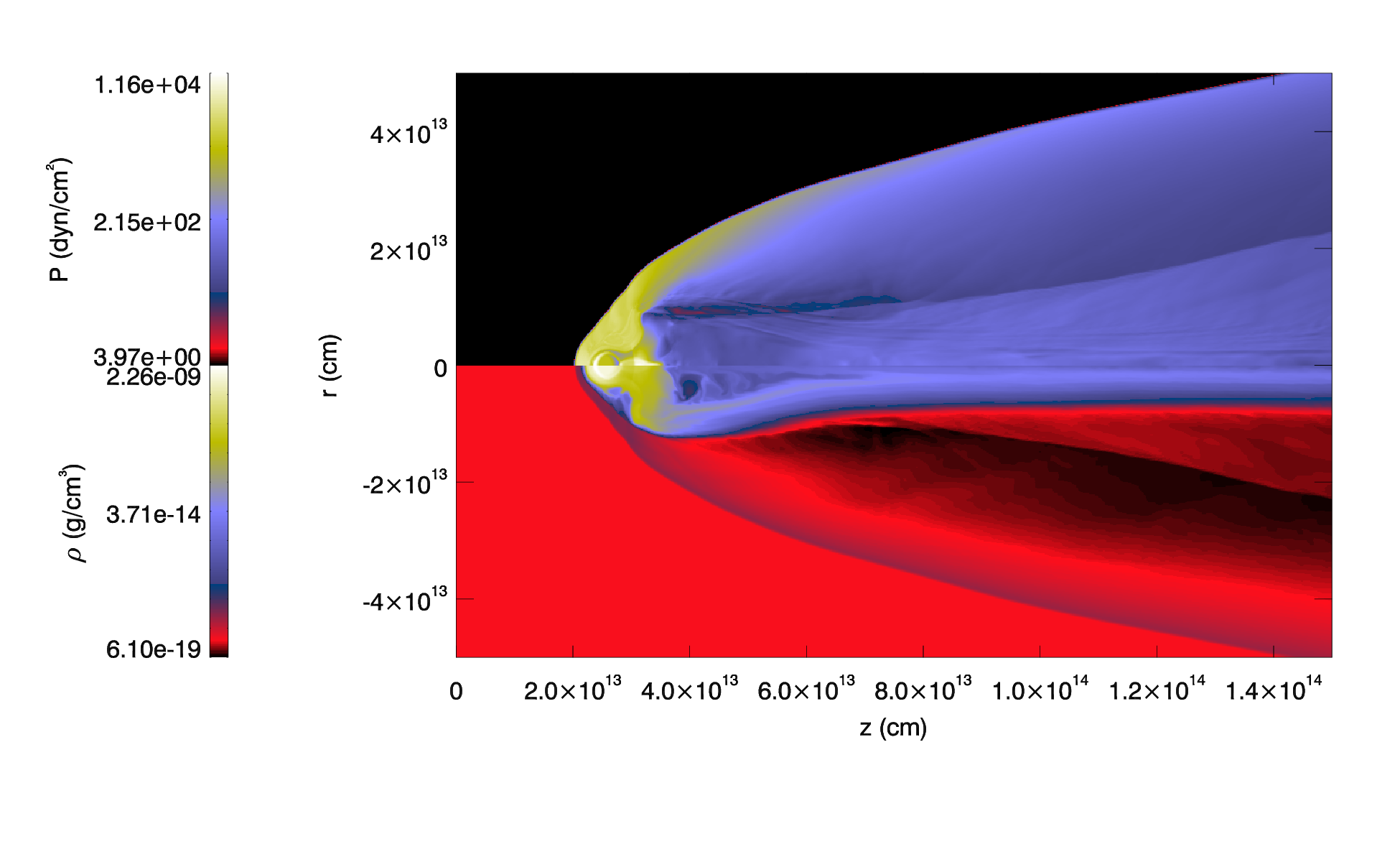}\\
  \includegraphics[clip,angle=0,width=0.48\textwidth]{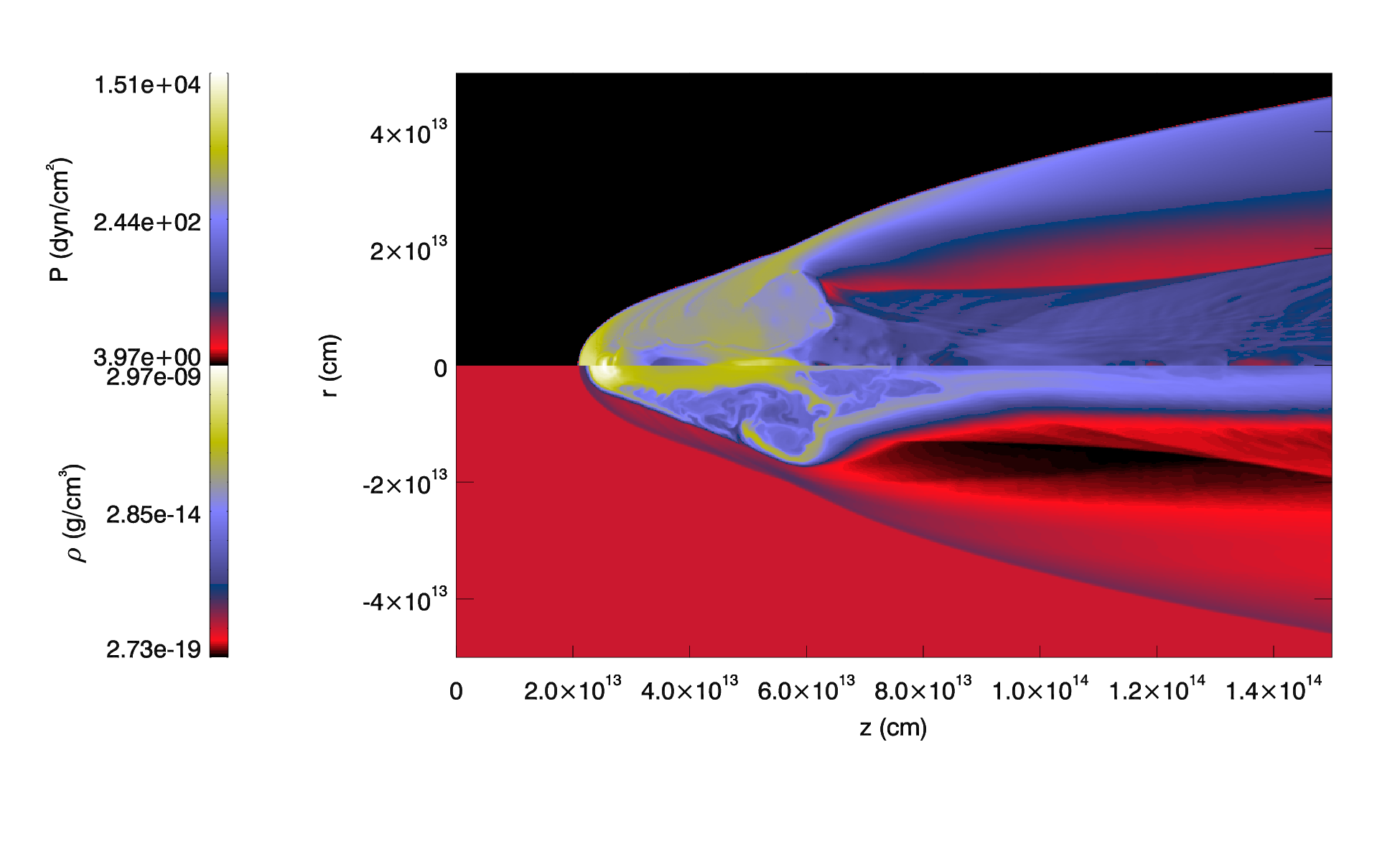}
 \includegraphics[clip,angle=0,width=0.48\textwidth]{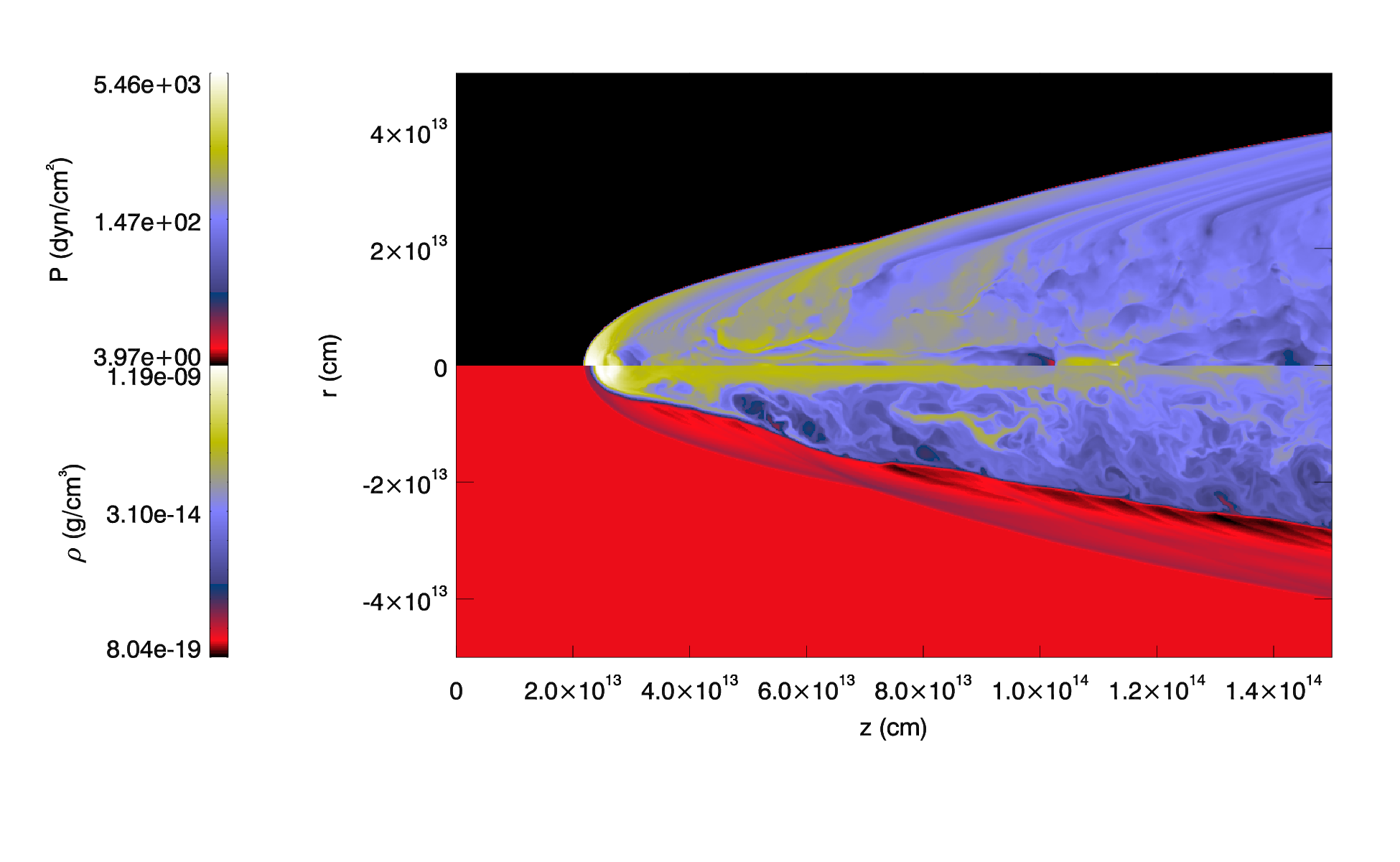}
  \caption{The upper panels show the same parameters as Fig.~\ref{fig:maps1.1} for simulation S3 at time 
$t\simeq5\times10^4\,$s. The central and bottom panels are similar to the left ones in Fig.~\ref{fig:maps1.1} at 
$t\simeq1.6\times10^5\,$s, $t\simeq4.3\times10^5\,$s, $t\simeq7.9\times10^5\,$s, and $t\simeq1.25\times10^6\,$s from top to bottom and left to right.}
  \label{fig:maps3.1}
  \end{figure*}

Simulation S3 reproduces a region of S2 with double resolution, and half of its grid size along and across, resulting in a physical grid size of $5\times10^{13}\,{\rm cm}\,
\times\,1.5\times10^{14}\,{\rm cm}$. From the results, we see that the apex pressure and temperature in S3 are very similar to those in S2. Figure~\ref{fig:maps3.1} shows different snapshots of the simulation, at the same
instants as for S2 but stopping at $t\simeq1.25\times10^6\,$s. At first sight, both simulations follow the same phases, i.e., a quasi-steady phase in which the outer layers of the
obstacle are being crossed by the shock and the mass-loss is smoothly increasing, and a second phase, starting when the outer region has been completely crossed by the shock, and a large
amount of gas is detached from the core and dragged downstream. However, the main difference, which can only be related to the numerical resolution, is that this detached clump of gas is
larger and produces some widening of the jet-obstacle interaction cross-section (compare Figs.~\ref{fig:maps2.1} and \ref{fig:maps3.1} at $t\simeq7.9\times10^5\,$s), after the whole outer layers of the obstacle have
been shocked at  $t\simeq4\times10^5\,$s. Thus, this material receives a larger transfer of energy from the jet, and is accelerated in a more efficient way, reaching
$z=1.5\times10^{14}\,{\rm cm}$ at $t\simeq10^6\,$s, compared to $t\simeq1.4\times10^6\,$s in S2, as shown in Fig.~\ref{fig:S3.2}.

\begin{figure*}[!t]
  \includegraphics[clip,angle=0,width=0.32\textwidth]{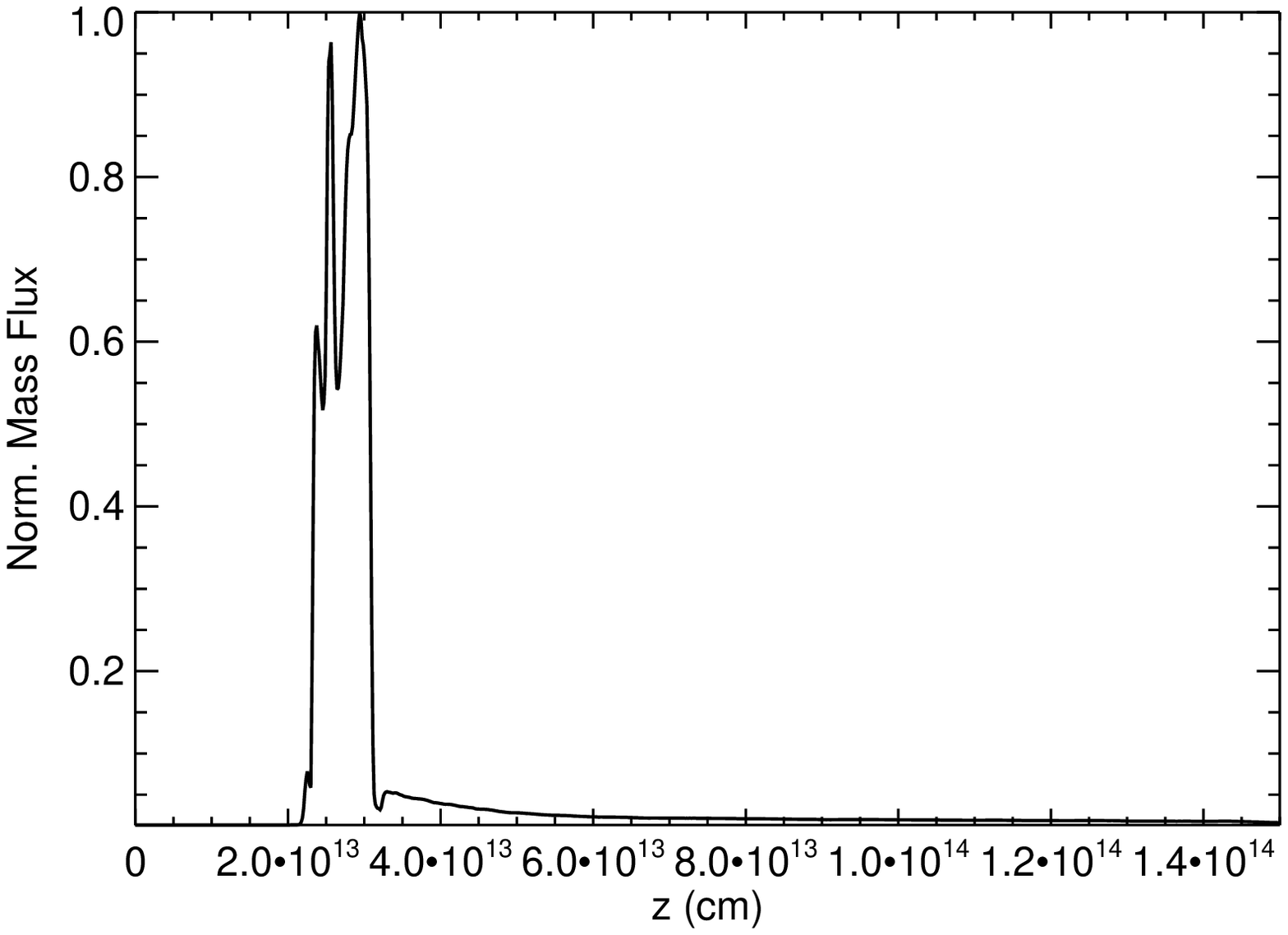}
  \includegraphics[clip,angle=0,width=0.32\textwidth]{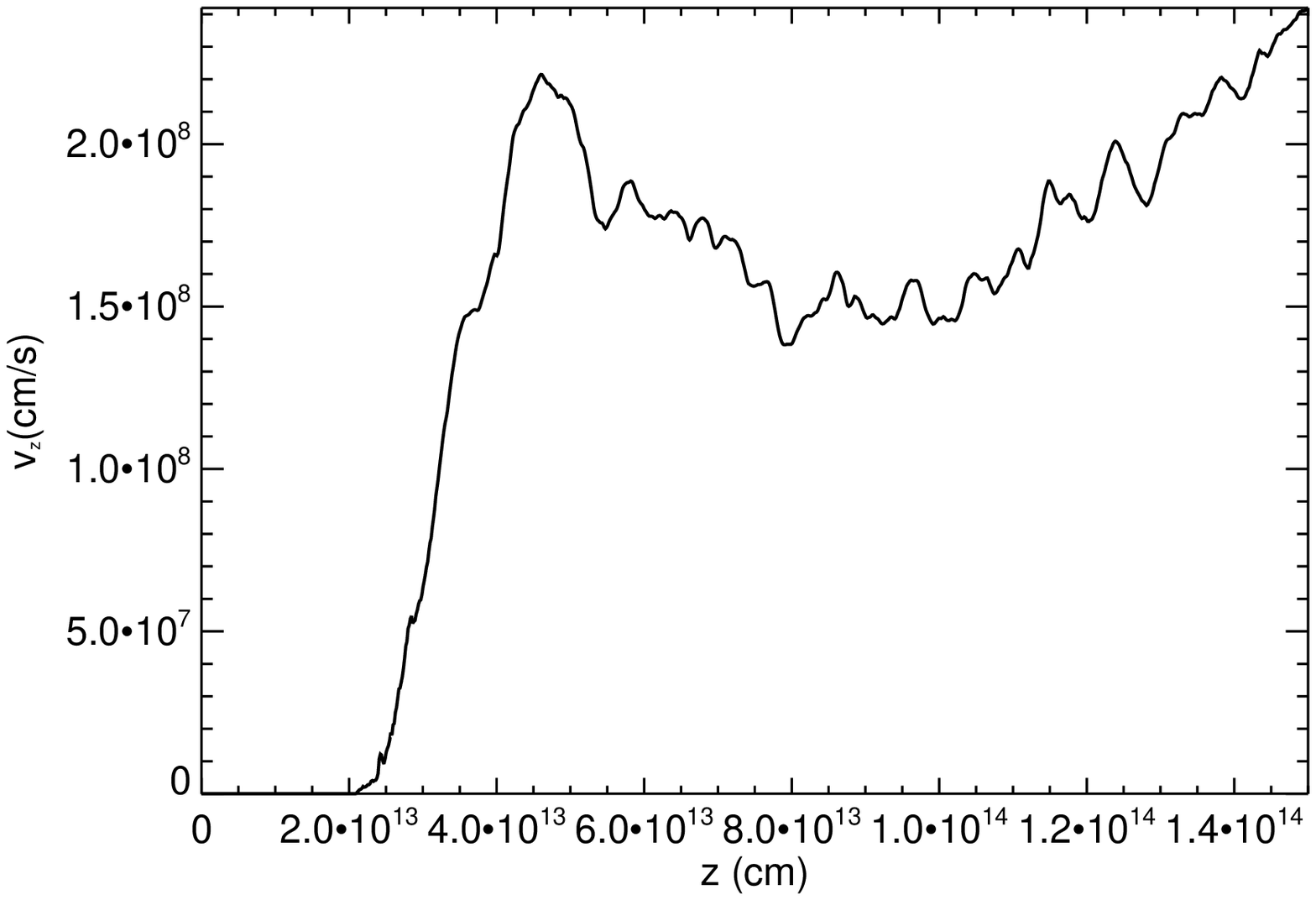}
  \includegraphics[clip,angle=0,width=0.32\textwidth]{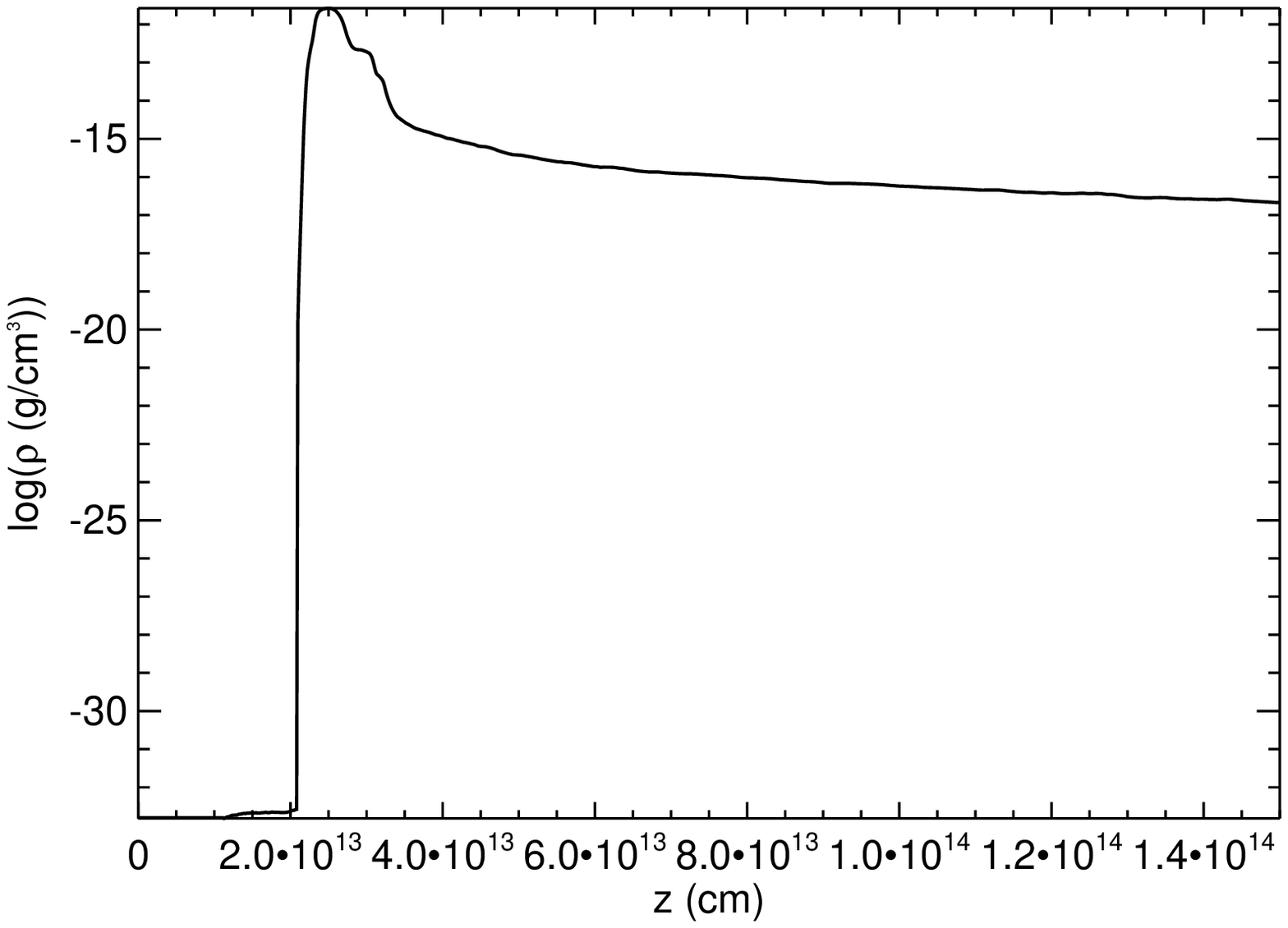}
  \caption{Axial cuts of different quantities for simulation S3 at $t\simeq 3.1\times 10^5\,$s, 
during the initial phase, before the shock completely crosses the obstacle. 
The left panel shows the normalized mass flux, normalized to its maximum value, $3.7\times10^{22}\,{\rm g/s}$. The central and right panels show 
the mean velocity and density in the obstacle gas.}
  \label{fig:S3.1}
  \end{figure*}

\begin{figure*}[!t]
  \includegraphics[clip,angle=0,width=0.32\textwidth]{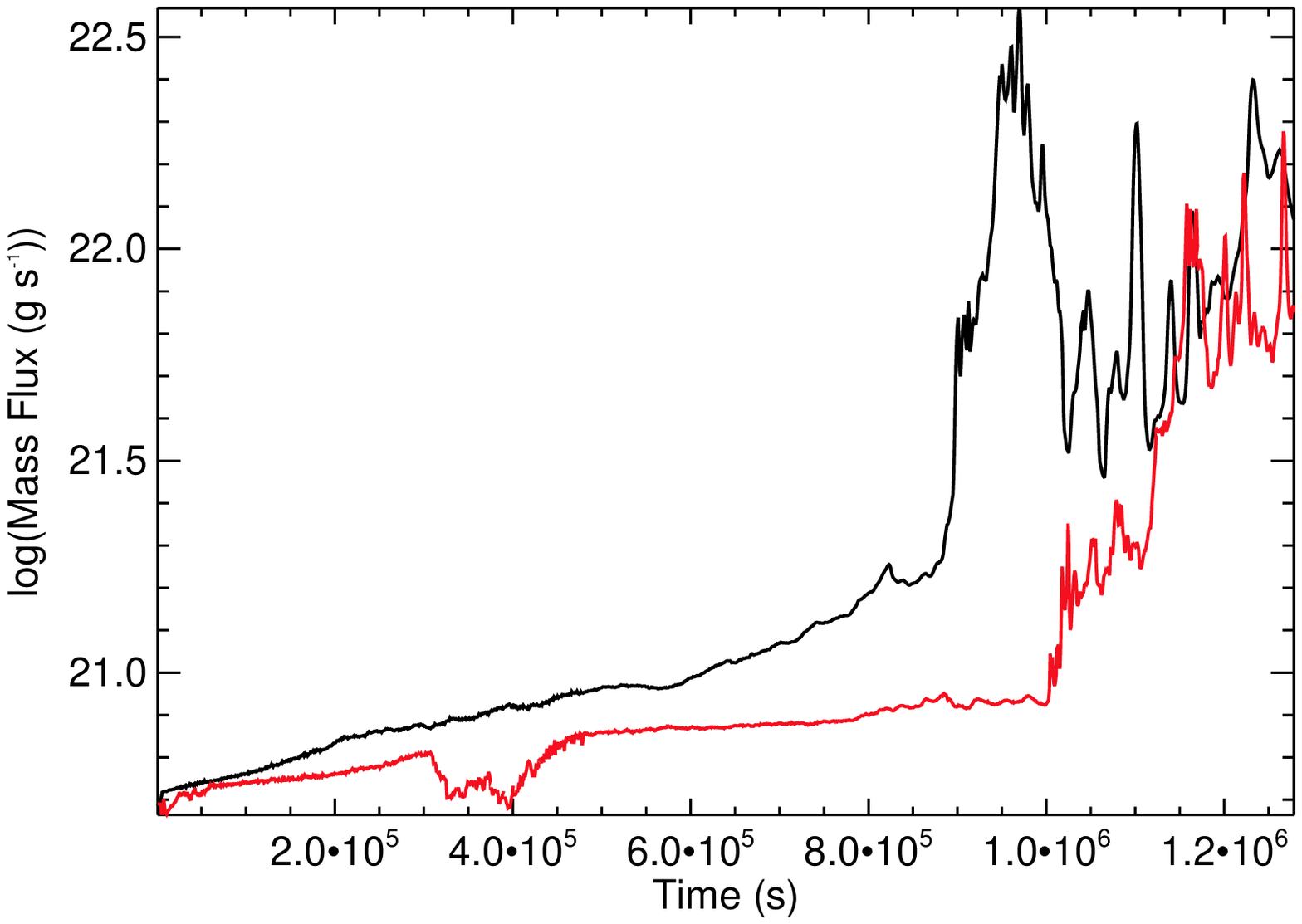}
  \includegraphics[clip,angle=0,width=0.32\textwidth]{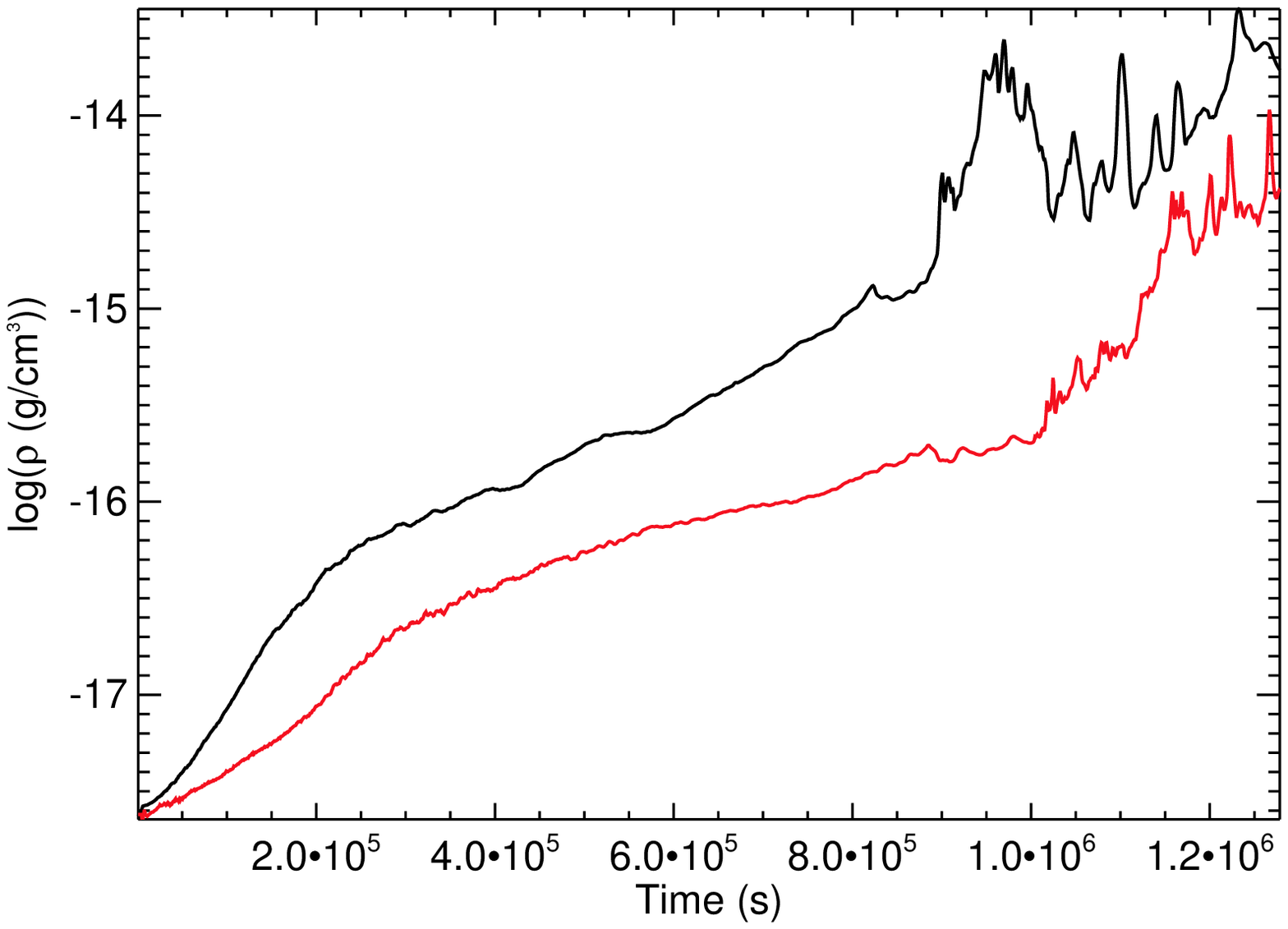}
  \includegraphics[clip,angle=0,width=0.32\textwidth]{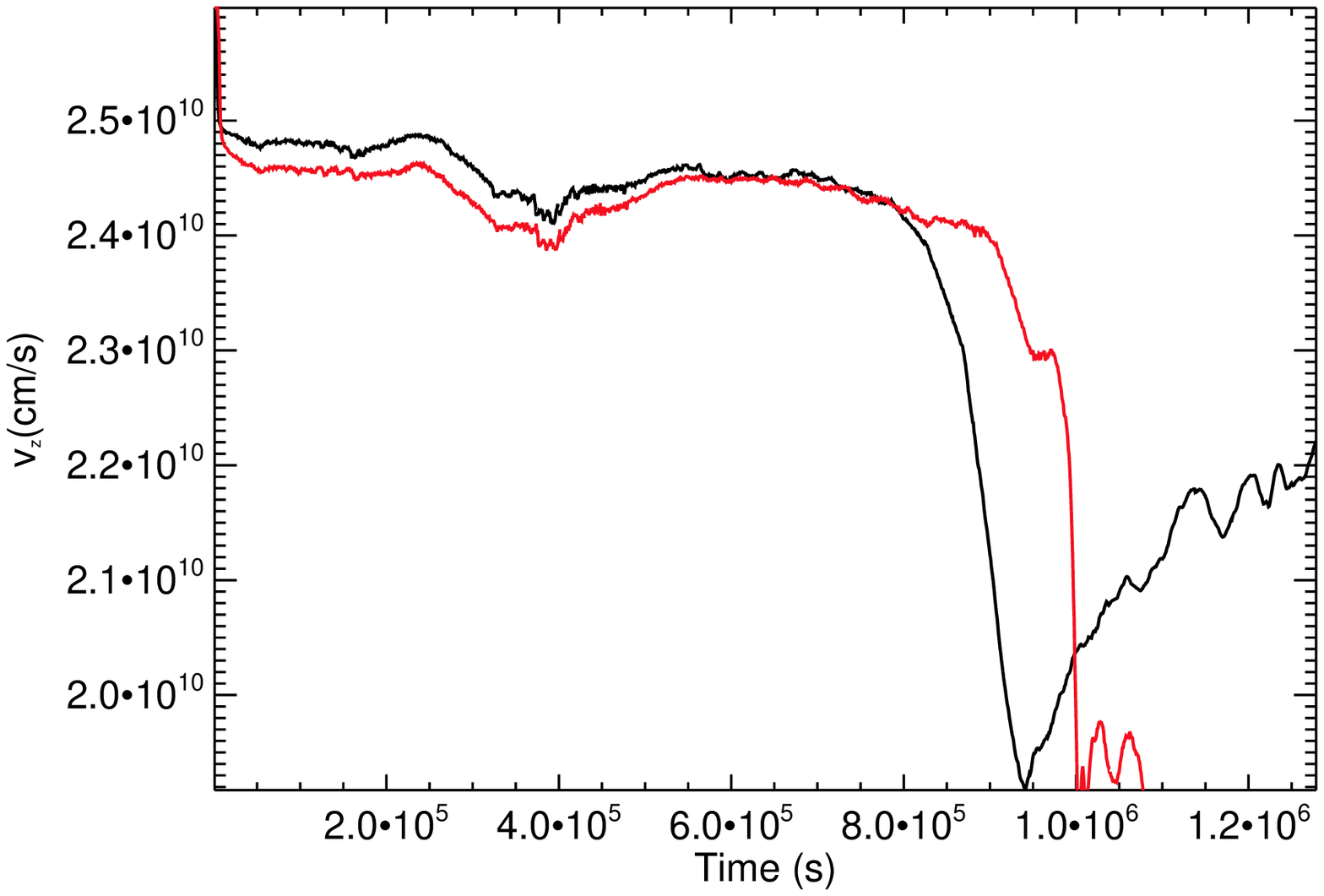}
  \caption{Mass flux (left), mean density (central), and mean velocity (right) versus time in simulation S3 at 
$z=7.5\times10^{13}$ 
(half grid, black lines) and $1.5\times10^{14}\,{\rm cm}$ (end of the grid, red lines).}
  \label{fig:S3.2}
  \end{figure*}

Figure~\ref{fig:S3.1} shows the mass flux across the simulated grid, and the mean obstacle gas velocity and density, versus distance along the axis at a time similar to that in
Fig.~\ref{fig:S2.1}. We have omitted the luminosity plot because it shows the same behavior in both S2 and S3. However, at the time of maximum obstacle expansion in S3, i.e. $t\approx 7.9\times 10^5$~s, the reprocessed luminosity is smaller than S1 at $t\approx 1.15\times 10^6$~s, but significantly larger than in S2. 
Regarding the mass flux, we observe that it keeps smaller relative (and absolute)
values in S3 than in S2. The tail is narrower in S3 than in S2. The mean obstacle gas density along the tail has similar values in both simulations, as well as the temperature. However, the
mean obstacle gas velocity presents larger velocities by a factor of 2 in S3 than in S2. The smaller value of the obstacle mass-loss during the initial phase could be responsible for this
difference, as indicated by the mean velocity of the obstacle gas right at the region of interaction, which is a factor two larger in S3 than in S2.

Figure~\ref{fig:S3.2} shows the mass flux, mean density and mean velocity at $z=7.5\times 10^{13}$ and $1.5\times 10^{14}$~cm versus time. Note that a direct comparison between S2 and S3
here cannot be done, as the physical cross section of the numerical grid is different in both simulations. However, Fig.~\ref{fig:S3.2} already gives interesting information regarding the
mass-load process. In S3, the mass flux, which should be similar to that in S2 when dominated by the outer material of the obstacle, presents smaller values during the initial quasi-steady phase and larger
values  when the aforementioned large clump is detached from the core. 
This implies that the obstacle mass-loss is slower in S3 during the initial phase and, thus, after 
the obstacle has been completely crossed by the initial shock, there is still a significant amount of material attached to the core. Unlike S3, in S2 the obstacle mass-loss is faster initially, and when
the obstacle is completely shocked, the smaller amount of gas left in the outer layers, when detached, is not enough to increase the obstacle cross-section.
To which extent a larger increase in the resolution could even change the results further in the same direction should be studied.  

The more efficient acceleration found in S3 and the higher values of the mass flux relies on the expansion of the obstacle gas after $t\simeq4\times10^5\,$s, as in S1. The expansion of the
obstacle gas in S3 is very different from that of S2, as clearly seen comparing, e.g., the middle right panel in Fig.~\ref{fig:maps2.1} and the bottom left one in Fig.~\ref{fig:maps3.1}.
When part of the expanded material in S3 enters the rarefied region surrounding the cylindrical tail behind the obstacle, it is accelerated, reaching mean velocities $v_{\rm
c}\simeq10^9\,{\rm cm/s}$. In S2, the expansion of the clump is not as important, so the pressure-driven acceleration has a smaller effect.

\subsection{Mass-loading}\label{ml}

On the axis, the acceleration of the obstacle material is basically linear with distance during the initial phase in all three simulations. A simple extrapolation of the linear growth
results in $1.25\times10^{16}\,{\rm cm}$ as the shortest distance to accelerate the obstacle gas to $\sim v_{\rm j}$ in S1. Taking the mean obstacle gas velocity, this distance becomes a
factor of ten larger. In the case of S2 and S3, the extrapolation of the linear acceleration on the axis implies  $6.5\times10^{15}\,{\rm cm}$, which becomes $3.25\times10^{16}\,{\rm cm}$
for S2 and $1.3\times10^{16}\,{\rm cm}$ for S3, taking mean obstacle gas velocities. However, we have seen that this process is critically dependent on the way in which the mass ablation is
produced. If ablation mass-load is produced at a smaller rate, the acceleration is more efficient for the ablated gas. In addition, if the extracted gas clumps widen enough to face the
rarefied jet region, the obstacle material can be accelerated by pressure gradients up to a 5--10\% of the jet velocity already at $\sim 10^{14}\,{\rm cm}$, as for S1 and S3. For S2, the
obstacle gas does not expand enough to favor acceleration in the rarefied jet gas region. This material is confined around the symmetry axis, although this seems to be an effect of
numerical dissipation. We remark that in either case, the dragged matter reaches $z$-velocities $>v_{\rm o}/\chi$ soon after being shocked, which confirms that it will remain in the jet as long
as $t_{\rm d}$ is at least few times shorter than $t_{\rm j}$ (see Section~\ref{phys}). 

Efficient mixing of shocked jet and obstacle material could enhance acceleration (as in S1 and S3), significantly reducing the distance needed by the entrained gas to reach the jet speed.
When the obstacle gas expands due to heating, efficient mixing may happen as a consequence of the growth of instabilities that disrupt the obstacle (S1), or due to the obstacle material expanding into
a region occupied by shocked jet flow (S3). However, if the mass-loss from the obstacle is too slow and/or the obstacle gas does not expand enough, all the ablated gas will flow through the
tail. In this case, instabilities in the layer between the tail and the shocked jet flow would be necessary to favor mixing and subsequent acceleration of the obstacle gas. On the one hand,
the tail is dense and surrounded by the rarefied shocked jet gas, which is a very stable configuration in terms of Kelvin-Helmholtz or shear-layer instabilities. This is so because of
the large relative inertia that the tail has under these conditions. On the other hand, any asymmetry in the medium surrounding the tail, which can be given by inhomogeneities in the jet or
simply by the obstacle transverse motion, will result in the triggering of helical perturbations that are very disruptive. The latter cannot be observed in 2D axisymmetric simulations as
those presented here, but should appear in 3D simulations. In conclusion, the distances given in the previous paragraph for obstacle material acceleration should be regarded as upper
limits. The fact that they are of the order of the interaction $z_{\rm o}$ in this work, $\sim 10^{16}$~cm, likely implies that the jet will not get significantly diluted during the process and will be able to fully incorporate the obstacle matter.

The total mass flux in the simulated jets (considering the total jet cross-section) is $\dot{M}\simeq 10^{23}\,{\rm g/s}\simeq10^{-3}\,{\rm M_\odot/yr}$. We have seen that, typically, the
mass flux during the initial phase is $\dot{M}\simeq8\times10^{20}\,-\,3\times10^{21}\,{\rm g/s}$, which represents 0.8-3\% of the jet mass flux. However, during the development of
instabilities and strong mass-loss, we found peaks of mass flux of  $\dot{M}\leq3\times 10^{22}\,{\rm g/s}$, i.e., 10-30\% of the total jet flux. These episodes of dramatic mass-load could
generate important deceleration and anisotropies in the jet flow (see also Sect.~\ref{ml2}). 

\section{Discussion}\label{disc}

\subsection{Mass-loading}\label{ml2}

Jet formation likely involves a magnetocentrifugal mechanism in the ergosphere of a Kerr black hole, with $e^\pm$-pairs as the jet matter content \citep[e.g.][]{ruf75,bla77,pun90,bip92,kom04}.
Despite of being accretion-sustained, the magnetic field in the black-hole ergosphere has a monopole-like structure that prevents accreted baryonic matter from entering into the jet.
However, the jet may be already matter dominated at some $10^3$~$R_{\rm Sch}$ from the black hole \citep[e.g.][]{sik05},  probably with an important content of protons \citep[e.g.][]{cel08}.
In addition, the jet velocity, expected to increase in the jet acceleration region \citep[e.g.][]{kom07,kom09}, may decrease at higher values of the axial coordinate \citep[e.g.][]{hom09}. All
this can be explained by a jet initially dominated by its Poynting flux, accelerated afterwards by magnetic-to-kinetic energy transfer, and finally slowed down due to mass-loading. 

Currently, the dominant mechanism and location of jet mass-loading are unknown. As shown in Sect.~\ref{ml}, clouds and RG can effectively introduce baryonic matter into the jet innermost
regions, with the latter acting as a sink (as long as it is not completely quenched by mass-loading; e.g. \citealt{hub06}). A rough estimate of $\dot{M}$ can be done for the RG and BLR
scenarios. In the former, adopting $z_{\rm o}\sim 10^{17}$~cm, one may expect about one RG per year crossing the jet, delivering $10^{28}-10^{29}$~g \citep{bar10}, i.e. $\dot{M}\sim
0.01-0.1\,\dot{M}_{\rm cr}$ (see Sect.~\ref{phys}) for the jet considered in this work. In the BLR scenario, assuming a BLR filling factor $\sim 10^{-6}$ and $\rho_{\rm o}\sim 1.7\times
10^{-14}$~g \citep[see, e.g.,][]{ara10}, one gets $\dot{M}\sim 0.1\,\dot{M}_{\rm cr}$. For a leptonic lighter jet than considered here, with equal $L_{\rm j}$ but say $\Gamma_{\rm j}\sim
10$, then $\dot{M}\rightarrow \dot{M}_{\rm cr}$. 
Mass-loading can be important for more powerful jets if $z_{\rm o}$ is larger, since $\dot{M}\propto z_{\rm o}^2$ (although this effect is partially compensated by the fact that $\dot{M}_{\rm cr}\propto L_{\rm j}$).
Note that in the case of stars, and neglecting the stellar wind-channel for mass-loading (i.e. RG or massive stars), $z_{\rm o}$ is limited by the tidal or the nuclear radiation intensity (see Sect.~\ref{intro}).
We conclude that, although a case-by-case analysis would be required, it seems clear that baryonic loading may be already important in AGN jets on pc scales.

\subsection{High-energy emission}\label{he}

One obstacle covers a fraction of the jet cross section; about $(R_{\rm o}/R_{\rm j})^2$ ($=\xi$) of the jet magnetic and kinetic energy can be converted into internal energy, and a
fraction of it, into non-thermal energy. The production of non-thermal accelerated particles in the considered regions can occur in shocks, turbulence, sheared flows or current sheets with
reconnecting magnetic fields \citep[e.g.][]{ree78,rie04,tam09,lyu10}. These non-thermal particles, either protons or electrons, can interact with the ambient magnetic, radiation and matter
fields, producing high-energy emission. For protons, and in the context of jet-obstacle interactions, possible high-energy radiation channels are proton-proton ($pp$) collisions in the entrained matter, photomeson production if dense
photon fields are present, and proton synchrotron for high magnetic fields and proton energies \citep[e.g.][]{dar97,ara10,bar10,bar11}. For electrons, relevant mechanisms would be
synchrotron radiation at low energies, and synchrotron self- (SSC) and external Compton (EC) at high energies \citep[e.g.][]{bed97,ara10}. Secondary leptons, electrons and positrons (and
also neutrinos) will result from pp interactions and photomeson production. At the jet base, the ambient photon fields can be also dense enough to absorb gamma rays, which will also result
in $e^\pm$-pair production. Therefore, processes of radiation attenuation and reprocessing cannot be neglected \citep[e.g.][]{bla95,bed97,aha08}, except for underluminous sources
\citep[e.g.][]{rie08,ara10,bar10}. 

The simulations presented here show that for relatively homogeneous obstacles, the fraction $\xi$ can grow by more than one order of magnitude in few $t_{\rm d}$ ($\sim 10^5-10^6$~s),
before the obstacle total destruction and mixing with the jet. It is seen in particular in Fig.~\ref{fig:S1.3}, in which about almost all the jet power within the grid is reprocessed,
whereas the initial $\xi$-value was only $\sim 2.5\times 10^{-3}$ (homogeneous grid). This was already predicted for instance in \cite{bar10}, and implies that the $pp$ emissivity can grow
very quickly while the obstacle $pp$ cooling time is shorter than the dynamical time, assuming effective confinement of protons within the obstacle. When this cooling time becomes longer
than the dynamical time, $pp$ emissivity sharply decreases. This mechanism has been proposed to explain the day-scale TeV flares observed in M87 \citep[see][]{bar10}. 

\begin{figure}[!t]
  \includegraphics[clip,angle=270,width=0.5\textwidth]{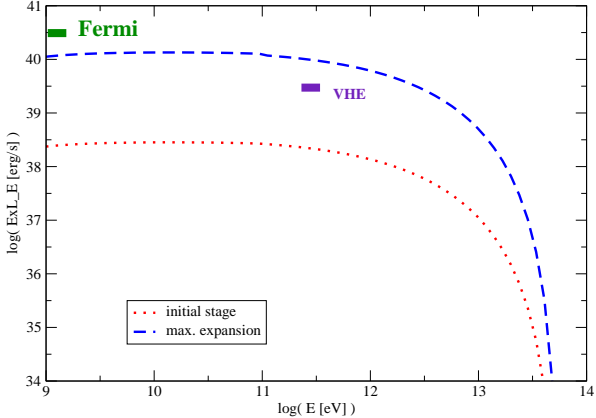}
  \caption{Spectral energy distribution beyond 1~GeV produced by relativistic protons accelerated in the bow shock and interacting with the obstacle gas via $pp$. The acceleration rate has been fixed to $0.01\,qBc$. The approximate sensitivities of {\it Fermi} and ground-based VHE instruments, for a source at $\sim 20$~Mpc, are also shown for comparison.}
  \label{pp}
  \end{figure}

\begin{figure}[!t]
  \includegraphics[clip,angle=0,width=0.5\textwidth]{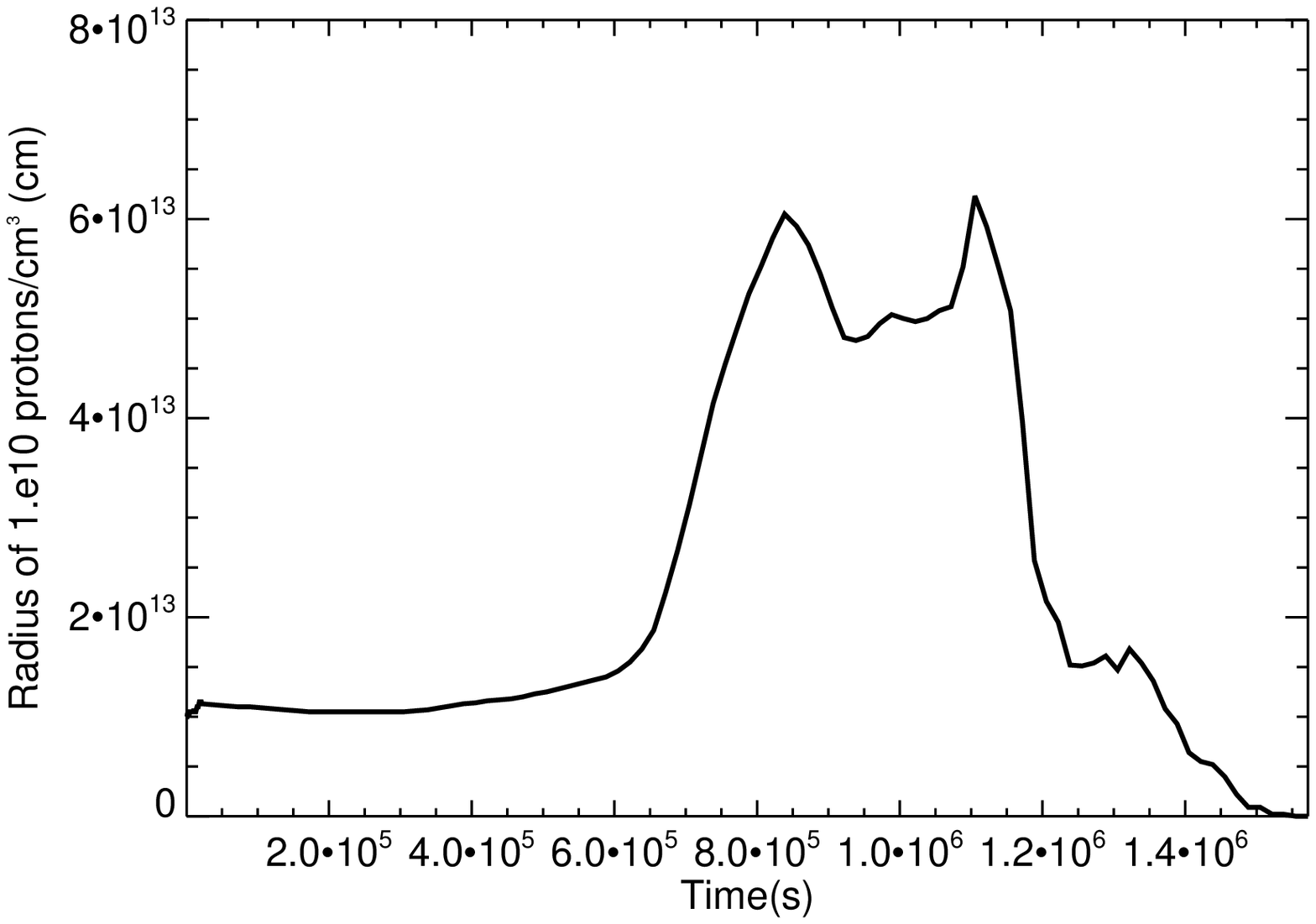}
  \includegraphics[clip,angle=0,width=0.5\textwidth]{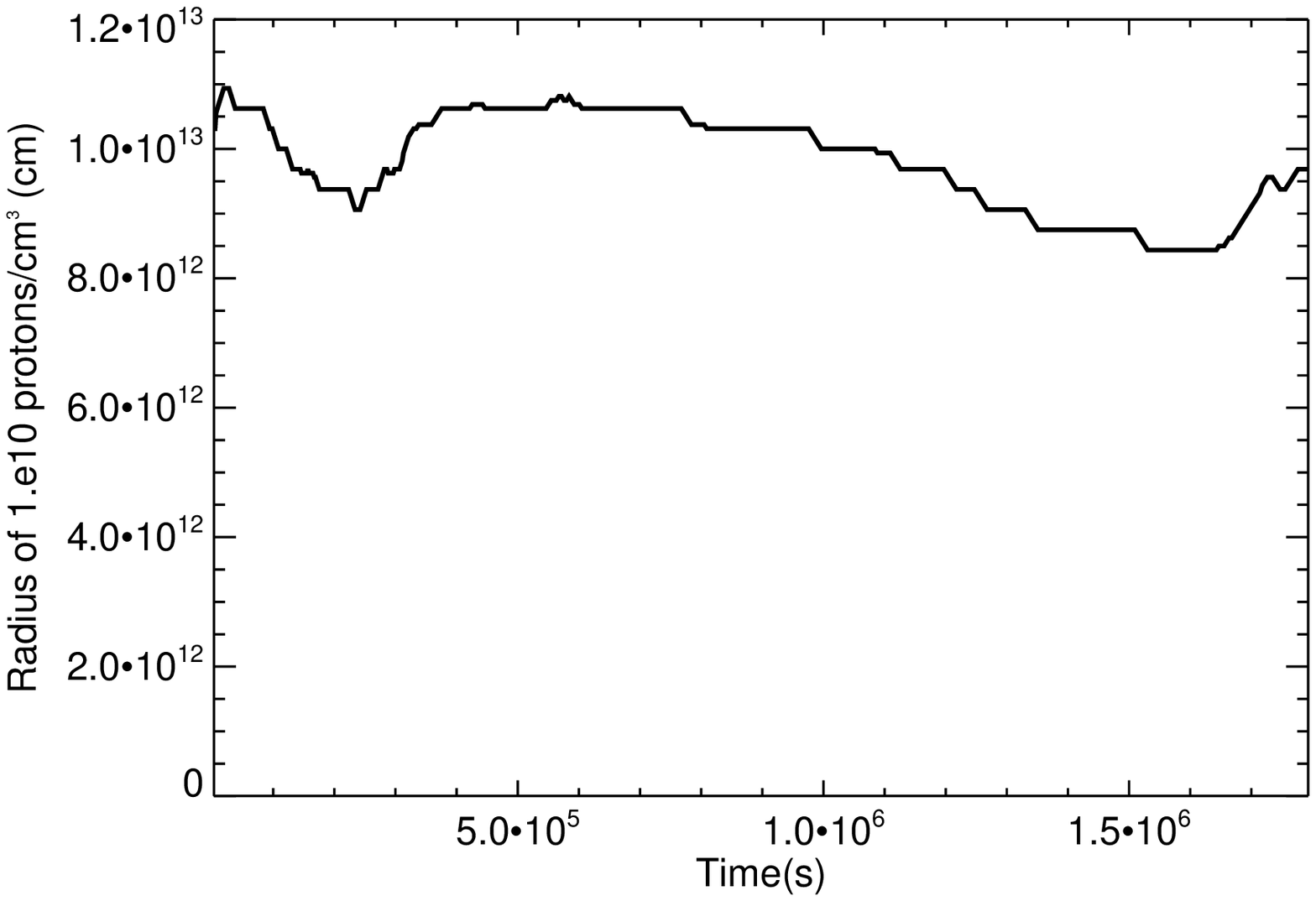}
  \includegraphics[clip,angle=0,width=0.5\textwidth]{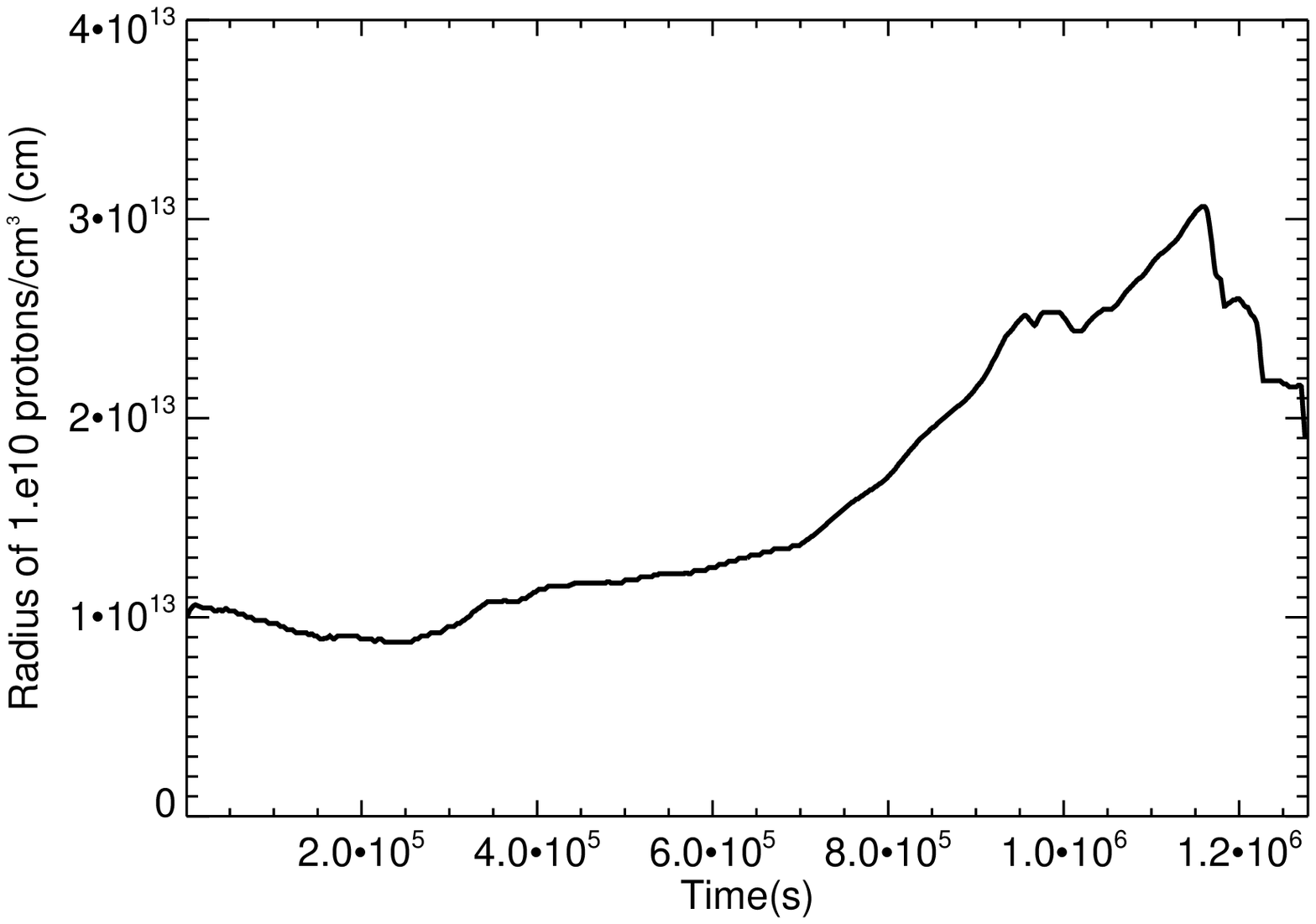}
  \caption{Evolution of the obstacle maximum radius enclosing densities $\ge 10^{10}$~cm$^{-3}$ (at least in some regions), for which the $pp$-cooling time is of the order of the dynamical time for S1 (top), S2 (middle) and S3 (bottom).}
  \label{pp2}
  \end{figure}

Figure~\ref{pp} shows the expected $pp$ gamma-ray spectrum for the times corresponding to Figs.~\ref{fig:S1.1} and \ref{fig:S1.3}, i.e. the beginning and the $\xi$-peak phases, assuming
that 10\% of the processed jet power (the internal energy luminosity) in the grid goes to relativistic protons that interact with the shocked obstacle gas. As seen in the figure, the
luminosity may grow by almost two orders of magnitude. In Fig.~\ref{pp2}, we show the evolution of the obstacle maximum radius enclosing densities $\ge 10^{10}$~cm$^{-3}$ (at least in some
regions), values for which the $pp$-cooling time is of the order of the dynamical time. For simplicity, we have taken a reference density of $10^{10}$~cm$^{-3}$, which yields a cooling time
$\sim 10^{5}$~s. The figures clearly show the difference in obstacle size evolution for the different cases discussed in the previous section, with a growth by a factor of six for S1,
a factor of three for S3, and no growth for S2. For S1 and to some extent S3, the results are consistent with those presented in \cite{bar10}, which were obtained using an analytical
approach for the obstacle dynamics. A more detailed account of the high-energy emission produced in RG-jet interactions will be presented elsewhere.

Scenarios different from $pp$, like proton synchrotron emission, SSC and EC, have been also considered to explain flaring as well as persistent emission from misaligned AGN and blazars, in
the context of jet-obstacle interactions \citep[e.g.][]{bed97,ara10,bar11}. In the case of proton synchrotron and EC, and also in photomeson production with external fields, the decay phase
of a flare would be longer than in the $pp$ and SSC scenarios (as long as particle acceleration runs). This is so because the former two processes are not so sensitive as the latter ones to
the obstacle size, since the magnetic and external radiation fields may keep relatively constant during the event. 

Interestingly, for power-law density profiles the obstacle ablated material is partially carried away in the form of a narrow tail behind a core, e.g. the RG itself. This, although to some
extent an effect of the simulation resolution, indicates that in the RG case, the degree of detachment and homogeneity of the external layers of the star can affect the value of $\xi$, and
thereby, the non-thermal luminosity. 

The quick obstacle acceleration by the jet can also have strong observational implications, separating two clear phases, one in which the emission is rather isotropic, during the first few
$t_{\rm d}$ \citep[e.g.][]{ara10,bar10}, and one in which the obstacle remains move relativistically but jet-driven bow shocks still surround them. In the latter situation, beaming effects
would play a role, making this phenomenon relevant also in the blazar context \citep[e.g.][]{bar11}.  

Regarding a persistent component from interactions of many obstacles with the jet, its maximum luminosity can be estimated from $N_{\rm o}\,<(R_{\rm o}/R_{\rm j})^2>\,L_{\rm j}\le L_{\rm
j}$, where $N_{\rm o}\,<(R_{\rm o}/R_{\rm j})^2>$ would be a typical jet cross section fraction covered by all the obstacles within $z_{\rm o}$ \cite[e.g.][]{ara10}. We note that RG or
occasional large clouds will be relatively infrequent, and thus contribute little to this quantity in the regions of interest (despite their occasional bright appearance). 

\section{Summary}

Dense clumps and stars are expected to interact with/enter into the jets in the crowded and inhomogeneous central regions of AGN. Such events are likely to have strong consequences in the
content, hydrodynamics and radiation of AGN jets. For typical jet luminosities and characteristics of the obstacle (BLR cloud, RG, etc.), significant amounts of mass are quickly dragged
downstream, mixing with the jet flow. The mass injected into the jet can load it with baryons and reduce its Lorentz factor already on pc scales. The fast growth of the obstacles due to
heating leads to a significant increase of the interaction cross section, enhancing the potential conversion of energy from the jet to non-thermal particles. This growth of the obstacle
cross section strongly feeds back itself, and potentiates disruption. The acceleration of the obstacle by the jet, due to which the speed of the former approaches $v_{\rm j}$ after several
$t_{\rm d}$, changes the beaming pattern of the emission with time. Transient and persistent activity, related to different types of obstacles and their evolution within the jets, yields a
rich and complex picture for the high energy emission of AGN, either misaligned or of the blazar-type.

\begin{acknowledgements}
The authors want to thank the referee, Jim Beall, for his open and encouraging attitude towards our work.
The research leading to these results has received funding from the European
Union Seventh Framework Program (FP7/2007-2013) under grant agreement
PIEF-GA-2009-252463. V.B.-R. acknowledges support by the Spanish 
Ministerio de Ciencia e Innovaci\'on
(MICINN) under grants AYA2010-21782-C03-01 and FPA2010-22056-C06-02.  
MP acknowledges
support by the Spanish ``Ministerio de Ciencia e Innovaci\'on''
(MICINN) grants AYA2010-21322-C03-01, AYA2010-21097-C03-01 and
CONSOLIDER2007-00050.
We acknowledge the use of the cluster Tirant, as well as the \emph{Servei d'Inform\`atica} of the \emph{Universitat de Val\`encia},
for the computational time allocated for the simulations. 

\end{acknowledgements}

\end{document}